\documentclass{aa} 
\usepackage{natbib}
\usepackage{graphicx}
\usepackage{rotating}
\usepackage{txfonts}
\usepackage{ulem}
\usepackage{threeparttable}
\usepackage{mathrsfs}
\usepackage{amsmath}
\usepackage{comment}
\usepackage{longtable}
\usepackage{hyperref}
\hypersetup{
    colorlinks=true,
    linkcolor=blue,
    citecolor=blue,
    urlcolor=green,
    }

\begin{document}

\title{The \textit{Planck} clusters in the LOFAR sky}
\subtitle{II. LoTSS-DR2: Recovering diffuse extended emission with LOFAR}

   \author{L. Bruno 
        \inst{1,2},
        G. Brunetti
        \inst{1},
        A. Botteon
        \inst{3,2,1},
        V. Cuciti
        \inst{4,1},
        D. Dallacasa
        \inst{2,1},
        R. Cassano
        \inst{1},
        R. J. van Weeren
        \inst{3},
        T. Shimwell
        \inst{5,3},
        G. Taffoni
        \inst{6},
        S. A. Russo
        \inst{6},
        A. Bonafede
         \inst{2,1},
        M. Br\"uggen
        \inst{4},  
        D. N. Hoang
        \inst{4},
        H. J. A. Rottgering
        \inst{3},
        C. Tasse
        \inst{7,8}
}

   \institute{Istituto Nazionale di Astrofisica (INAF) - Istituto di Radioastronomia (IRA), via Gobetti 101, 40129 Bologna, Italy
   \and
   Dipartimento di Fisica e Astronomia (DIFA), Universit\`a di Bologna, via Gobetti 93/2, 40129 Bologna, Italy
             \and
    Leiden Observatory, Leiden University, PO Box 9513, 2300 RA Leiden, The Netherlands
             \and
             Hamburger Sternwarte, Universit\"at Hamburg, Gojenbergsweg 112, D-21029 Hamburg, Germany
             \and
             ASTRON, Netherlands Institute for Radio Astronomy, Oude Hoogeveensedijk 4, 7991 PD, Dwingeloo, The Netherlands
           \and
           Istituto Nazionale di Astrofisica (INAF) - Astronomical Observatory of Trieste, Trieste, Italy
           \and
       GEPI \& USN, Observatoire de Paris, Université PSL, CNRS, 5 Place Jules Janssen, 92190 Meudon, France and 
        \and 
   Department of Physics and Electronics, Rhodes University, PO Box 94, Grahamstown, 6140, South Africa
             \\
\email{luca.bruno4@unibo.it}
}


 
  \abstract
   {Extended radio sources in the sky require a dense sampling of short baselines to be properly imaged by interferometers. This problem arises in many areas of radio astronomy, such as in the study of galaxy clusters, which may host Mpc-scale diffuse synchrotron sources in the form of radio halos. In clusters where no radio halos are detected, owing to intrinsic absence of emission or extrinsic (instrumental and/or observational) effects, it is possible to determine upper limits.}
   {We consider a sample of \textit{Planck} galaxy clusters from the Second Data Release of the LOFAR Two Meter Sky Survey (LoTSS-DR2) where no radio halos are detected. We use this sample to test the capabilities of LOFAR to recover diffuse extended emission and derive upper limits.}
   {Through the injection technique, we simulate radio halos with various surface brightness profiles. We then predict the corresponding visibilities and image them along with the real visibilities. This method allows us to test the fraction of flux density losses owing to inadequate \textit{uv}-coverage and obtain thresholds at which the mock emission becomes undetectable by visual inspection. }
   {The dense \textit{uv}-coverage of LOFAR at short spacings allows to recover $\gtrsim90\%$ of the flux density of targets with sizes up to $\sim 15'$. We find a relation that provides upper limits based on the image noise and extent (in terms of number of beams) of the mock halo. This relation can be safely adopted to obtain upper limits without injecting when artifacts introduced by the subtraction of the discrete sources are negligible in the central region of the cluster. Otherwise, the injection process and visual inspection of the images are necessary to determine more reliable limits. Through these methods, we obtain upper limits for 75 clusters to be exploited in ongoing statistical studies.}
   {}

   \keywords{radiation mechanisms: non-thermal  -- galaxies: clusters: general -- instrumentation: interferometers}
   
\titlerunning{The \textit{Planck} clusters in the LOFAR sky (II)}
\authorrunning{Bruno et al.}
   \maketitle
%

\section{Introduction}

Radio sources with large extents in the sky may be not properly recovered by radio interferometers with an insufficient number of short baselines. A poorly sampled \textit{uv}-coverage at short spacings will thus cause unavoidable flux density losses \citep[e.g.][]{wilner&welch94,deo&kale17}. This problem arises in many fields of radio astronomy, such as the study of extended radio galaxies, supernova remnants, and diffuse emission in galaxy clusters. 

Among the various diffuse synchrotron sources found in galaxy clusters, radio halos are the most extended ones. They are centred on the cluster core region, extend up to Mpc scales roughly following the distribution of the intra-cluster medium (ICM), and are characterised by steep ($\alpha \gtrsim 1$) radio spectra\footnote{We define the spectral index $\alpha$ through $S_{\rm \nu}\propto \nu^{-\alpha}$, where $S_{\rm \nu}$ is the radio flux density at the frequency $\nu$.} \citep[][for a review]{vanweeren19}. The origin of radio halos is associated with the re-acceleration of (primary or secondary) particles via stochastic Fermi II-type processes; these are driven by turbulence in the ICM induced by cluster mergers \citep{brunetti01,petrosian01,brunetti&lazarian07,beresnyak13,miniati15,brunetti&lazarian16}. Nevertheless, the complex energy transfer mechanisms operating from large ($\sim$Mpc) to smaller scales are still unclear \citep[see][for a review]{brunettijones14}.

Radio halos are not ubiquitous in galaxy clusters. They are mainly found in massive\footnote{$M_{500}$ is the mass within $R_{500}$, which is the radius enclosing $500\rho_{\rm c}(z)$, where $\rho_{\rm c}(z)$ is the critical density of the Universe at a given redshift.} ($M_{\rm 500} \gtrsim 5\times10^{14} \; M_\odot$) and dynamically disturbed clusters, with increasing detection fraction with the host mass \citep[e.g.][]{cassano13,kale13,cuciti15,cuciti21b}. The non-detections of radio halos may be intrinsically due to absence of radio emission (off-state clusters), or result from extrinsic instrumental and/or observational limits, namely a combination of insufficient sensitivity and poor sampling of short baselines. 

In galaxy clusters where no radio emission is detected, upper limits on the radio power of a possible halo can be determined. Upper limits are important in statistical studies to constrain theoretical models of formation and evolution of the diffuse sources \citep[e.g.][]{brunetti07}. In this respect, upper limits are necessary to understand whether off-state and on-state clusters belong to two distinct populations (i.e. relaxed and disturbed, respectively), and thus obtain constraints on the origin of the non-thermal emission. To this aim, it is necessary to compare the radio powers of detected halos and limits as a function of the mass of the host cluster \citep[e.g.][]{cassano13,cuciti21b}. Moreover, deep upper limits can be exploited to test the level at which purely hadronic \citep{dennison80,blasi&colafrancesco99,dolag&ensslin00} radio halos may be detected or test models of dark matter interactions \citep{storm17}. Various methods have been adopted to determine upper limits. The most widely used method follows the `injection' technique first exploited for radio halos by \cite{brunetti07} and \cite{venturi08}. It consists of modelling simulated radio halos, whose predicted visibilities are added to the observed ones, and then regularly processed to obtain images to check for possible detections at a given flux density threshold. Since then, this technique has been commonly adopted on data of galaxy clusters from different facilities \citep[e.g.][]{kale13,bonafede17,johnston-hollitt17,cuciti21a,george21,osinga21,duchesne22}.

LOFAR (Low Frequency Array) is currently mapping the whole northern sky with unprecedented sensitivity and resolution at low frequencies through the LOFAR Two-Meter Sky Survey \citep[LoTSS;][]{shimwell17} with the High Band Antenna (HBA) operating at 120-168 MHz. Observations of galaxy clusters with LOFAR are promising to detect new radio halos, which are brighter at low frequencies owing to their steep synchrotron spectrum. The region of the sky covered by the Second Data Release of LoTSS \citep[LoTSS-DR2;][]{shimwell22LoTSS}  includes 309 galaxy clusters in the Second Planck Sunyaev-Zel'dovich (PSZ2) catalogue \citep{planckcollaborationB16}, thus providing the largest sample of mass-selected clusters observed at low radio frequencies to date. This sample is extensively described in \cite{botteon22LoTSS}, showing the large variety of diffuse radio sources that were found\footnote{\url{https://lofar-surveys.org/planck_dr2.html}}. In the present work, we focus on the sub-sample of 140 clusters where diffuse emission was not detected. By means of the injection technique, we generate simulated radio halos to test the capabilities of LOFAR to recover diffuse extended emission and obtain upper limits on the radio power of these clusters. These upper limits will be exploited in ongoing statistical analyses \citep[][Cuciti et al., in prep.]{zhang22,cassano23}, where the properties of clusters with detected or undetected radio halos will be compared to investigate the origin of these populations.

This paper is organised as follows: in Sect. 2 we summarise the processes adopted to calibrate the LoTSS-DR2 data of \textit{Planck} galaxy clusters that we will exploit in this work. In Sect. 3 we describe the procedures of injection. In Sect. 4 we simulate mock halos to test the capabilities of LOFAR to recover extended diffuse emission. In Sect. 5 we discuss the methods to obtain radio upper limits to non-detections  of radio halos in galaxy clusters. In Sect. 6 we carry out simulations on other interferometers to be compared with the performances of LOFAR. In Sect. 7 we summarise our work. We adopt a standard $\Lambda$CDM cosmology with $H_0=70\;\mathrm{km\; s^{-1}\; Mpc^{-1}}$, $\Omega_{\rm M}=0.30$, and $\Omega_{\rm \Lambda}=0.70$.

\section{LoTSS-DR2 data}
\label{sect:LoTSS-DR2 data}

LoTSS-DR2 covers the $27\%$ of the northern sky observed in the range 120-168 MHz (the nominal central frequency is 144 MHz). Each LoTSS pointing is 8 hr long, the typical resolution is $\sim 6''$, and the median noise is $\sim 0.08 \; {\rm mJy \; beam^{-1}}$ \citep{shimwell22LoTSS}.

We refer to \cite{shimwell19LOTSS,tasse21,shimwell22LoTSS} for a complete description of the data processing by means of the Survey Key Project (SKP) pipelines. They include direction-independent and direction-dependent calibration and imaging through {\tt PREFACTOR} \citep{vanweerenDDFACET16,williams16,degasperin19} and {\tt DDF-pipeline}, which makes use of {\tt DDFacet} \citep{tasse18} and {\tt KillMS} \citep{tasse14a,tasse2014b,smirnov15}, and finally the `extraction \& re-calibration' scheme \citep{vanweeren21} to further refine the quality of the images in the direction of the target. The final fully calibrated \textit{uv}-dataset includes only sources within the extraction region, and can thus be easily manipulated in further analysis. Imaging is carried out by excluding baselines shorter than $80\lambda$; these spacings are typically more challenging to calibrate and sample the possible emission on angular scales larger than $\sim 40'$ from our Galaxy that needs to be filtered.

In this work, we focus on a sample of 140 non-detections (`NDE') of diffuse emission from the ICM in \textit{Planck} clusters belonging to LoTSS-DR2 \citep[see][for details]{botteon22LoTSS}. According to our definition, diffuse emission from the ICM does not include emission associated with radio galaxies (e.g. lobes, filaments, tails, fossil bubbles), which may be present in the NDE clusters. We exploit these datasets to test the capabilities of LOFAR to recover extended emission and obtain upper limits to the presence of possible radio halos.

\section{Injection algorithm}
\label{sect:injection algorithm}

\begin{figure*}
	\centering
	\includegraphics[width=0.31\textwidth]{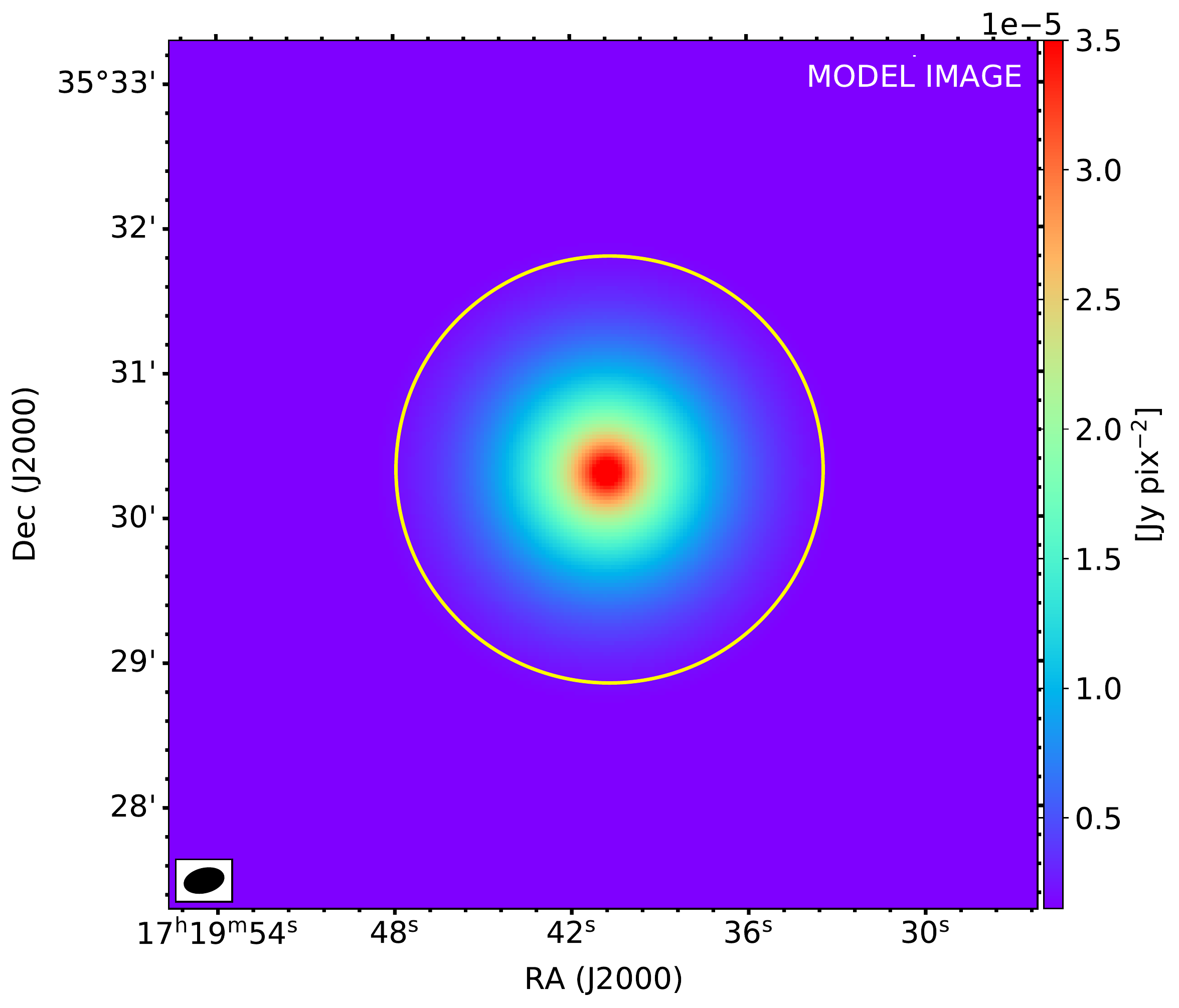}
	\includegraphics[width=0.32\textwidth]{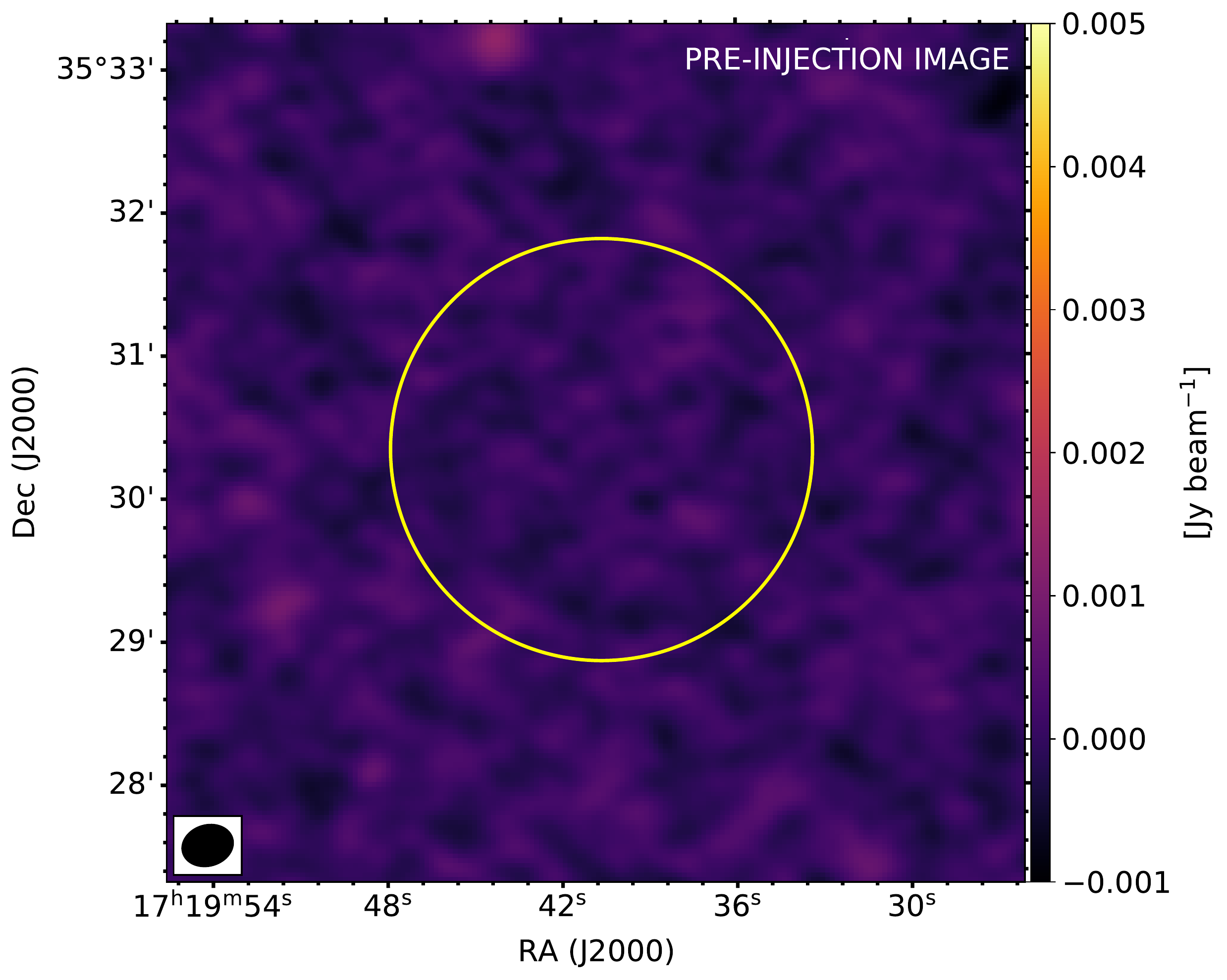}
	\includegraphics[width=0.32\textwidth]{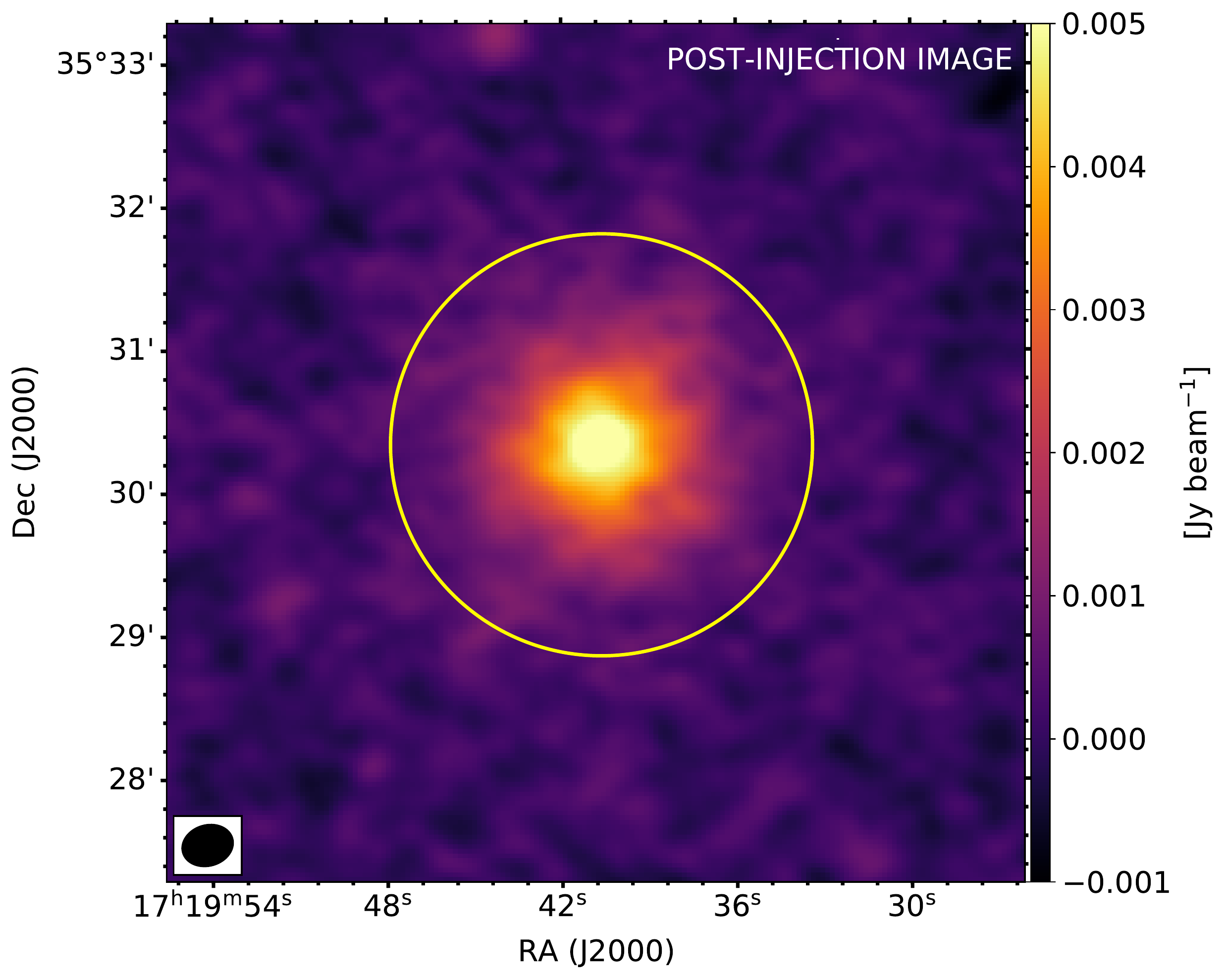}
	\caption{Example of the injection process. The yellow circle is centred on (${\rm RA_{\rm inj}}$, ${\rm DEC_{\rm inj}}$) and has a radius $r=3r_{\rm e}$. \textit{Left}: model image at 144 MHz of the mock halo following the exponential surface brightness profile of Eq. (\ref{SBprofile}). \textit{Middle}: pre-injection image. \textit{Right}: post-injection image. }
	\label{example_MOCK059}%
\end{figure*}   

To test the capabilities of LOFAR and determine upper limits for the NDE clusters, we followed the injection technique. We first derived the visibilities from the Fourier inversion of a set of models of mock halos (`prediction' step) sampling a range of flux densities and angular sizes, and then added these mock visibilities to the dataset of each observation (`injection' step). Imaging and Fourier transforms were carried out by means of {\tt WSClean} v. 2.10 \citep{offringa14,offringa17}. The main steps of the procedure can be summarised as below: 
\begin{enumerate}
\item The coordinates (${\rm RA_{\rm inj}}$; ${\rm DEC_{\rm inj}}$) of the centre of the injection and the total injected flux density at 144 MHz ($S_{\rm inj,tot}$) are required as inputs. 
\item Frequency-dependent model images of the mock halo that follow an exponential surface brightness profile are built (see Sect. \ref{surfacebightnessSECTION}). We assumed a spectral index $\alpha=1.3$ in the frequency range 120-168 MHz, as typical of radio halos. This is computed through the {\tt channels-out 6} parameter in {\tt WSClean}, which produces model images at 6 sub-bands spaced by 8 MHz each. 
\item The model images are Fourier transformed by means of the {\tt predict} function in {\tt WSClean} to obtain the corresponding mock visibilities, which are then added to the real \textit{uv}-data. 
\item The updated (real plus mock visibilities) datasets are imaged by adopting the multi-scale (fixed scales of [0, 4, 8, 16, 32, 64] pixels) and multi-frequency ({\tt channels-out 6}) cleaning algorithm. To enhance the diffuse emission, the baselines are tapered by a Gaussian function ({\tt taper-gaussian}) to lower resolutions.
\end{enumerate}
An example of this process is shown in Fig. \ref{example_MOCK059}, where we injected a bright ($S_{\rm inj,tot}=100$ mJy) mock halo into a source-free region in the field of PSZ2 G059.18+32.91. 

To investigate the capabilities of LOFAR to recover extended emission and derive upper limits, we need to repeat the steps 2-4 by gradually decreasing $S_{\rm inj,tot}$ at step 1 until some given detection criteria (e.g. recovered flux density or sizes, signal-to-noise ratio, inspection by eye) are not fulfilled anymore, as discussed further in Sect. \ref{sect:Upper limit calculation}. Dealing with LOFAR data is resource-intensive and requires proper computing and storage resources \citep[e.g.][]{taffoni22}. We ran all the analyses on the {\tt HOTCAT} High Performance Computing (HPC) cluster at INAF Trieste \citep{bertocco20,taffoni20}, by accessing nodes with 40 CPU cores and 6 GB of RAM each. Interacting with these nodes when running the pipelines is necessary to efficiently inspect intermediate results. To this aim, extra side-work in terms of setting up tunneling and virtual displays forwarding is required; moreover, advanced pipelines (such as those for LOFAR analyses) rely on complex dependencies which often require some system-dependant set up steps. To overcome these issues, we made use of {\tt Rosetta} \citep{russo2022}, a container-centric science platform for interactive data analysis, which has been recently made available for accessing the {\tt HOTCAT} cluster. {\tt Rosetta} can automatically set up interactive analysis environments (such as remote desktops and Jupyter Notebooks) on HPC cluster nodes by using software containers, thus allowing us to interactively access the cluster nodes and efficiently manage the dependencies of our pipelines.

\subsection{Modelling of the surface brightness profile}
\label{surfacebightnessSECTION}

\cite{murgia09} showed that a simple exponential law provides a good representation of the observed surface brightness distribution of a number of radio halos, with few free parameters. Following a common approach adopted for both real and mock halos \citep[e.g.][]{murgia09,bonafede17,boxelaar21,osinga21,botteon22LoTSS,hoang22}, we assumed a spherically-symmetric exponential law to model the surface brightness profiles of our mock halos: 
\begin{equation}
I(r)=I_{0}e^{-\frac{r}{r_{\rm e}}} \; ,
\label{SBprofile}
\end{equation}
where $I_{0}$ is the central surface brightness and $r_{\rm e}$ is the \textit{e}-folding radius.

The flux density is obtained by integrating Eq. (\ref{SBprofile}) in circular annuli up to a certain radius $\hat{r}$:
\begin{equation}
S_\nu=2\pi \int_{0}^{\hat{r}} I(r)rdr=2\pi I_{\rm 0}[r_{\rm e}^2-(\hat{r}r_{\rm e}+r_{\rm e}^2)e^{-\frac{\hat{r}}{r_{\rm e}}}] \; .
\label{flusso-I0}
\end{equation}
Eq. (\ref{flusso-I0}) can be simply expressed as $S_\nu=2\pi f(\hat{r}) I_{\rm 0}r_{\rm e}^2$, where $f(\hat{r})=[r_{\rm e}^2-(\hat{r}r_{\rm e}+r_{\rm e}^2)e^{-\frac{\hat{r}}{r_{\rm e}}}]$ is the fraction of flux density within the integration radius $\hat{r}$ to the total one ($f(\hat{r})=1$ when $\hat{r}=+\infty$). Radio halos do not extend indefinitely, thus their emission is typically measured up to $\hat{r}=3r_{\rm e}$, which provides a fraction $f(\hat{r})=0.8$ of the total flux density. 

For the injection process, we derive the central brightness from the total injected flux density as $I_{0}=S_{\rm inj,tot}/{2\pi r_{\rm e}^2}$. In the following, we will refer to the injected diameter of the mock halo as being $D=6r_{\rm e}$ (i.e. we assume a radius $R=3r_{\rm e}$).

\subsection{Schemes of injection}
\label{sect:Schemes of injection}
\begin{table*}
\centering
	\caption{Comparison between injected and recovered parameters within $3r_{\rm e}$ of the mock halos in Fig. \ref{MOCK120}. Column 1 reports the adopted algorithm scheme discussed in the text. Columns 2, 3 report the location of the injection and the flux density measured in the halo area before the injection ($S_{\rm pre}$). Columns 4, 5, 6 report the injected ($S_{\rm inj}=0.8S_{\rm inj,tot}$), the measured by hand ($S_{\rm meas}$), and the fitted ($S_{\rm fit}$) flux density. Columns 7, 8, 9, 10 report the injected and fitted $I_{\rm 0}$ and $r_{\rm e}$.}
	\label{testMOCK120}
	\begin{tabular}{cccccccccc}
	\hline
	\noalign{\smallskip}
	Scheme  & Location & $S_{\rm pre}$ & $S_{\rm inj}$ & $S_{\rm meas}$ & $S_{\rm fit}$ & $I_{\rm 0, inj}$ & $I_{\rm 0, fit}$ & $r_{\rm e, inj}$ & $r_{\rm e,fit}$  \\  
	& & (mJy) & (mJy) & (mJy) & (mJy) & ($\mu$Jy$\;$arcsec$^{-2}$) & ($\mu$Jy$\;$arcsec$^{-2}$) & (arcsec) & (arcsec)  \\  
	\hline
	\noalign{\smallskip} 
	& &  & 80.0 & $81.1\pm8.8$ & $84.0\pm5.3$ & 9.6 & $10.1^{+0.7}_{-0.7}$ &  & $40.7^{+2.4}_{-2.2}$ \\
	 Inj. & ${\rm Centre}$  & 0.4 & 40.0 & $40.6\pm5.3$ & $42.9\pm4.4$ & 4.8 & $5.2^{+0.6}_{-0.5}$ & 40.7 & $40.7^{+3.9}_{-3.4}$   \\
    & & & 16.0 & $16.6\pm3.8$ & $18.8\pm4.1$ & 1.9 & $2.3^{+0.6}_{-0.5}$ &  & $40.5^{+9.3}_{-7.8}$ \\
    \\
    &	 &  & 80.0 & $79.8\pm8.7$ & $80.0\pm4.4$ & 9.6 & $9.5^{+0.6}_{-0.6}$ &  & $40.7^{+2.0}_{-2.0}$ \\
	Inj. & ${\rm Off}$-${\rm centre}$ &  $-0.8$ & 40.0 & $39.7\pm5.2$ & $39.6\pm3.3$ & 4.8 & $4.6^{+0.4}_{-0.4}$ & 40.7 & $41.2^{+3.2}_{-2.9}$   \\
    & & & 16.0 & $15.7\pm3.7$ & $13.0\pm3.0$ & 1.9 & $2.1^{+0.6}_{-0.5}$ &  & $35.1^{+8.0}_{-7.1}$ \\ \\
    	 & & & 80.0 & $81.7\pm8.8$ & $82.6\pm4.9$ & 9.6 & $10.5^{+0.7}_{-0.7}$ &  & $39.5^{+2.2}_{-2.0}$ \\
	Sub. \& Inj. & ${\rm Centre}$ &  0.7 & 40.0 & $41.0\pm5.2$ & $41.3\pm3.8$ & 4.8 & $5.8^{+0.6}_{-0.6}$ & 40.7 & $37.5^{+3.2}_{-2.9}$   \\
    & &  & 16.0 & $16.8\pm3.7$ & $15.2\pm3.1$ & 1.9 & $3.5^{+1.0}_{-0.7}$ &  & $29.5^{+6.3}_{-5.6}$ \\
    \\
    &	 &  & 80.0 & $79.9\pm8.6$ & $81.0\pm3.9$ & 9.6 & $9.8^{+0.5}_{-0.5}$ &  & $40.5^{+1.7}_{-1.7}$ \\
	Sub. \& Inj. & ${\rm Off}$-${\rm centre}$ &  $-0.8$ & 40.0 & $39.6\pm5.1$ & $39.8\pm3.4$ & 4.8 & $5.0^{+0.5}_{-0.5}$ & 40.7 & $40.0^{+2.9}_{-2.9}$   \\
   & &  & 16.0 & $15.9\pm3.6$ & $13.7\pm2.7$ & 1.9 & $2.6^{+0.6}_{-0.5}$ &  & $32.7^{+6.1}_{-5.6}$ \\
    \\ 
    &	 &  & 80.0 & $81.9\pm8.8$ & $82.9\pm4.9$ & 9.6 & $10.5^{+0.7}_{-0.6}$ &  & $39.7^{+2.2}_{-2.0}$ \\
	Inj. \& Sub. & ${\rm Centre}$ &  0.4 & 40.0 & $40.8\pm5.2$ & $41.0\pm3.9$ & 4.8 & $5.8^{+0.6}_{-0.6}$ & 40.7 & $37.3^{+3.4}_{-3.2}$   \\
   & &  & 16.0 & $16.5\pm3.7$ & $15.1\pm3.1$ & 1.9 & $3.5^{+1.0}_{-0.7}$ &  & $29.3^{+6.6}_{-5.6}$ \\
    \\
    &	 &  & 80.0 & $79.9\pm8.6$ & $80.7\pm4.0$ & 9.6 & $9.9^{+0.5}_{-0.5}$ &  & $40.2^{+1.7}_{-1.7}$ \\
	Inj. \& Sub. & ${\rm Off}$-${\rm centre}$ &  $-0.8$ & 40.0 & $39.7\pm5.1$ & $40.0\pm3.5$ & 4.8 & $4.9^{+0.5}_{-0.5}$ & 40.7 & $40.2^{+3.2}_{-2.9}$   \\
  &  &  & 16.0 & $15.9\pm3.6$ & $14.4\pm2.8$ & 1.9 & $2.4^{+0.6}_{-0.5}$ &  & $34.6^{+6.6}_{-5.9}$ \\
	\noalign{\smallskip} \noalign{\smallskip} \noalign{\smallskip}

	\noalign{\smallskip}
	\hline
	\end{tabular}
   		\begin{tablenotes}
\item    {\small \textbf{Notes}. Errors on $S_{\rm meas}$ are calculated as $\Delta S_{\rm meas}= \sqrt{ \left( \sigma^2 \cdot N_{\rm beam} \right) + \left(  \xi_{\rm cal} \cdot S_{\rm meas} \right) ^2}$, where $\sigma$ is the noise of the restored image, $N_{\rm beam}$ is the number of beams within the considered region, and $\xi_{\rm cal}$ is the flux density scale uncertainty; we adopted $\xi_{\rm cal}=10\%$ \citep{shimwell22LoTSS}. Errors on $S_{\rm fit}$ take into account the fitting uncertainties only, which depend on $\Delta r_{\rm e,fit}$ and $\Delta I_{\rm 0,fit}$.  } 
 \end{tablenotes}	
\end{table*}

\begin{figure*}
    \centering
    \includegraphics[width=0.265\textwidth] {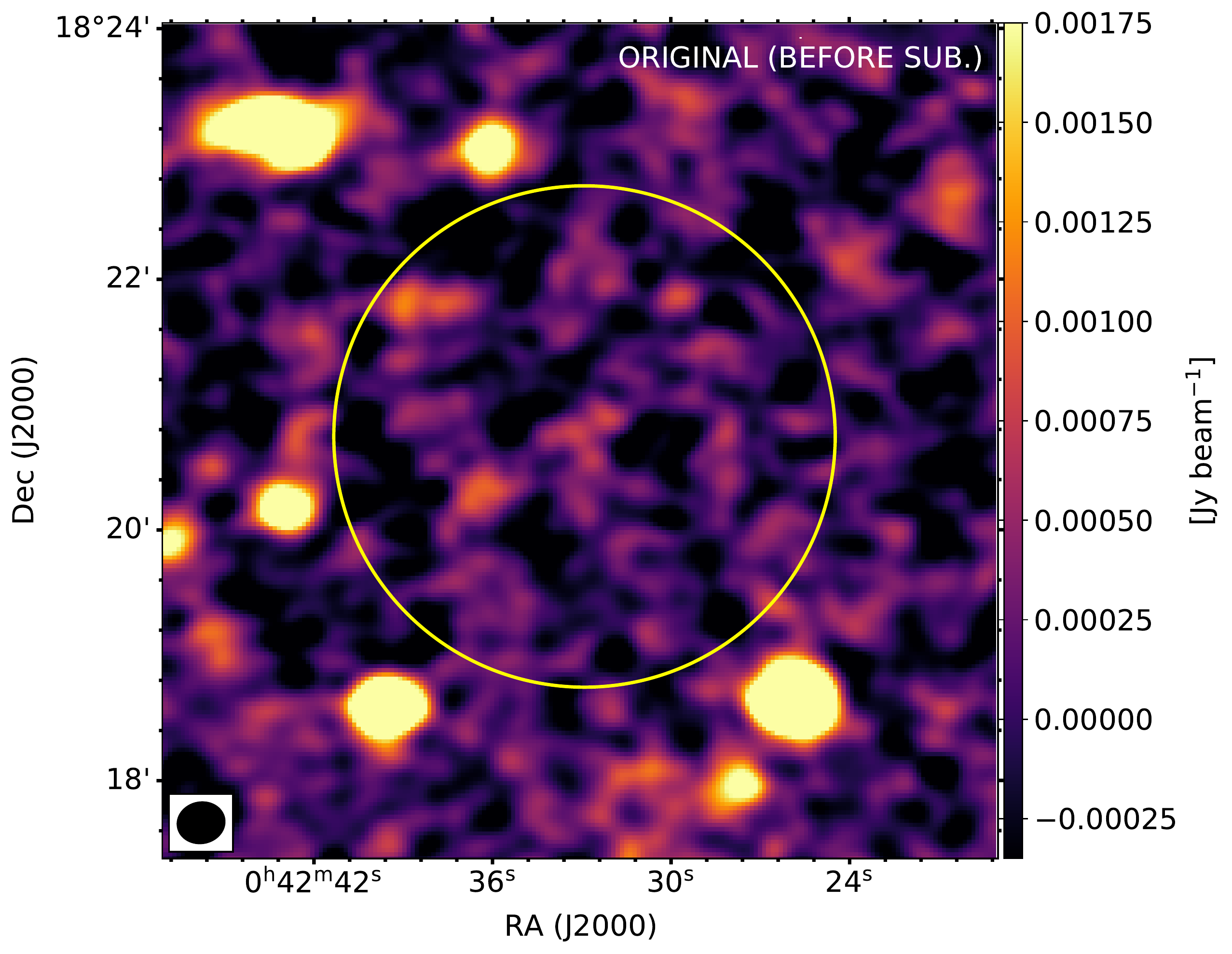}
    \includegraphics[width=0.265\textwidth] {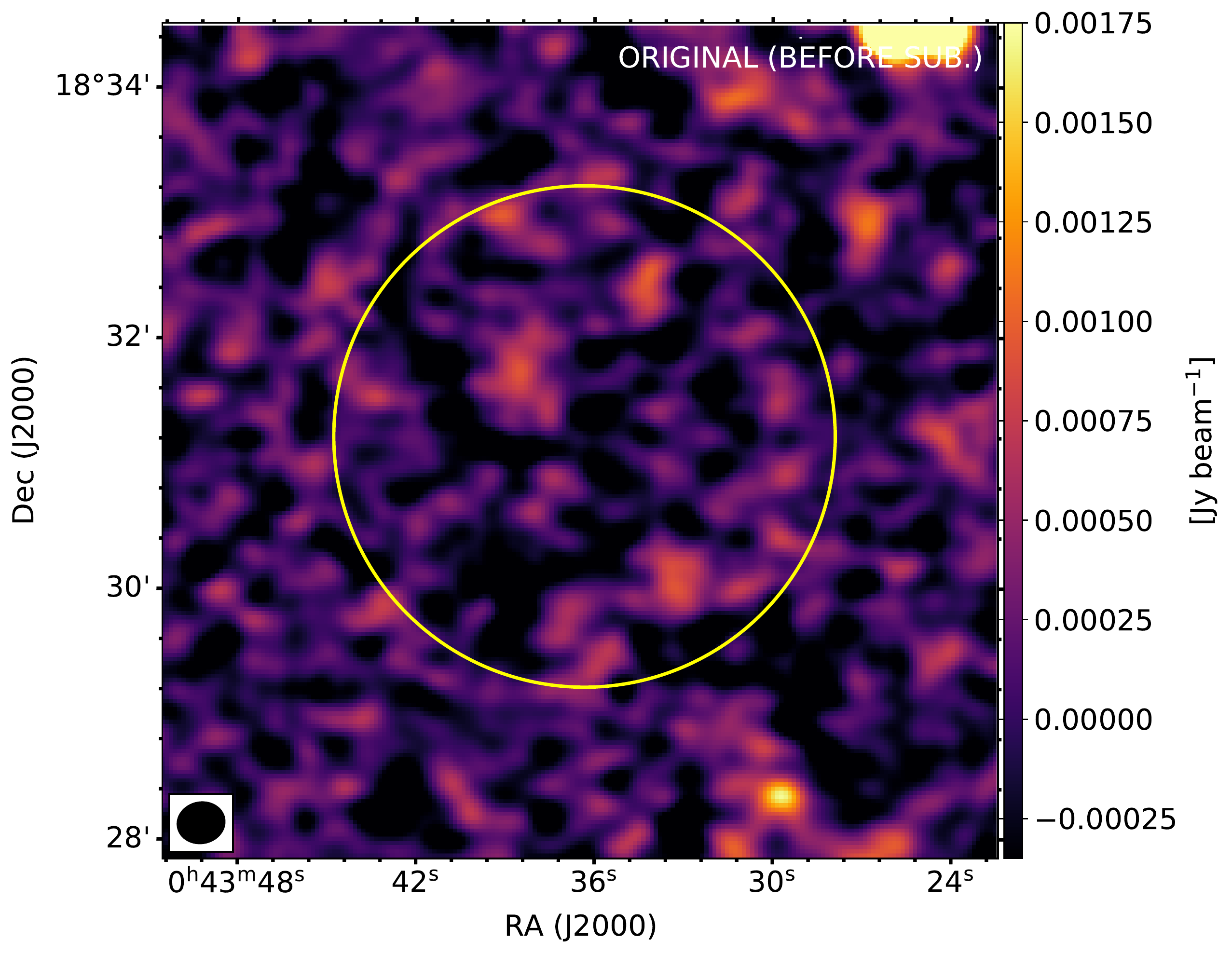} \\
    \includegraphics[width=0.265\textwidth] {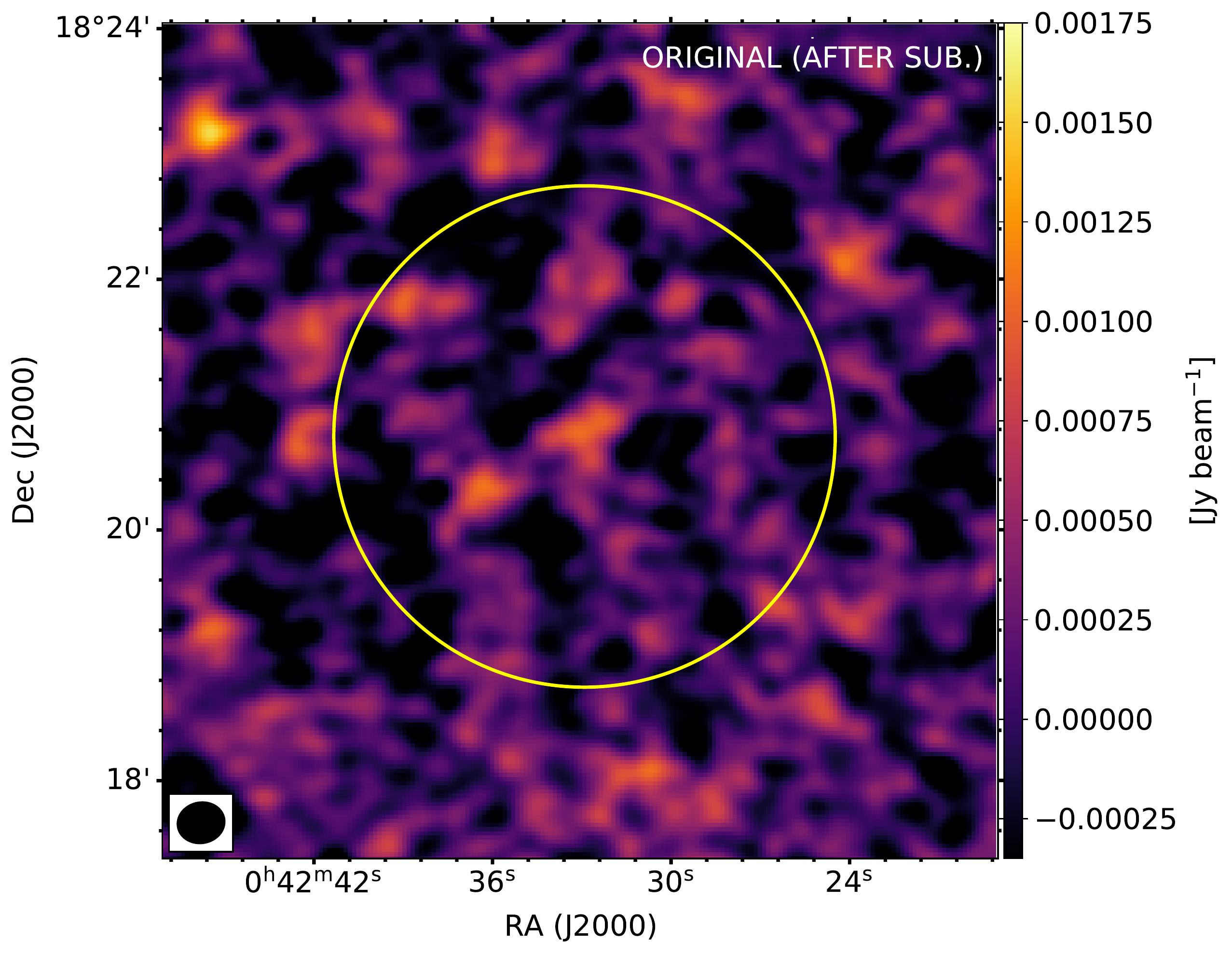} 
    \includegraphics[width=0.265\textwidth] {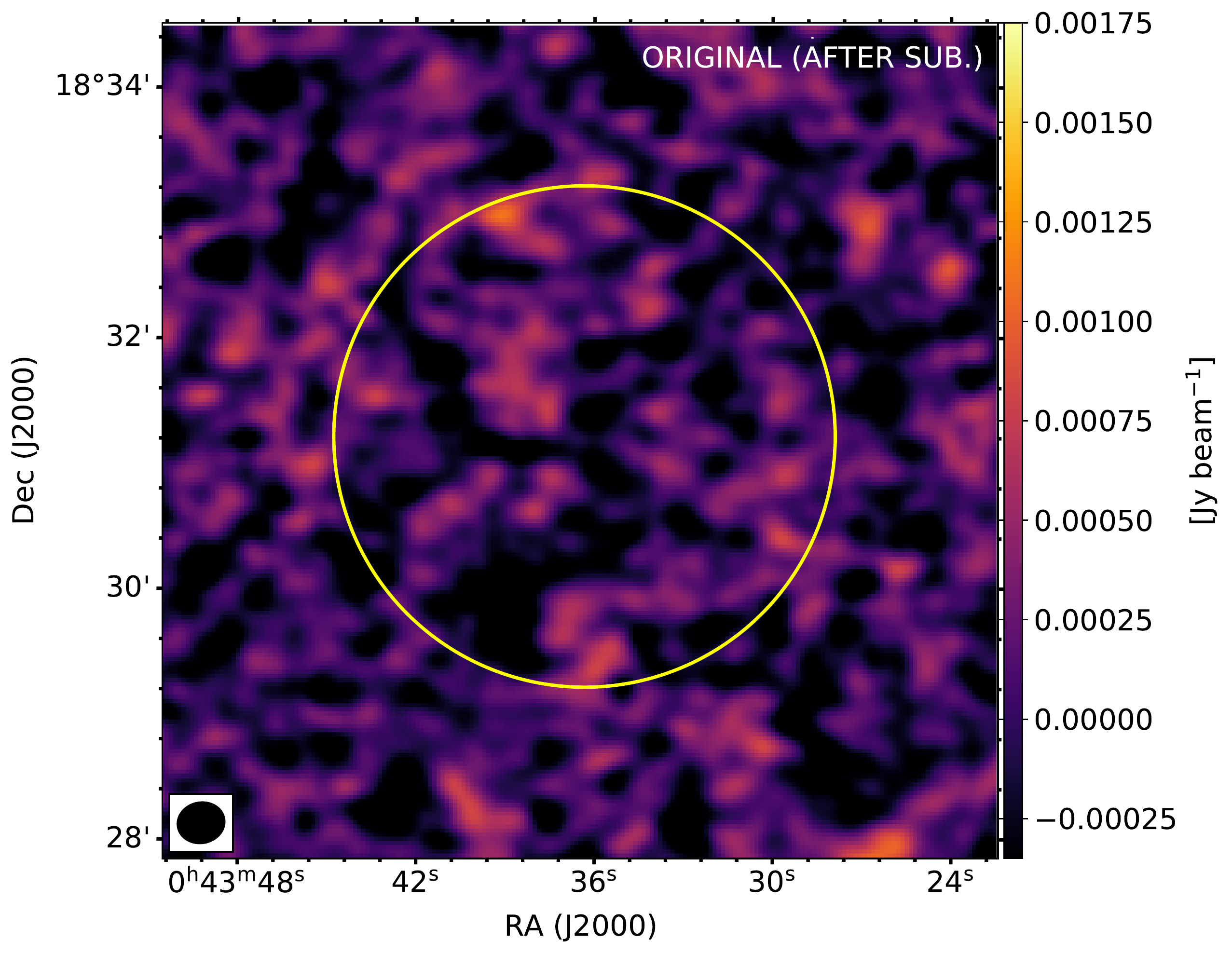}
    \caption{Restored images before injections in the centre (left panels) and periphery (right panels) of PSZ2 G120.08-44.41 before (upper panels) and after (lower panels) subtraction of discrete sources. }
    \label{MOCK120_ORIGINAL}
\end{figure*}

\begin{figure*}
	\centering

\includegraphics[width=0.265\textwidth] {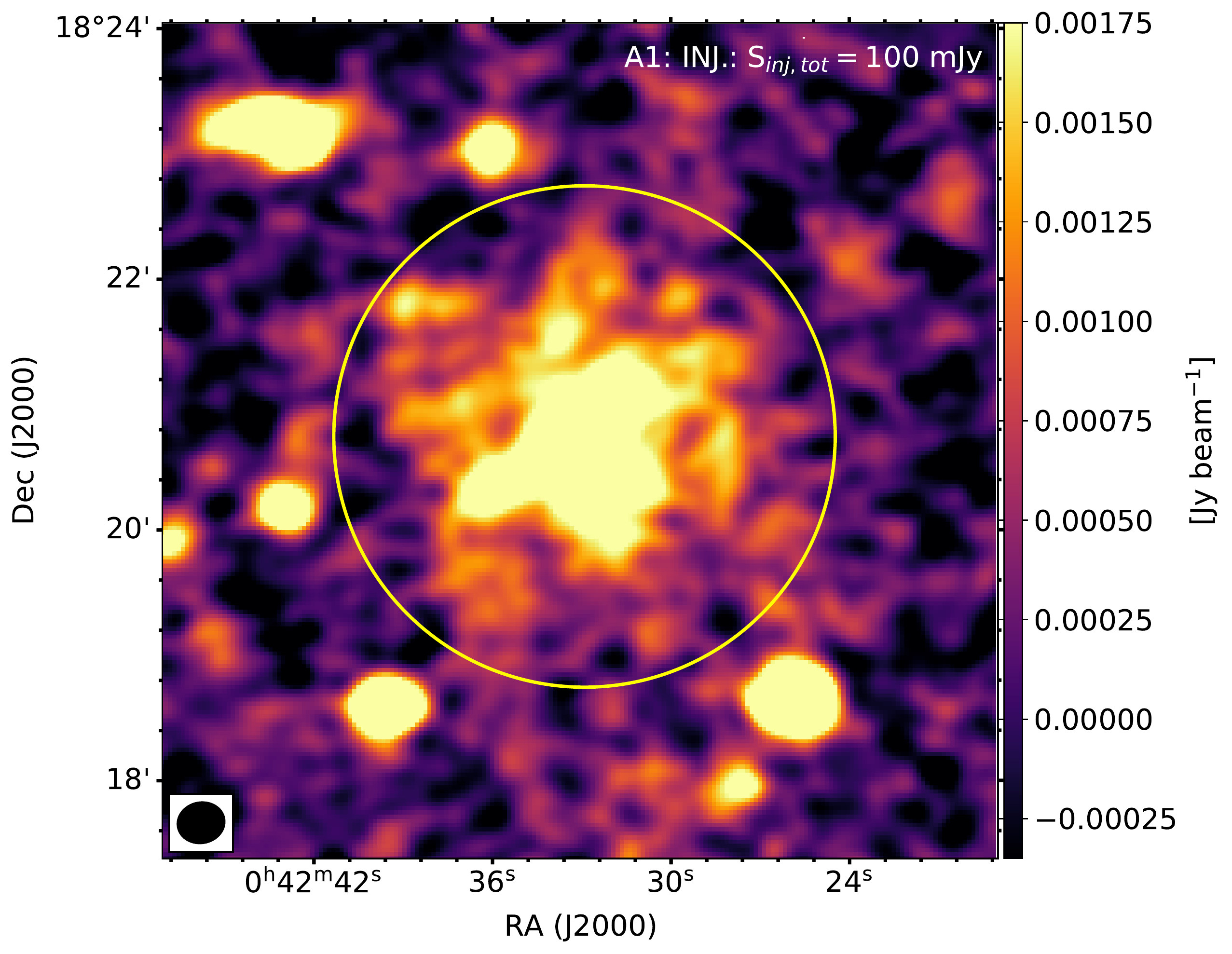}
\includegraphics[width=0.265\textwidth] {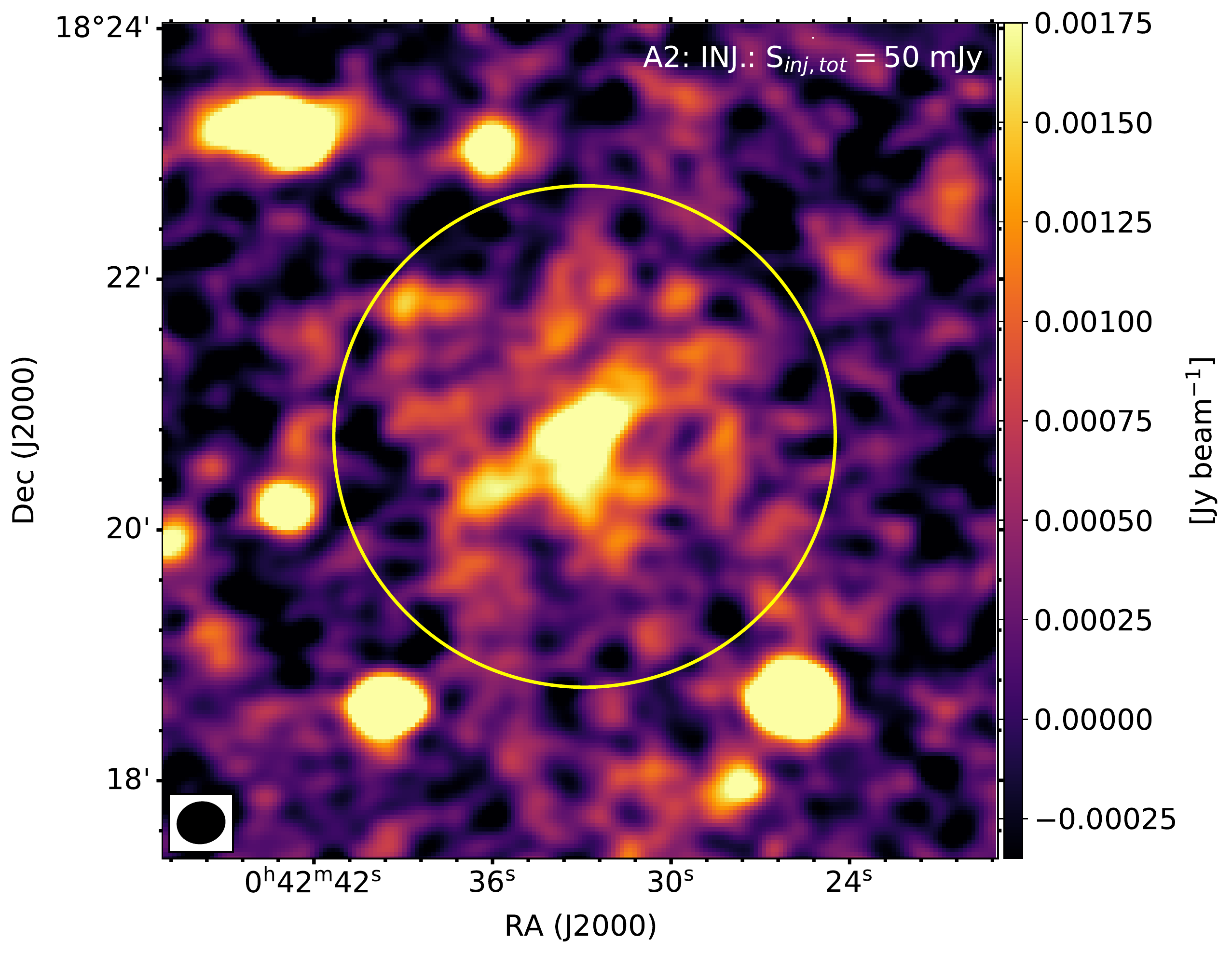}
\includegraphics[width=0.265\textwidth] {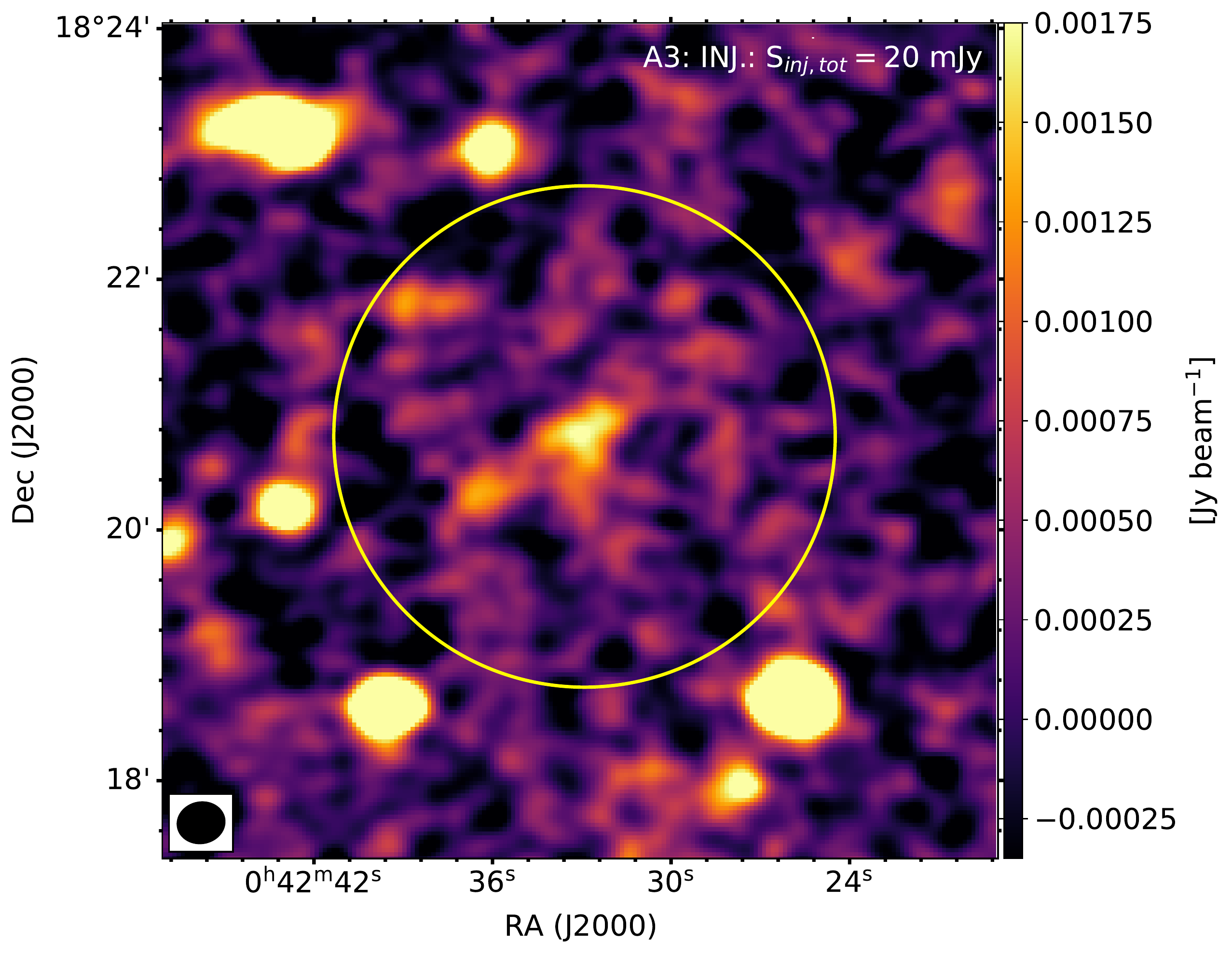}

\includegraphics[width=0.265\textwidth] {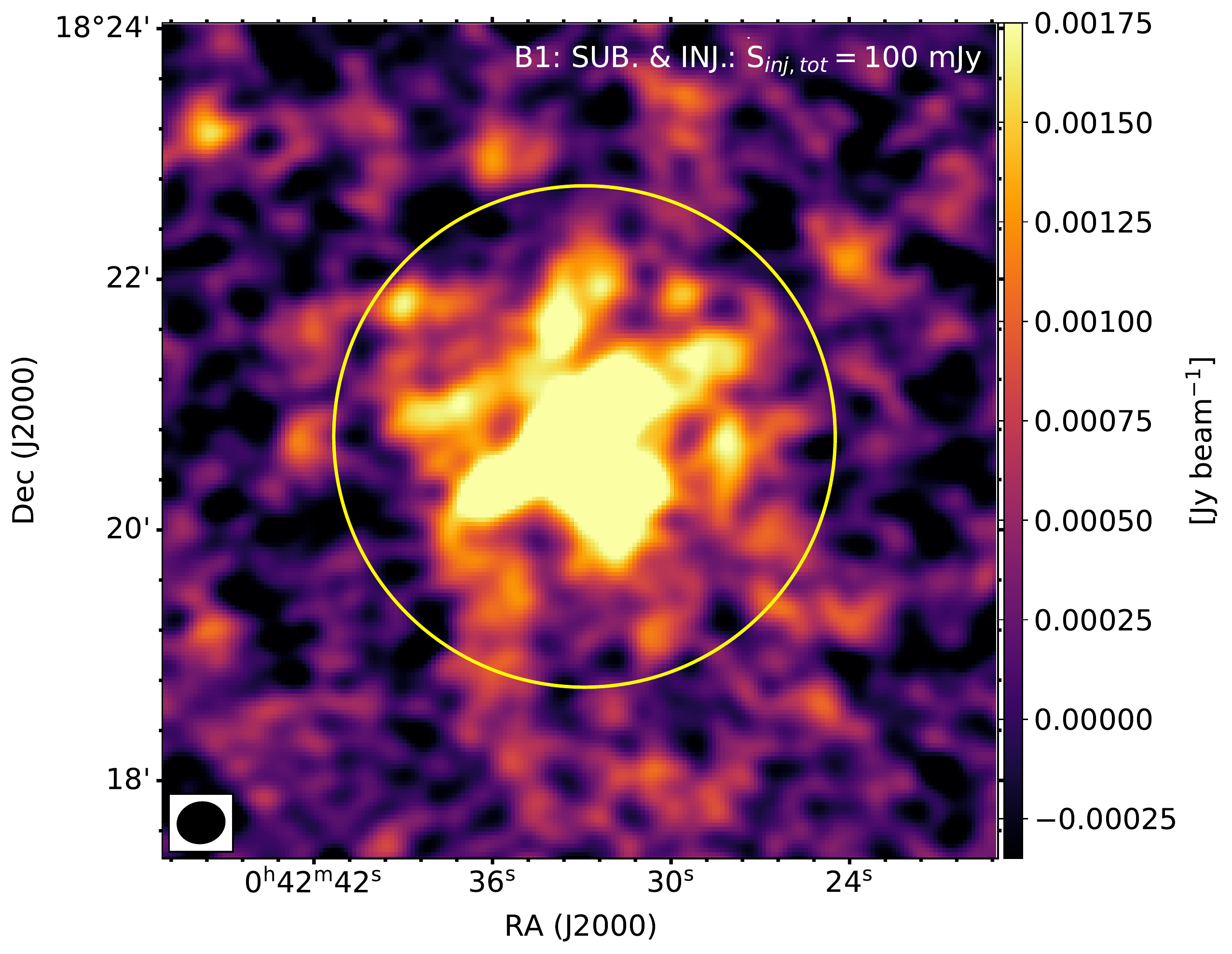}
\includegraphics[width=0.265\textwidth] {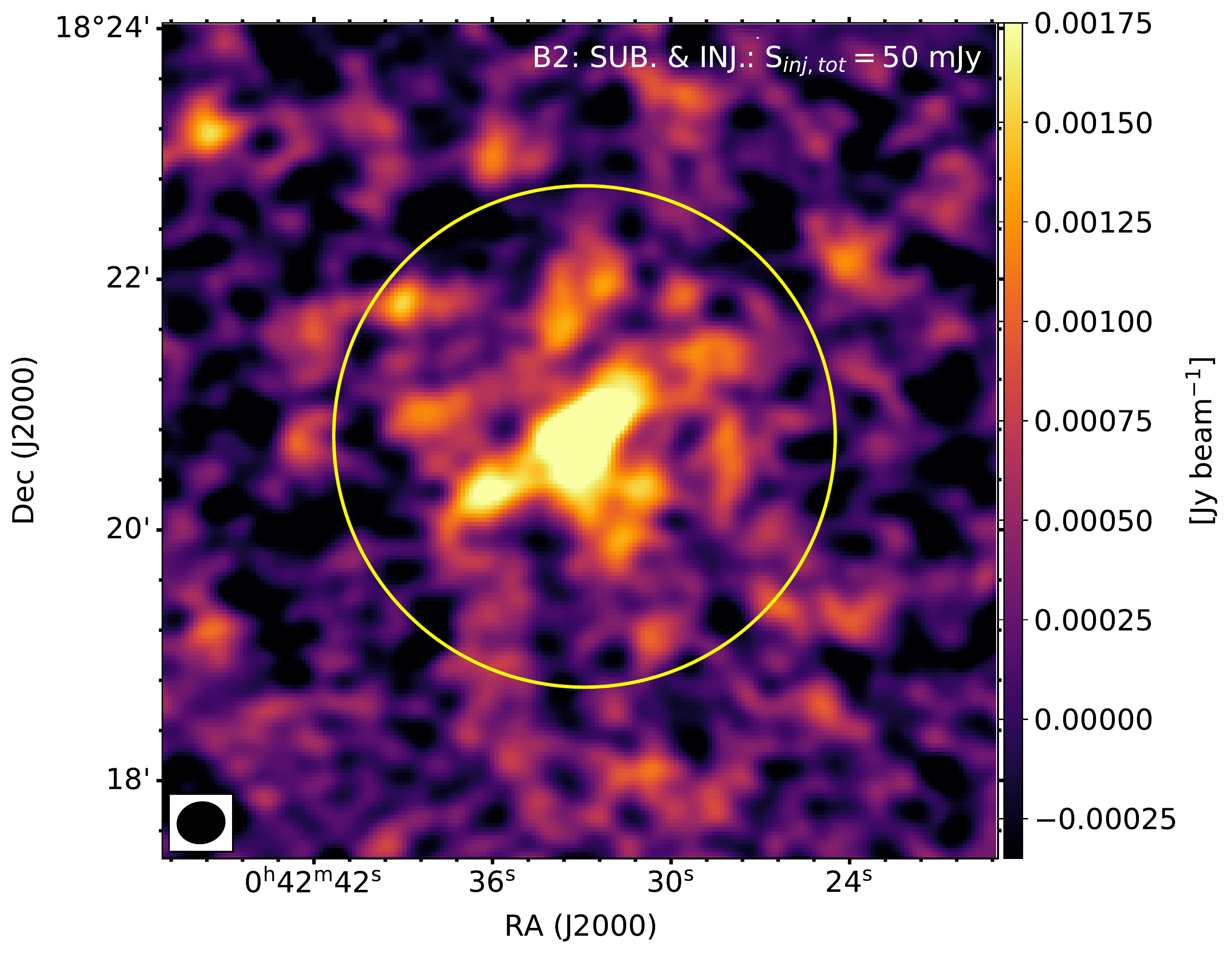}
\includegraphics[width=0.265\textwidth] {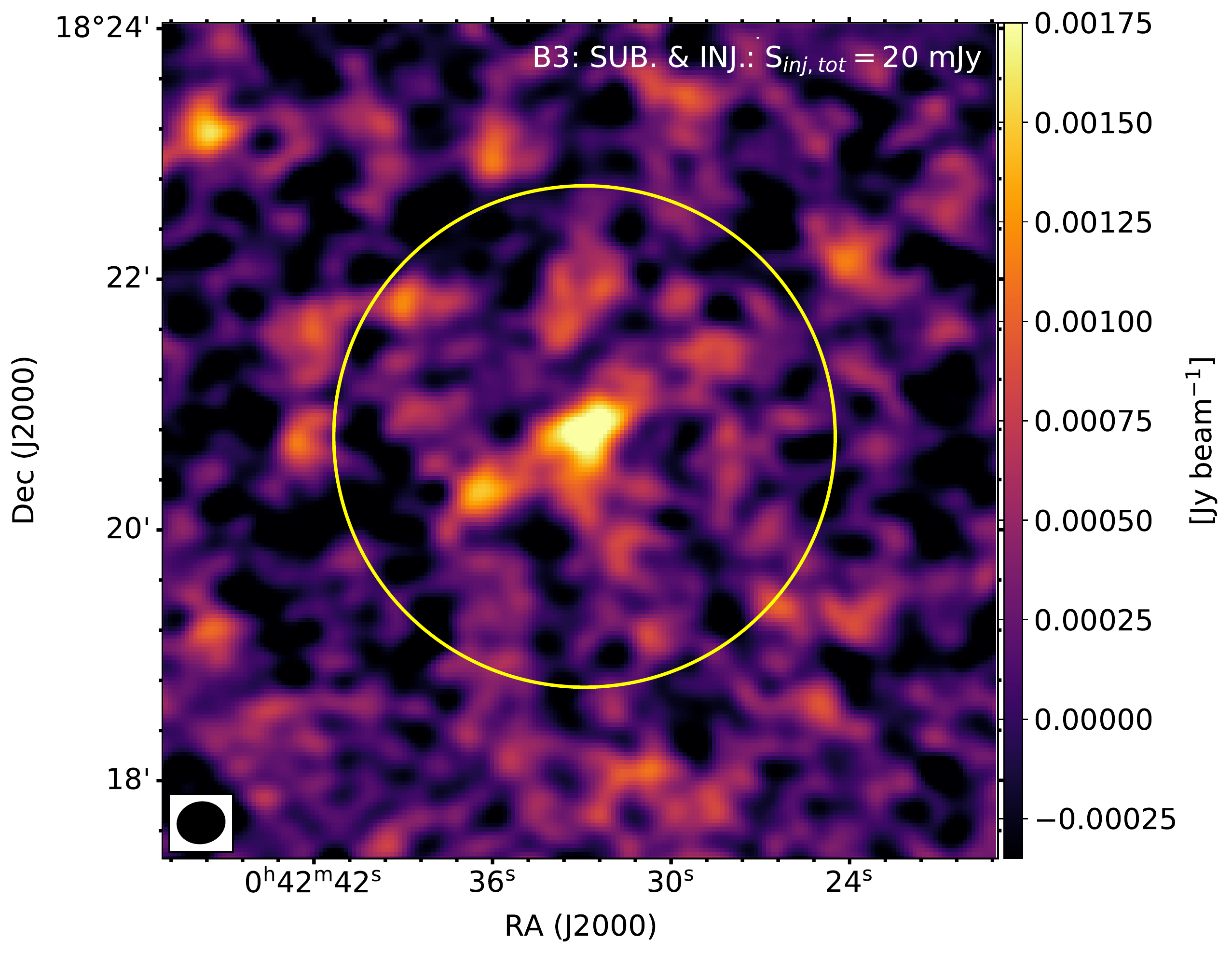}

\includegraphics[width=0.265\textwidth] {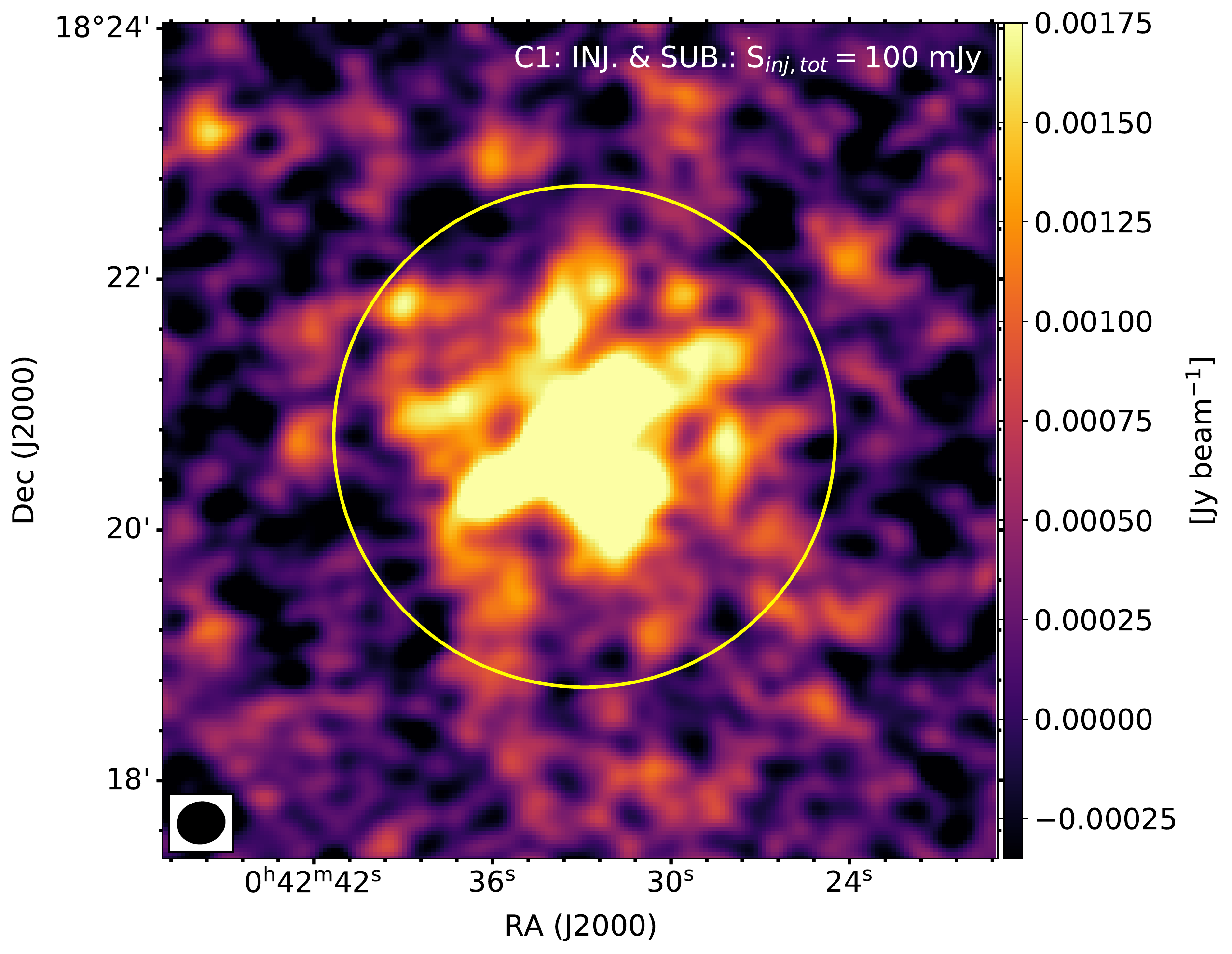}
\includegraphics[width=0.265\textwidth] {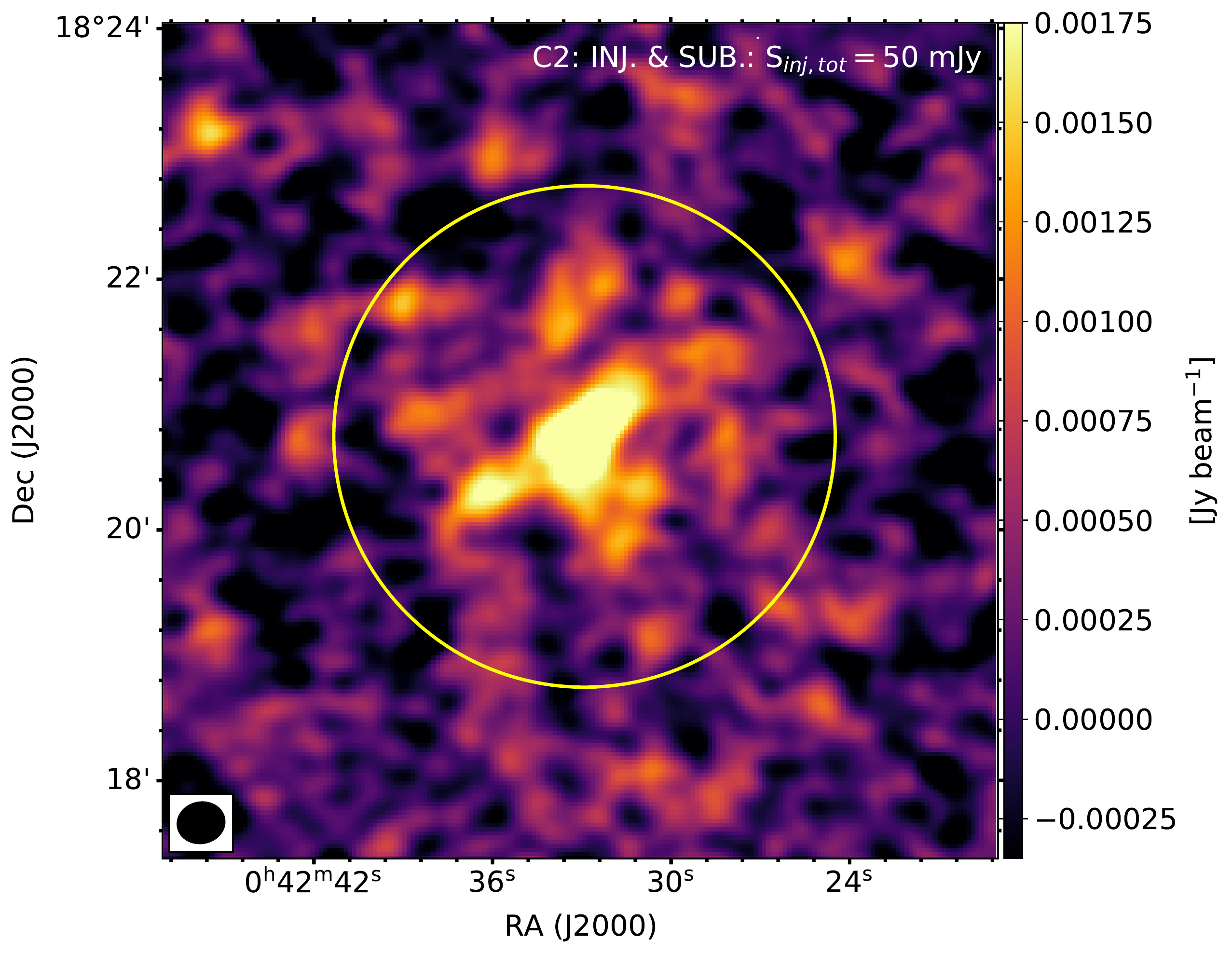}
\includegraphics[width=0.265\textwidth] {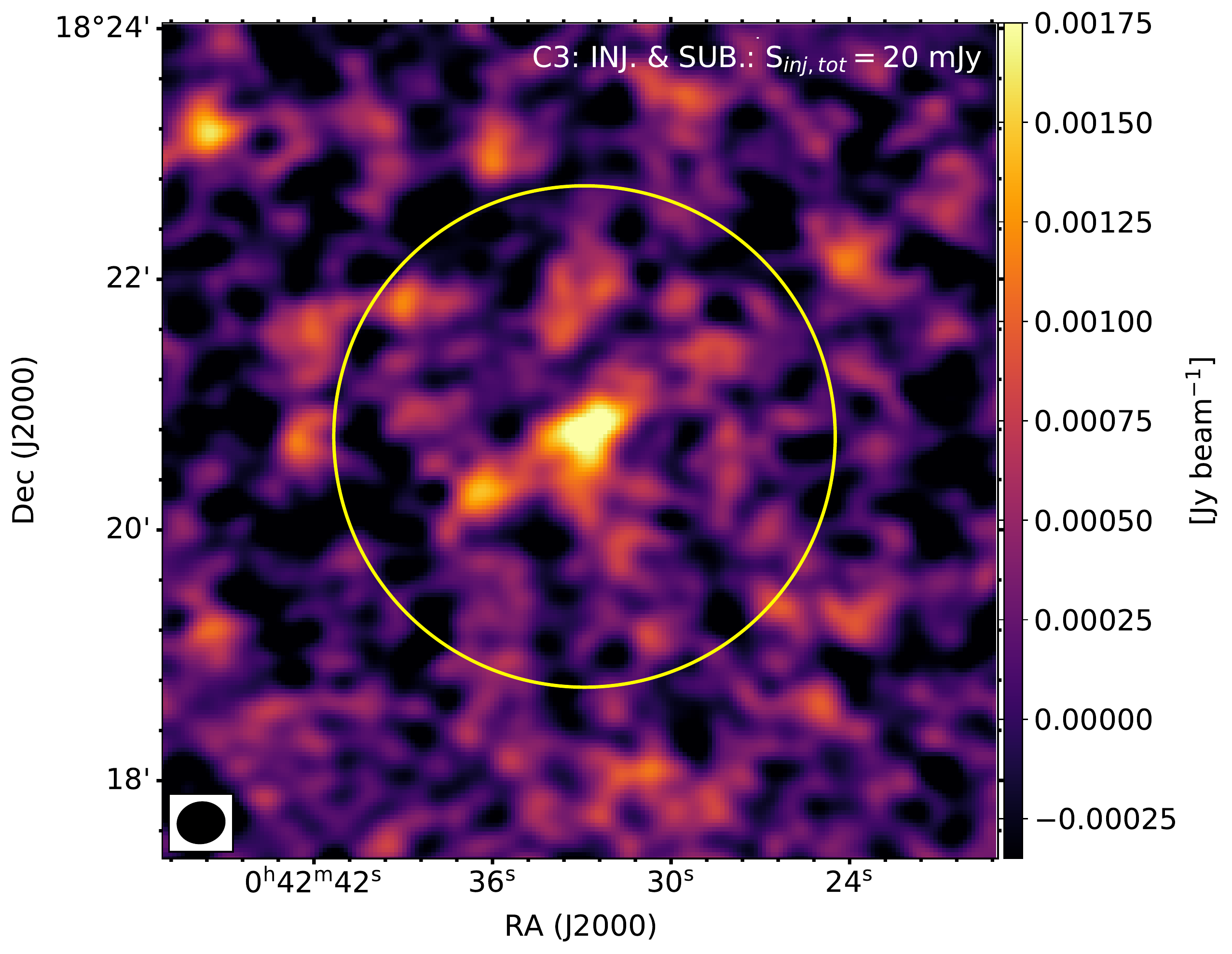}

\includegraphics[width=0.265\textwidth] {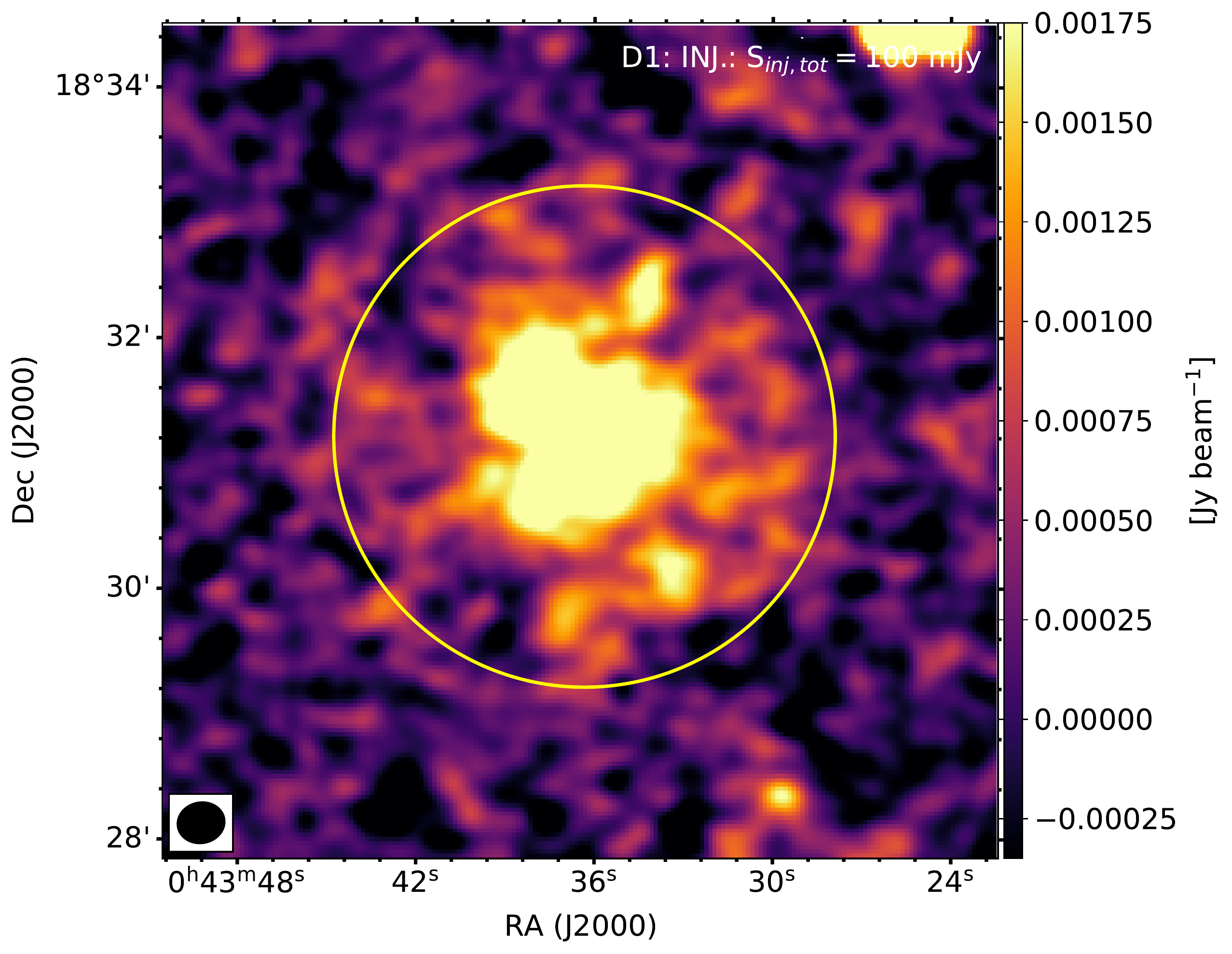}
\includegraphics[width=0.265\textwidth] {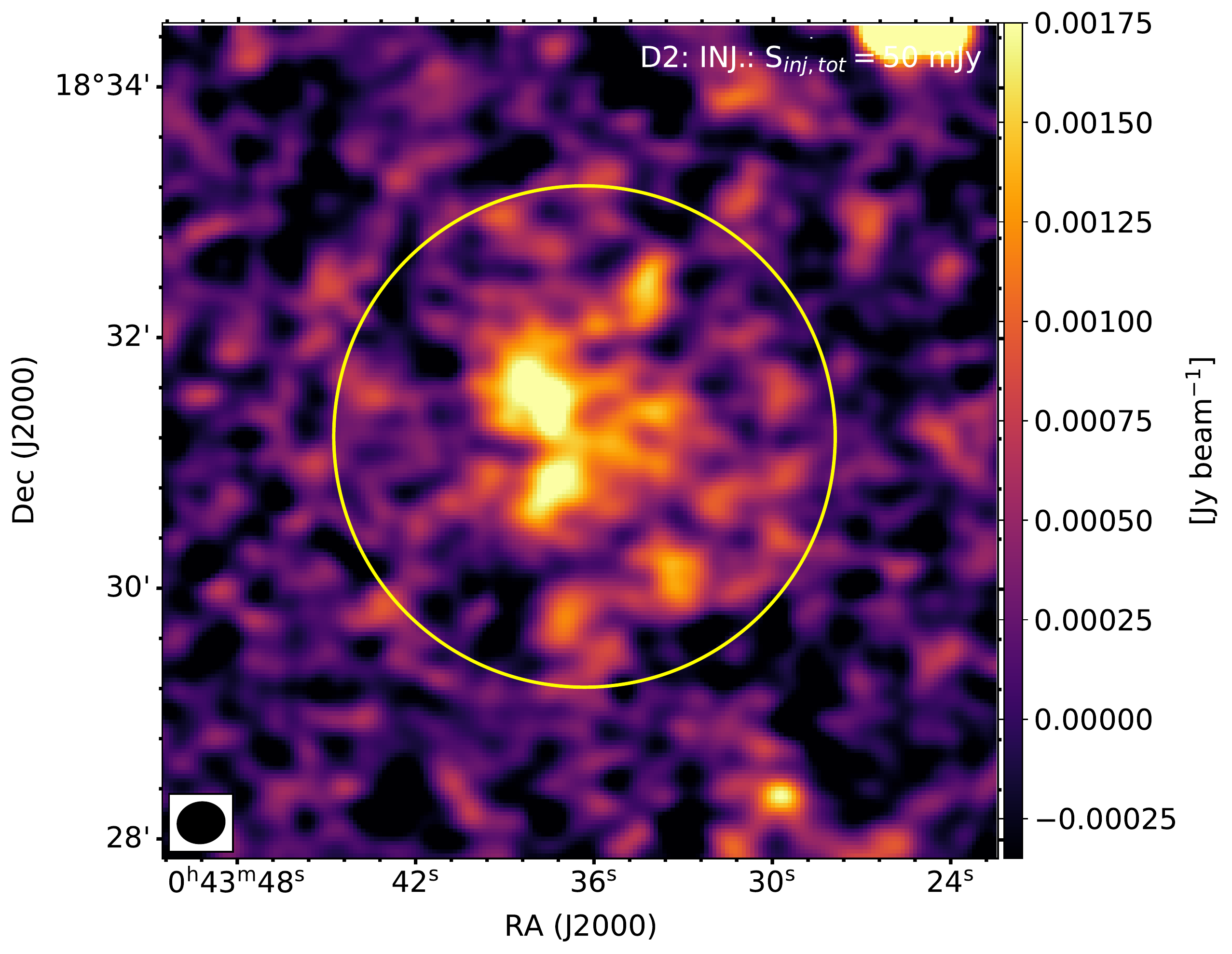}
\includegraphics[width=0.265\textwidth] {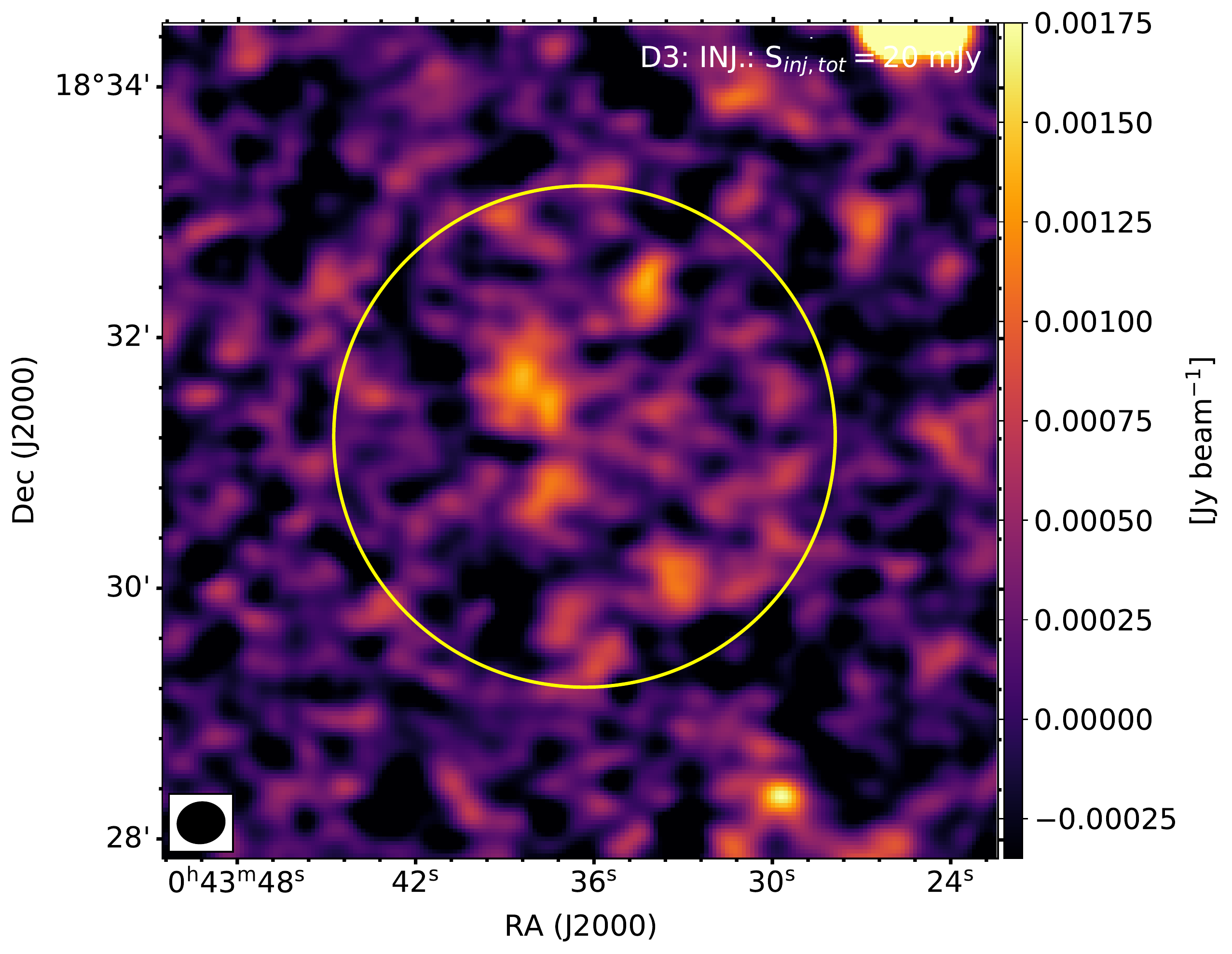}

\includegraphics[width=0.265\textwidth] {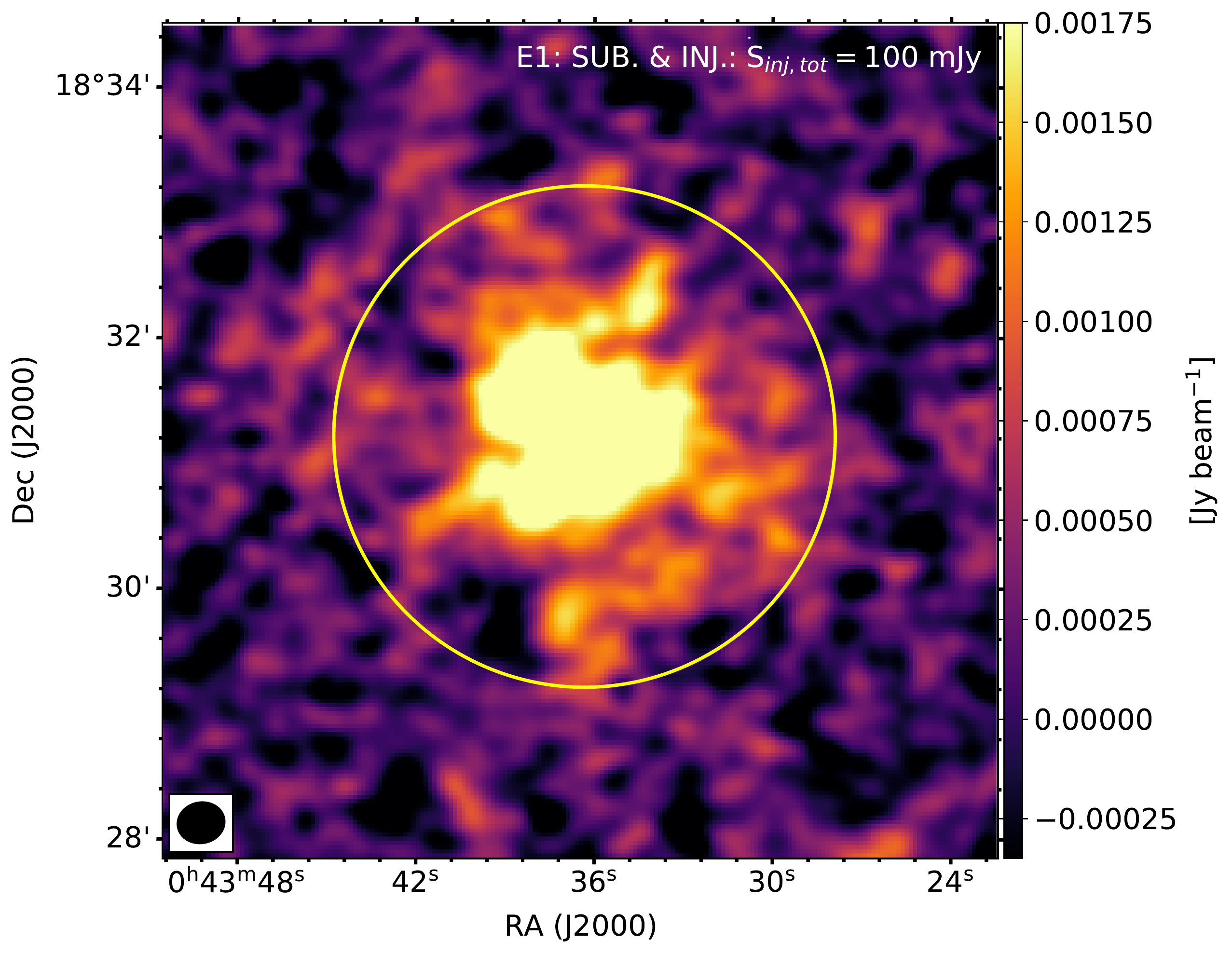}
\includegraphics[width=0.265\textwidth] {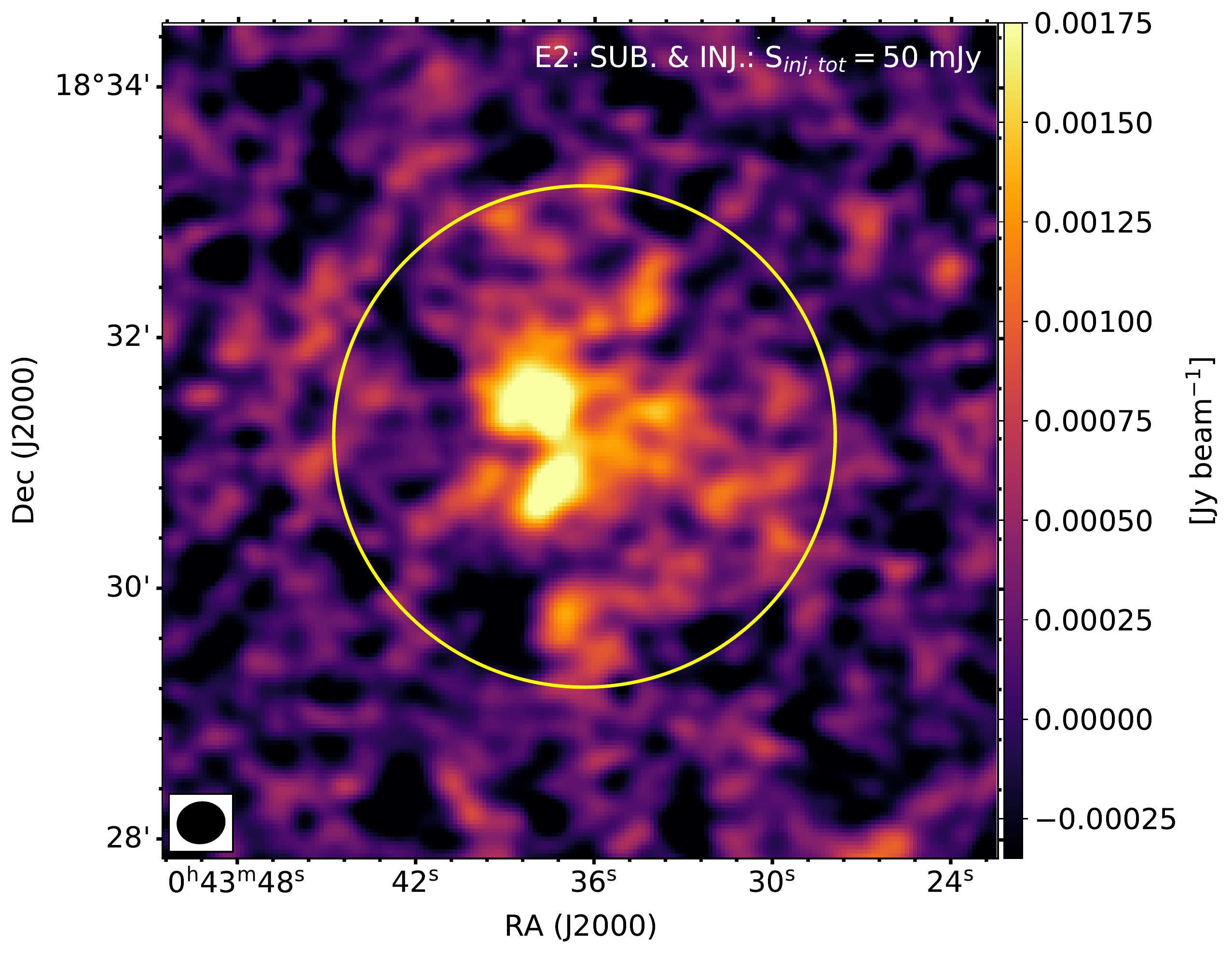}
\includegraphics[width=0.265\textwidth] {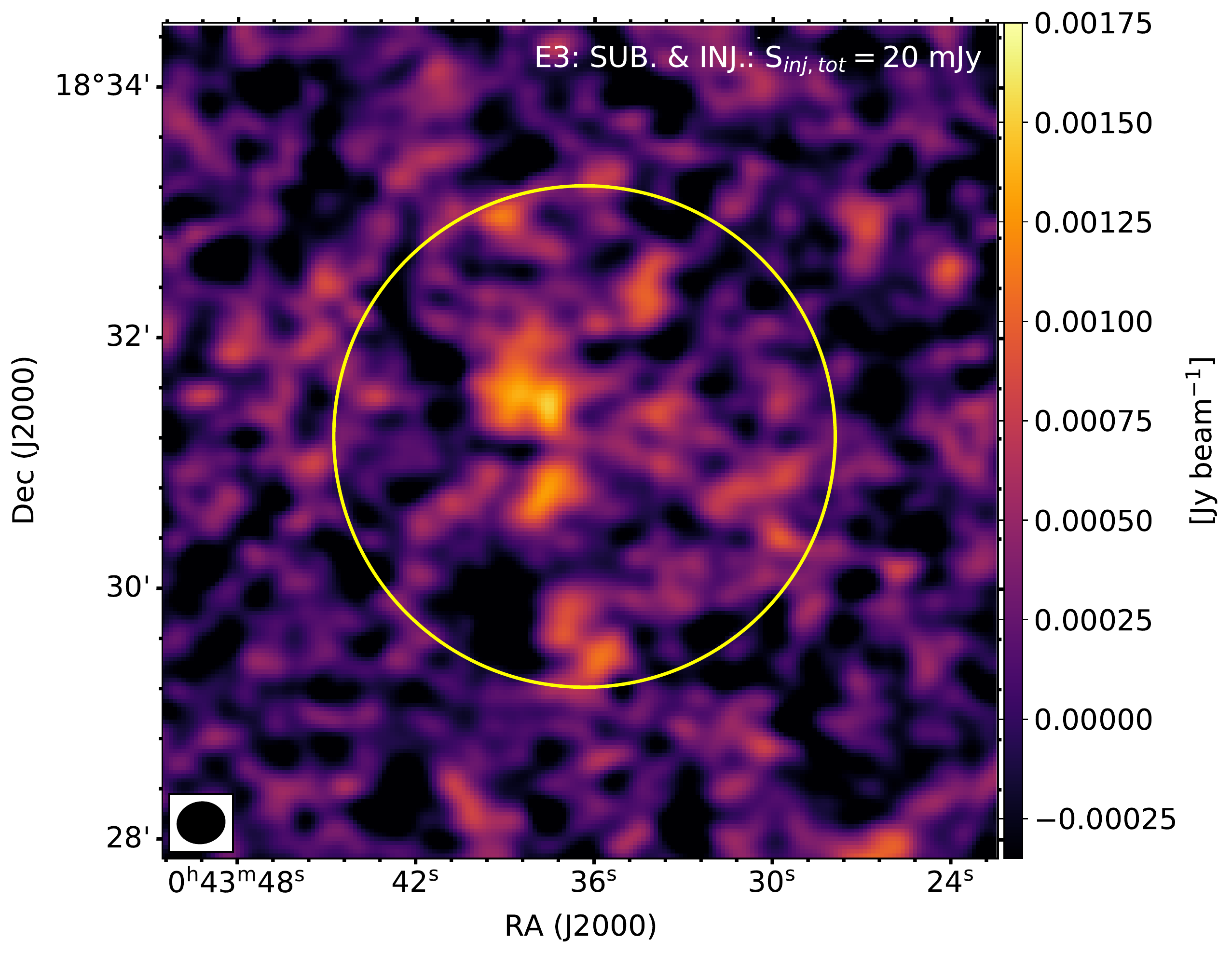}

\includegraphics[width=0.265\textwidth] {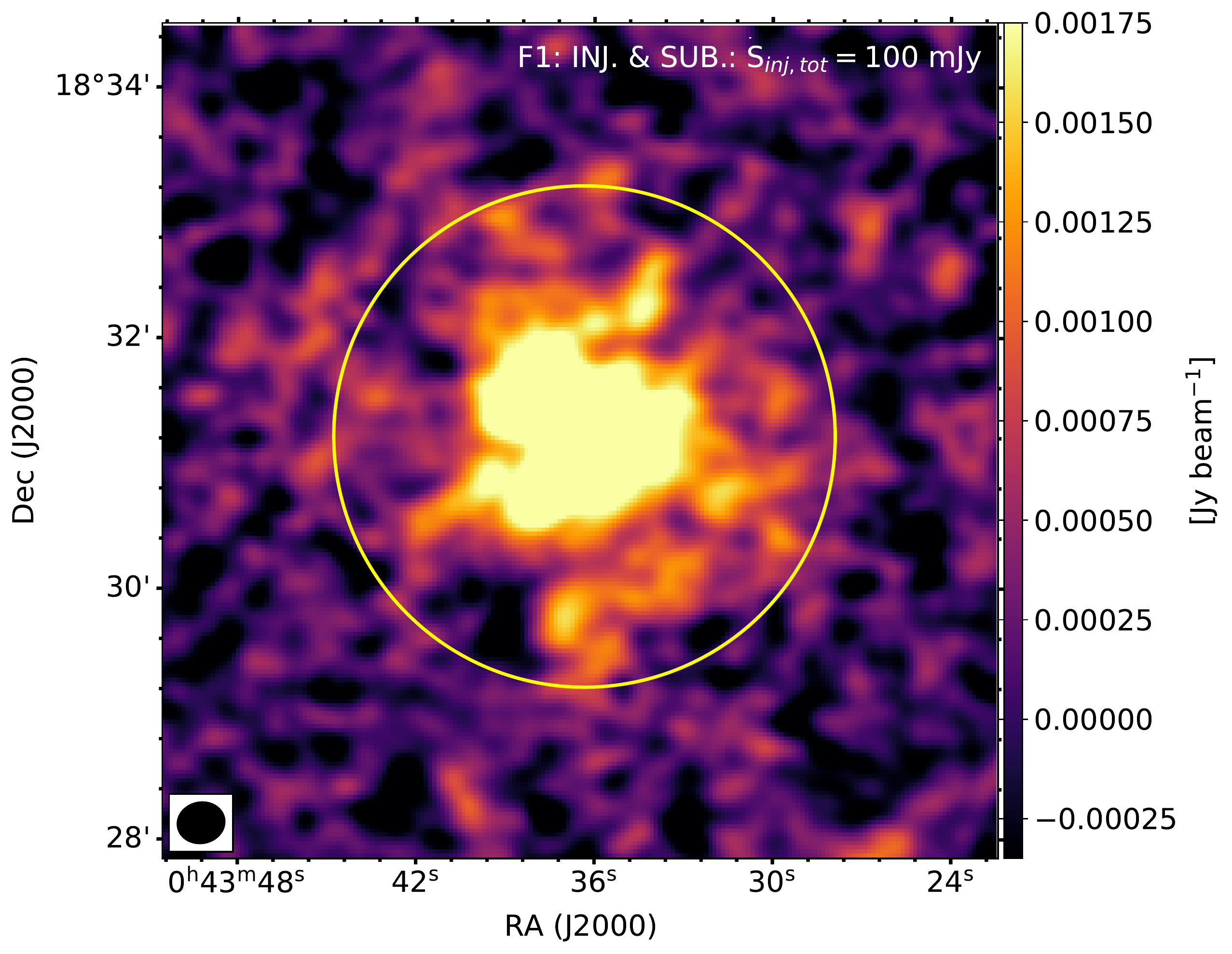}
\includegraphics[width=0.265\textwidth] {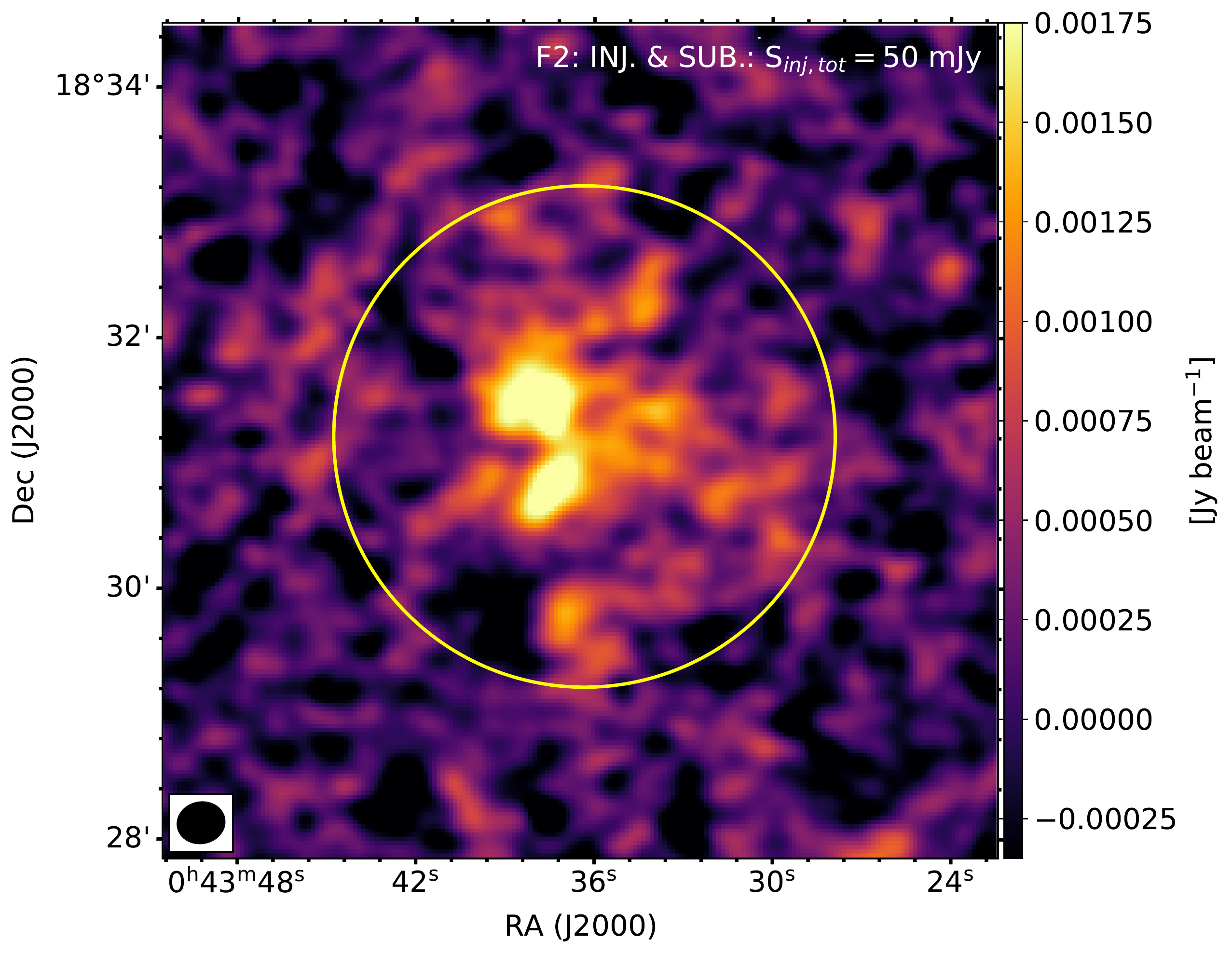}
\includegraphics[width=0.265\textwidth] {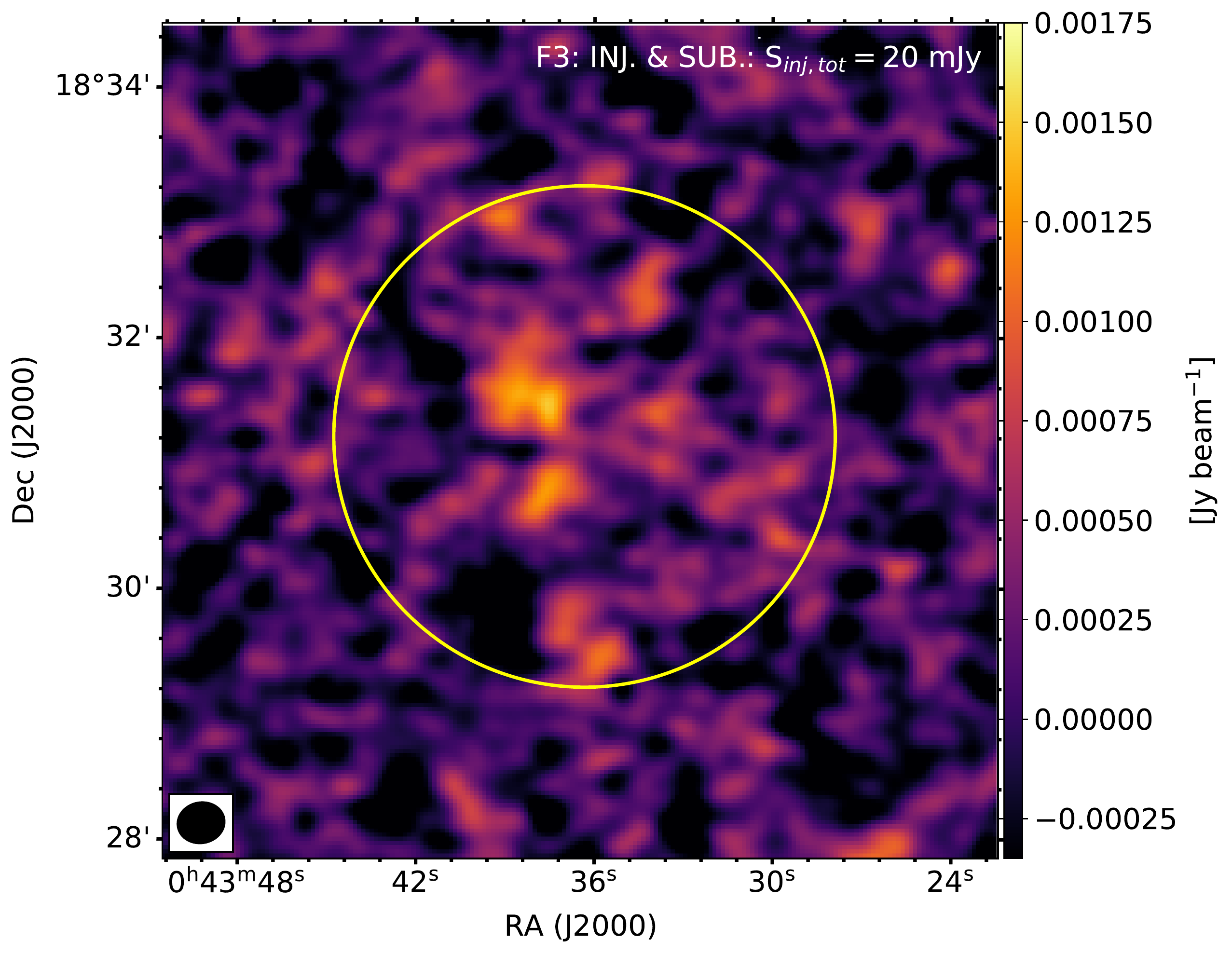}

	\caption{Restored images after injections in the centre (panels A, B, C) and periphery (panels D, E, F) of PSZ2 G120.08-44.41 with different schemes: injection in the original dataset (`INJ.', panels A, D), injection in the discrete source-subtracted dataset (`SUB. $\&$ INJ.', panels B, E), and injection in the original dataset and subsequent subtraction of the discrete sources (`INJ. $\&$ SUB.', panels C, F). The yellow circle (centred on ${\rm RA_{\rm inj}}$, ${\rm DEC_{\rm inj}}$) has a radius of $3r_{\rm e}$ and contains $S_{\rm inj}=0.8S_{\rm inj,tot}$, where $S_{\rm inj,tot}=100,\; 50,\; 20\;$ mJy, for panels in columns 1, 2, 3, respectively.}
	\label{MOCK120}%
\end{figure*}

\begin{figure}
    \centering
    \includegraphics[width=0.35\textwidth] {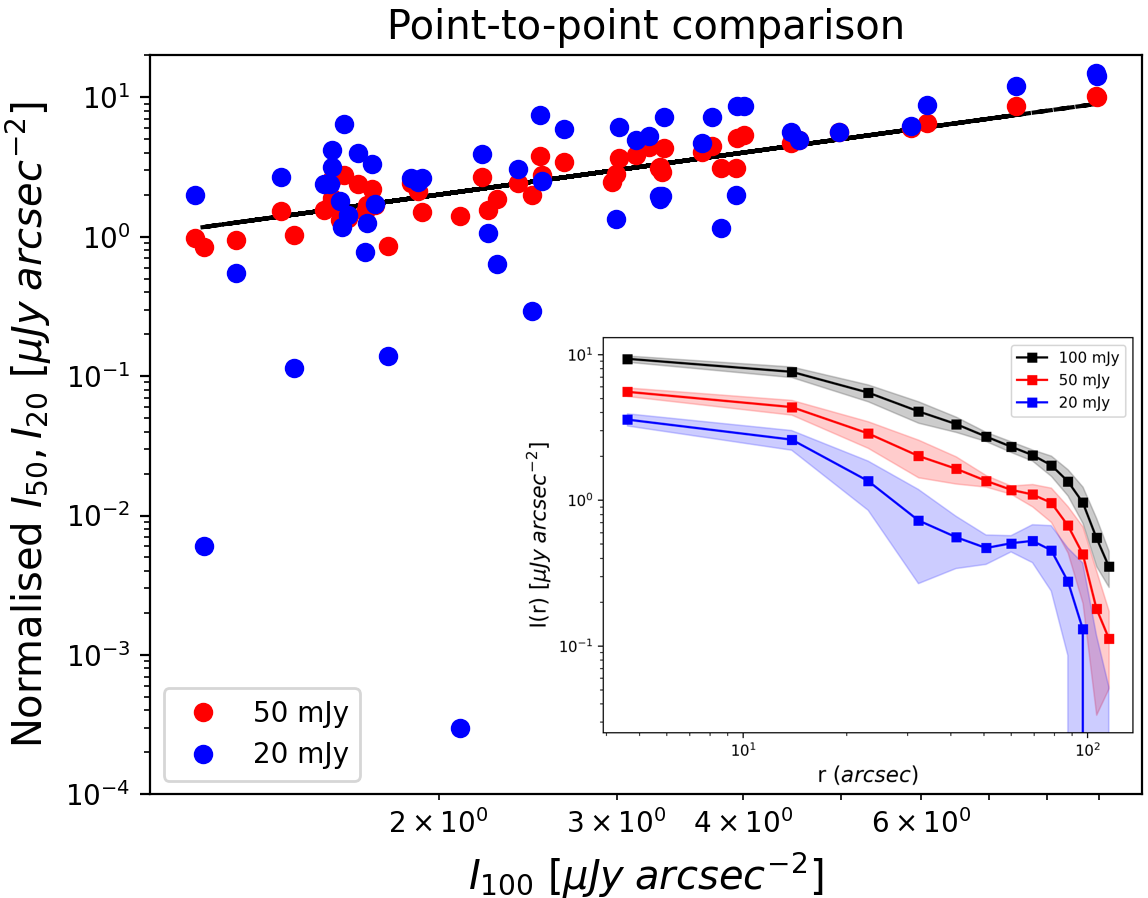}
    \includegraphics[width=0.35\textwidth] {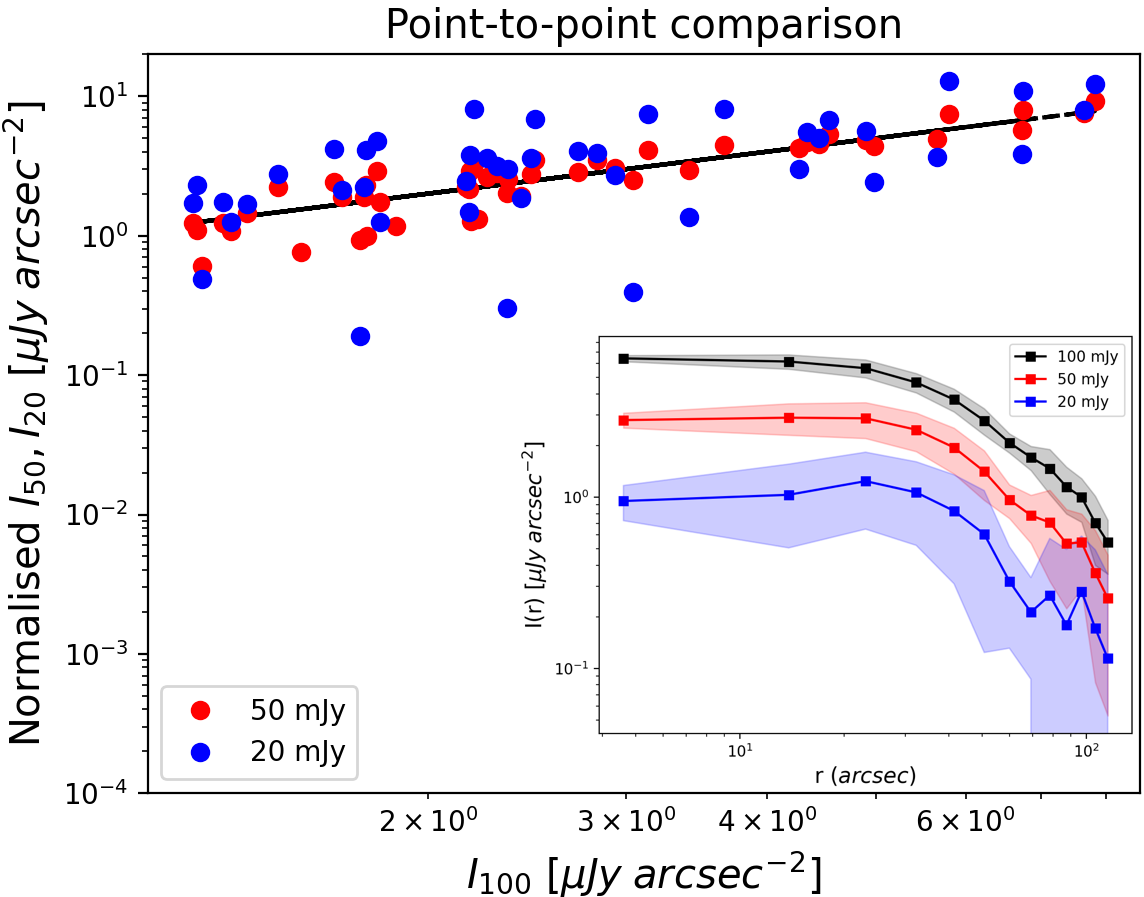}
    \caption{Point-to-point comparison of the surface brightness of mock halos in PSZ2 G120.08-44.41 (`Sub. \& Inj.' scheme; central and peripheral injections are shown in the upper and lower panel, respectively). Values at 20 and 50 mJy are multiplied by a factor 5 and 2, respectively, to match the injection at 100 mJy. The one-to-one line is plotted in black. Insets show the corresponding radial profiles, whose points are obtained as the average brightness of four $90^{\rm o}$-wide sectors, whereas the shaded region represents their standard deviation. The point-to-point plots show that the lower the brightness and the injected flux density, the higher the scatter around the one-to-one line, indicating the progressive arise of patches of emission. Analogously, the radial profiles indicate that the scatter increases at larger radii and lower brightness.  }
    \label{ptp}
\end{figure}

As examples, in Fig. \ref{MOCK120} we show the injections of mock halos with $r_{\rm e}=40.7''$ and $S_{\rm inj,tot}=$ 20, 50, 100 mJy that we performed both close to the pointing centre of PSZ2 G120.08-44.41 and far from it (at a distance of $\sim 19'$), by using the extracted \& re-calibrated datasets (`Inj.' scheme); images of this field before the injections are shown in Fig. \ref{MOCK120_ORIGINAL} with the same color scale for a comparison. As preliminary checks of our procedure, we fitted the surface brightness of the mock halos (discrete sources close to the mock emission were masked) by means of the {\tt Halo-FDCA}\footnote{\url{https://github.com/JortBox/Halo-FDCA}} code \citep{boxelaar21}, which was used for radio halos in LoTSS as well \citep{vanweeren21,botteon22LoTSS,hoang22}, and measured the flux density in circles of radius $\hat{r}=3r_{\rm e}$, where a fraction $f(\hat{r})=0.80$ of the total injected flux density is expected according to Eq. (\ref{flusso-I0}). The images in Fig. \ref{MOCK120} show that injecting the same flux density in different locations generally provides visually-different mock halos, because of the presence of discrete sources and local fluctuations of the noise, contaminating the morphology of the mock emission. Nevertheless, as reported in Table \ref{testMOCK120}, the flux densities measured by hand ($S_{\rm meas}$) are in good agreement for all the corresponding injections in the centre and far from it. The flux density obtained through {\tt Halo-FDCA} is computed as $S_{\rm fit}\propto I_{\rm 0,fit}r_{\rm e,fit}^2$, thus it is strongly dependent on the fitted \textit{e}-folding radius and its associated error; in particular, the less the significance of the emission of mock or real halos, the higher the errors on $r_{\rm e,fit}$. Despite this, all the corresponding injected and fitted parameters are consistent within the fitting errors. These tests indicate that our simulations are barely dependent on the position of the injection, meaning that the response of the instrument can be considered uniform, at least across the extraction regions.

Background, foreground, and embedded sources contaminate the faint diffuse emission of radio halos. A model of the discrete sources can be obtained by selecting only the longest baselines, depending on the angular scales that need to be filtered out, and their contribution can be then directly subtracted from the \textit{uv}-data. We compared the results of the `Inj.' scheme with the injection in source-subtracted datasets (`Sub. \& Inj.' scheme) and the effects of subtraction after injecting in the original datasets (`Inj. \& Sub.' scheme). To this aim, following the approach described in \cite{botteon22LoTSS}, we selected the baselines corresponding to projected sizes $<250$ kpc at the cluster redshift, and removed discrete sources in our targets. As reported in Table \ref{testMOCK120}, the measured and fitted flux densities are consistent for all the three schemes. We notice, however, differences in $I_{\rm 0,fit}$ and $r_{\rm e,fit}$ in case of the injection with the lowest flux density, i.e. $S_{\rm inj,tot}=20$ mJy. As previously mentioned, this is due to the low significance of the diffuse emission with respect to the background, and in these cases, 
residuals from the source subtraction process that were not masked during the fit could have a non-negligible impact on the fit with {\tt Halo-FDCA}. By comparing the `Sub. \& Inj.' and `Inj. \& Sub.' schemes, we find an almost perfect agreement in both the measures and fitting of the corresponding cases, meaning that the mock diffuse emission is not included in the subtraction model. This latter result has a major practical utility for our work; indeed, the subtraction can be performed only once, thus allowing us to save a huge amount of computing time, and different injections can be carried out in the same source-subtracted dataset, with no need to subtract sources after each injection (see also Sect. \ref{sect:Upper limit calculation}). 

 We conclude this section by commenting on the recovered morphology of mock halos. For very bright mock halos, the spherical and smooth profile is typically recovered. On the other hand, in the case of less bright mock halos, the roundish and smooth shape can be easily perturbed depending on the local noise pattern and effective sensitivity of the observations. By gradually decreasing the injected flux density, the symmetry is progressively broken, and the mock halo will appear as patches of emission around the peak. This behaviour can be seen in Fig. \ref{MOCK120}. Even though the recovered morphology depends on the specific dataset and injection position, we can quantify the relative deviations as a function of radius and injected flux density through a point-to-point comparison. To this aim, we considered our brightest injections at 100 mJy (with the `Sub. \& Inj.' scheme), and sampled the corresponding mock halos with a grid of beam-size square boxes down to the $2\sigma$ contour level. The same grid was used to sample the mock halos of $S_{\rm inj,tot}=50$ and 20 mJy as well. The surface brightness at 20 and 50 mJy were normalised to match the injection at 100 mJy (i.e. they are multiplied by factor of 5 and 2, respectively), and the corresponding measures were reported in Fig. \ref{ptp} as a function of the brightness at 100 mJy. These plots show that progressively increasing deviations from the one-to-one line are found from higher to lower brightness (i.e. outwards from the peak), and that the scatter is larger for the injections at 20 mJy than at 50 mJy. This can be also observed in the insets of Fig. \ref{ptp}: the reported radial profiles are obtained by averaging profiles extracted in four $90^{\rm o}$-wide sectors for each mock halo. Both the point-to-point plot and radial profiles show that the scatter of the brightness is larger in the outermost regions of the mock halos, where the faint emission consists of asymmetric patches and less flux density is recovered. Moreover, the differences between the injections in the centre and periphery are associated with the local noise conditions, which notably contribute to the recovered morphology of the mock halo. By decreasing $S_{\rm inj,tot}$ and approaching the noise level, the scatter will further increase.

\section{LOFAR performances}
\label{sect:LOFAR performances}

In this section, we explore in detail the capabilities of LOFAR to recover extended emission, by simulating mock radio halos with a wide range of flux densities and angular sizes in some clusters of our reference sample. By adopting the standard \textit{uv}-cut for LOFAR HBA, baselines $<80\lambda$ will be not employed in our analysis.

\subsection{The role of the \textit{uv}-coverage}
\label{sect: The role of the uv-coverage}

\begin{figure*}
	\centering
	\includegraphics[width=1.0\textwidth]{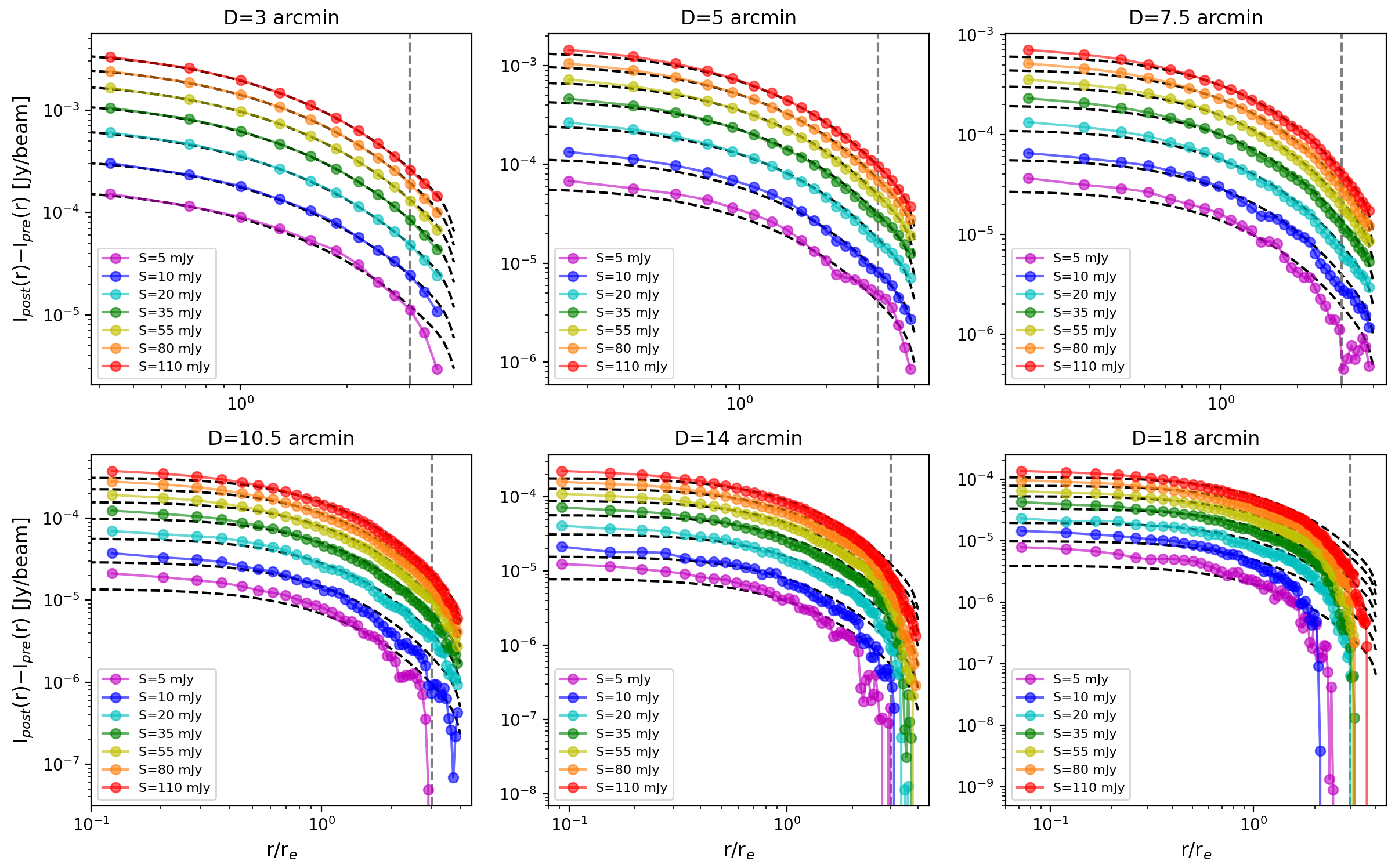} 
	\caption{Net azimuthally-averaged surface brightness profiles of the images shown in Appendix \ref{appendix:mock98} for PSZ2 G098.62+51.76. The dashed black lines represent the theoretical injected profiles. The sampled profiles ($S_{\rm inj,tot}$ is reported in the legend) are obtained by subtracting the pre-injection from the post-injection contribution. The grey vertical line indicates $r=3r_{\rm e}$.}
	\label{MOCK98_netprofiles}%
\end{figure*}   

\begin{figure}
	\centering
	\includegraphics[width=0.49\textwidth] {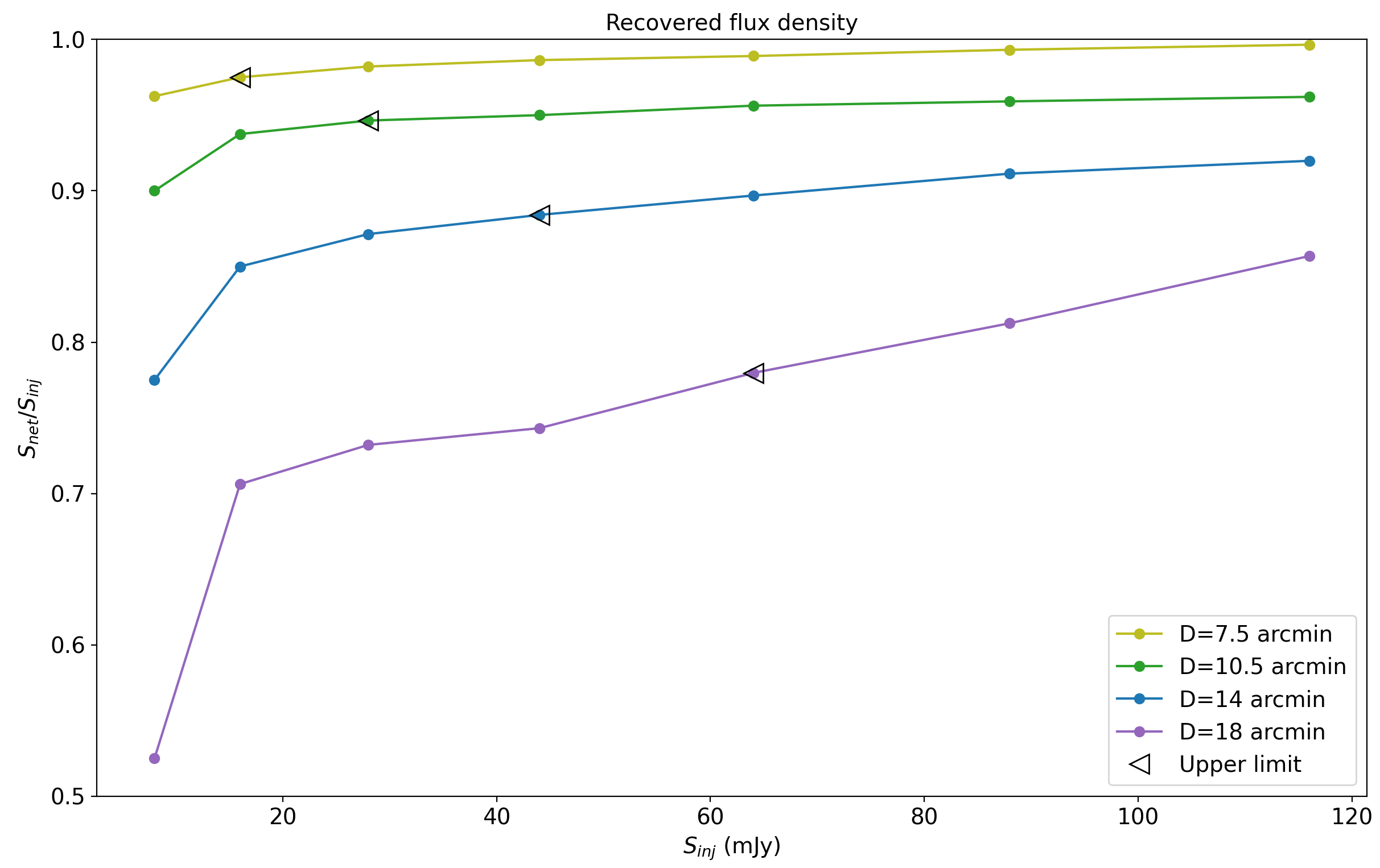}
	\caption{Fraction of recovered net flux density as a function of the injected flux density (within $3r_{\rm e}$) at various angular scales. Triangular markers indicate the flux density threshold at which the mock emission is not visible based on images in Appendix \ref{appendix:mock98}. Non-negligible flux density losses ($>15\%$) are expected for sources with angular sizes $D\gtrsim15'$ only.}
	\label{fractionSloss}%
\end{figure}

When dealing with extended radio emission, the role played by the \textit{uv}-coverage needs to be carefully taken into account. Indeed, interferometers lacking short baselines will not be able to fully recover extended sources in the sky. As a consequence, non-detections might arise from insufficient sampling of the \textit{uv}-coverage at short spacings. 
In previous works, LOFAR \textit{uv}-coverage was tested on the basis of injections on specific observations \citep{hoang18,botteon20b}. Here we aim to systematically investigate the associated flux density losses by performing $\sim 500$ injections in $\sim10$ cluster fields characterised by different data quality and noise. In each of these clusters, we injected mock halos with various flux densities (ranging from 5 to 145 mJy) and angular diameters (3, 5, 7.5, 10.5, 14, 18 arcmin, corresponding to baseline lengths of $\sim 1150{\rm k\lambda}, \; 690{\rm \lambda}, \; 460{\rm \lambda}, \; 330{\rm \lambda}, \; 250{\rm \lambda}, \; 190{\rm \lambda}$, respectively). As a representative example of one of these clusters, in Appendix \ref{appendix:mock98} we show a collection of mock halos injected in the field of PSZ2 G098.62+51.76, which was used to produce the plots discussed throughout this and next section. For each mock halo, we obtained the corresponding azimuthally-averaged surface brightness profile $I_{\rm post}(r)$ by adopting sampling annuli of width equal to half of the FWHM of the restoring beam of the images. The same analysis was performed on the pre-injection images, whose profiles $I_{\rm pre}(r)$ include the contribution of both radio sources and noise.

In Fig. \ref{MOCK98_netprofiles} we show the net profiles computed as the difference in each bin between $I_{\rm post}$ and $I_{\rm pre}$, overlaid on the injected profiles after being convolved with the restoring beam; since the peak of the profiles is always recovered, the first sampling annulus is not reported. The recovered net profiles are in agreement with the injected profiles for mock halos with sizes up to $D=10.5'$; this result is a consequence of the dense \textit{uv}-coverage of LOFAR on short spacings. On the other hand, deviations are visible in the cases of $D=14'$ and $D=18'$.

To estimate the losses in case of extended sources, in Fig. \ref{fractionSloss} we report the fraction of recovered flux density as a function of the injected flux density for each angular scale $D\geq7.5'$.  Overall, the recovered fractions are approximately constant over a wide range of injected flux densities up to $D=14'$. To further constrain the effective losses, we inspected the images shown in Appendix \ref{appendix:mock98} and determined the threshold among the injected flux densities $S_{\rm inj}$ at which the mock emission cannot be distinguished anymore from the local noise. Based on these flux density levels (indicated by triangular markers in Fig. \ref{fractionSloss}), on average flux density losses associated with the \textit{uv}-coverage are $\lesssim5\%$ up to $D=10.5'$, $\sim10\%$ for $D=14'$, and $\gtrsim20\%$ for sources with $D=18'$. In summary, we conclude that flux density measurements of sources observed by LOFAR are weakly dependent on the \textit{uv}-coverage up to $D\sim 15'$, meaning that the instrument is still well sampled at spacings $\sim250\lambda$, whereas non-negligible losses should be taken into account for sources with larger angular sizes. In the latter case, the use of baselines $<80\lambda$ should be considered to mitigate the flux density losses.

It is worth to mention that the recovered net profiles can appear higher than the injected profiles. This effect is described in details in \cite{shimwell22LoTSS} and is due to uncleaned components during the imaging step: when the peak of a cleaning component is lower than the cleaning threshold, it will be not correctly deconvolved and will be enhanced in the restored image. As seen in Fig. \ref{MOCK98_netprofiles}, this cleaning bias increases for wider (large $r_{\rm e}$) and fainter (low $I_{\rm 0}$) profiles. In the worst cases, i.e. $S_{\rm inj,tot}=5$ mJy, the recovered peaks are biased by factors of $\sim$1.2, 1.4, 1.6, 1.6, and 2, for $D=5', \; 7.5', \; 10.5', \; 14',\; 18'$, respectively, in agreement with \cite{shimwell22LoTSS} who found maximum factors of 2 as well through injections of Gaussian profiles. For a comparison, with $S_{\rm inj,tot}=10$ mJy and $D=18'$, the maximum cleaning bias factor is  $\sim$1.5, and further decreases for higher injected flux densities. Therefore, even if this systematic effect is present, the global results we drawn in this section still hold. 

This work focuses on exponential profiles as they can reproduce the observed brightness of radio halos, but losses with different injected models can be derived with the same approach. \cite{shimwell22LoTSS} probed the fraction of recovered flux density in LoTSS-DR2 by injecting Gaussian profiles, and found that this is of the order of $\sim95\%$ for standard deviations $\leq 2'$. Gaussian profiles are narrower than exponential profiles of the same width. Therefore, higher flux density fractions are expected to be recovered because of the increasing contribution of longer baselines. A systematic comparison between exponential and Gaussian profiles is beyond the aim of the present paper. However, as a sanity check, we injected bi-dimensional Gaussian profiles in the form of $I(r)=I_{\rm 0}e^{-\frac{r^2}{2 r_{\rm e}^2}}$ (where $r_{\rm e}$ now represents the standard deviation of the Gaussian and $S_{\rm inj,tot}=2\pi I_{\rm 0}r_{\rm e}^2$) in PSZ2 G098.62+51.76. We found losses $\lesssim10\%$ up to $D=18'$ (i.e. $r_{\rm e}=3'$), in agreement with results of \cite{shimwell22LoTSS}.

\subsection{Dependencies of the upper limits}
\label{sect:Upper limits dependencies}
\begin{figure*}
	\centering
	\includegraphics[width=1.0\textwidth] {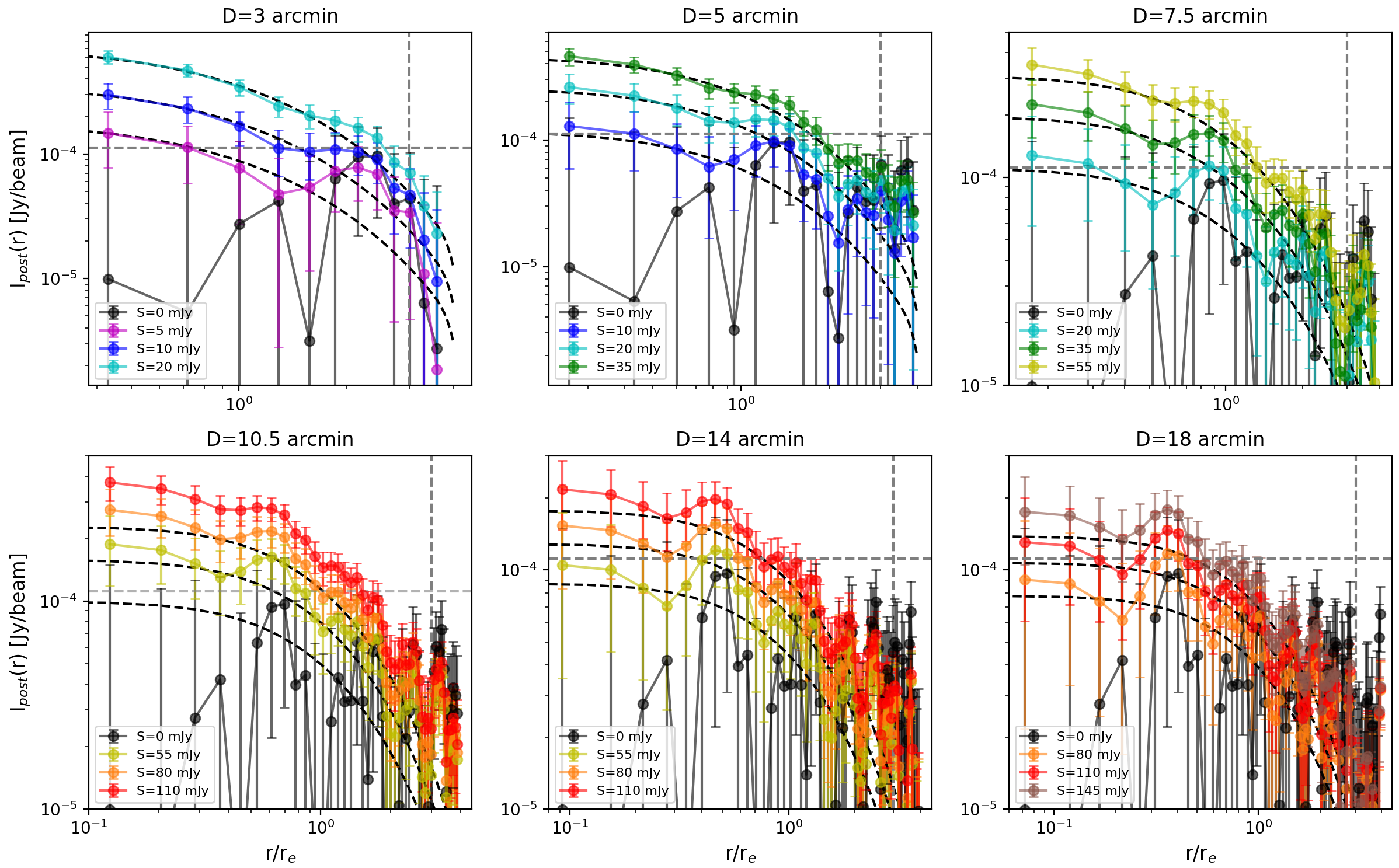} 
	\caption{Azimuthally-averaged profiles of the images shown in Appendix \ref{appendix:mock98} for PSZ2 G098.62+51.76. The dashed black lines represent the theoretical injected profiles. The sampled pre-injection and post-injection profiles are shown with black and coloured dots (see the total injected flux density in the legend), respectively. The grey vertical line indicates $r=3r_{\rm e}$. The grey horizontal line indicates the global $1\sigma$ noise level. }
	\label{MOCK98_grossprofiles}%
\end{figure*}   

\begin{figure}
	\centering
	\includegraphics[width=0.45\textwidth] {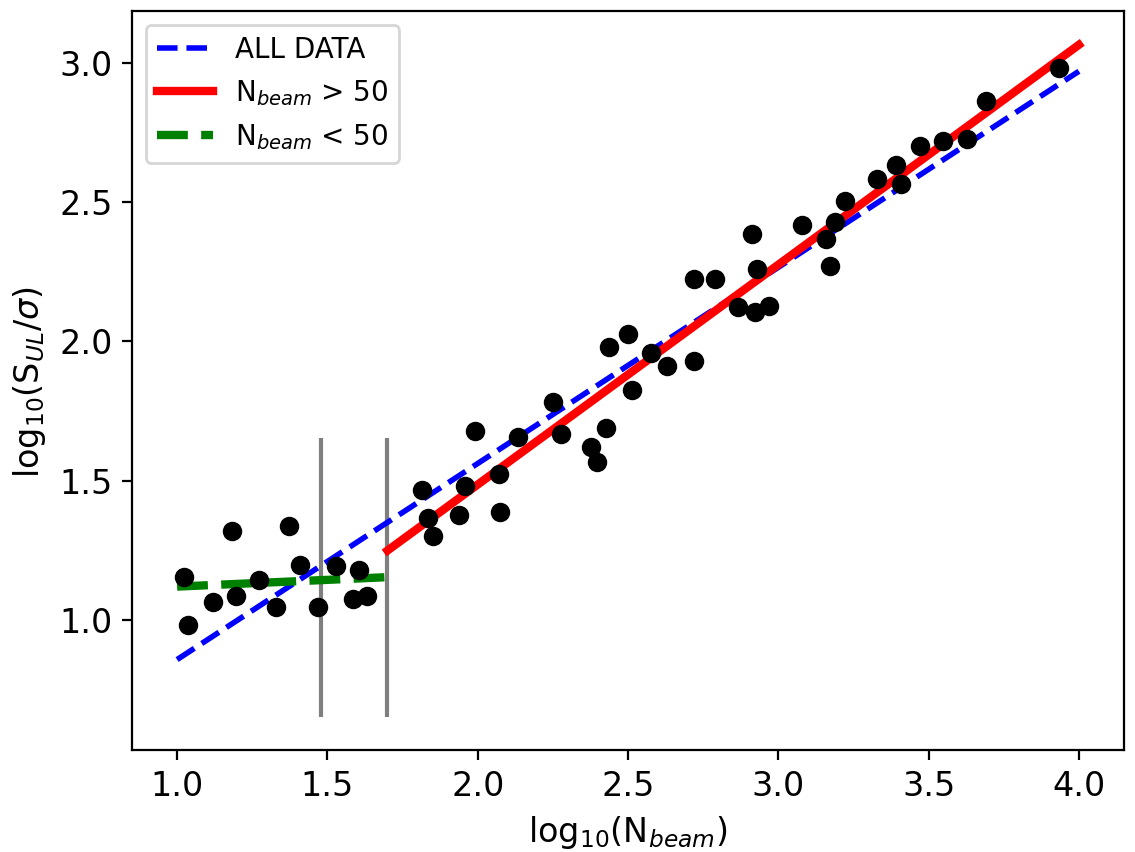} 
	\caption{Logarithmic plot of the ratio of the upper limit flux density against the noise of the map ($S_{\rm UL}/\sigma$) as a function of the number of beams ($N_{\rm beam}$) within the injected halo. The blue, red, and green lines are the fitted linear regressions (see Tab. \ref{table:fit}) obtained with no cut in $N_{\rm beam}$, with $N_{\rm beam}>50$, and with $N_{\rm beam}<50$, respectively. The grey vertical lines are drawn at $N_{\rm beam}=30$ and $N_{\rm beam}=50$.  }
	\label{ULcorrelation}%
\end{figure}  

\begin{table}
\centering
	\caption{Slope ($m$) and intercept ($q$) of the linear regressions shown in Fig. \ref{ULcorrelation} and discussed in the text.}
	\label{table:fit}
	\begin{tabular}{ccc}
	\hline
	\noalign{\smallskip}
	Data & $m$ & $q$  \\  
	\hline
	\noalign{\smallskip} 
	 All data & $0.703\pm 0.023$  & $0.155\pm0.057$ \\
	$N_{\rm beam}>50$ &  $0.789 \pm 0.029$ & $-0.091 \pm 0.081$ \\
	$N_{\rm beam}>30$ &  $0.794\pm 0.024$ & $-0.109\pm 0.066$ \\
	$N_{\rm beam}<50$  & $0.048\pm 0.141$ & $1.072\pm 0.192$ \\

	\noalign{\smallskip} 
	\hline
	\end{tabular}
\end{table}

Through our simulations, limited to baselines $>80\lambda$ only, we found that flux density losses are negligible and we can consider both detections and non-detections of extended emission independent on the \textit{uv}-coverage of LOFAR up to $D\sim15'$ (see also Sect. \ref{sect: Comparison} for a comparison with other facilities). Therefore, we expect that non-detections of radio halos in clusters are intrinsic or depend only on the sensitivity due to the depth of a specific observation. As a consequence, this suggests that our upper limits should rely on stringent parameters, that we aim to determine.

In Fig. \ref{MOCK98_grossprofiles} we both show the post-injection ($I_{\rm post}(r)$) and pre-injection ($I_{\rm pre}(r)$) surface brightness profiles (bright discrete sources were masked for this step). Uncertainties on the flux density of each sampling annulus are:
\begin{equation}\label{eq:errFLUXanelli}
\Delta S_{\rm bin}=\sigma\sqrt{N_{\rm beam,bin}} \; ,
\end{equation}
where $\sigma$ is the noise (in beam area units) of the radio image and $N_{\rm beam,bin}$ is the number of beams within each annulus. Therefore, the reported errors in Fig. \ref{MOCK98_grossprofiles} are computed as:
\begin{equation}\label{eq:errANELLI}
 \Delta I_{\rm bin}=\frac{\Delta S_{\rm bin}}{N_{\rm beam,bin}}=\frac{\sigma}{\sqrt{N_{\rm beam,bin}}} \; .
\end{equation}
We inspected our images and, as a rule of thumb, we found that the mock halos are still visible if at least 2-3 bins (as mentioned, the peak is excluded) are above the local noise level (i.e. $I_{\rm post}>I_{\rm pre}$) in our plots. Otherwise, the mock emission cannot be distinguished from the noise, allowing us to define the upper limit in combination with visual inspection.

For each set of injections, we determined the upper limit by eye, guided by the azimuthally-averaged surface brightness profiles. We found that limits depend on a combination of the noise ($\sigma$) and resolution ($\theta$) of the restored image, and on the angular size of the mock halo. In particular, the flux density of the upper limit correlates with the noise and the number of beams $N_{\rm beam}\sim D^2/\theta^2$ within the injected mock halo. As shown in Fig. \ref{ULcorrelation}, we performed a simple linear regression of the points ($\sim 50 $ upper limits among our injections) in a logarithmic plane as:
\begin{equation}\label{eq:fitUL}
 \log \left( {\frac{S_{\rm UL}}{\sigma}} \right) =m\log \left( {N_{\rm beam}} \right)+q\; ,
\end{equation}
where $m$ and $q$ are the fitted slope and intercept, respectively. 

Possible faint diffuse emission can be more easily identified by visual inspection if it is spread over larger areas, i.e. when $N_{\rm beam}$ is large, upper limits are guided by a surface brightness criterion; conversely, they follow a flux density criterion for smaller $N_{\rm beam}$. If our limits were determined based on a surface brightness criterion only, the flux density would scale with the area of the source as $S\propto r_{\rm e}^2 \propto N_{\rm beam}$, and thus we would expect a slope $m=1$ in Eq. (\ref{eq:fitUL}), whereas a flatter slope $m=0.5$ is expected if limits were exclusively driven by a flux density criterion \citep[see also Fig. 3 in][]{brunetti07}.  

These two regimes can be observed in Fig.
\ref{ULcorrelation}, with a flattening that roughly occurs for $N_{\rm beam}<50$. We performed different linear regressions by using all the points (blue line), points with $N_{\rm beam}>50$ (red line), and points with $N_{\rm beam}<50$ (green line); the results of these fits are reported in Table \ref{table:fit}. The average slope $m=0.703$ that we obtained by fitting all the points is in line with the competing trends predicted by the two criteria. By considering only points with larger $N_{\rm beam}$, the fitted slope $m=0.789$ steepens, in agreement with the behaviour predicted by the surface brightness criterion. We notice that the same (red) line can interpolate points down to $N_{\rm beam}\sim 30$ as well. Indeed, by considering points with $N_{\rm beam}>30$ we obtained a similar slope $m=0.794$, which is consistent with that obtained with $N_{\rm beam}>50$ within the fitting errors.

For fixed $\sigma$ and $D$, this positive correlation shows that deeper upper limits can be obtained for injections of lower $N_{\rm beam}$, which are achievable by decreasing the resolution of the restored image. Even though worse $\sigma$ are obtained by tapering the baselines, we found that images tapered to resolutions of $90''$ have typical rms which are factors $\lesssim2.5$ only with respect to those of images tapered to resolutions of $30''$. According to these results, the depth of the limit is primarily driven by $N_{\rm beam}\propto \theta^{-2}$ and it generally benefits from the decrease of the resolution.

\section{Upper limits for PSZ2 clusters in LoTSS-DR2}

In this section we describe the procedures adopted to obtain the upper limits for the NDE clusters of our sample, and compare them with the flux densities of the detected radio halos in LoTSS-DR2.

\subsection{NDE cluster sample}

\begin{figure}
	\centering
	\includegraphics[width=0.45\textwidth]{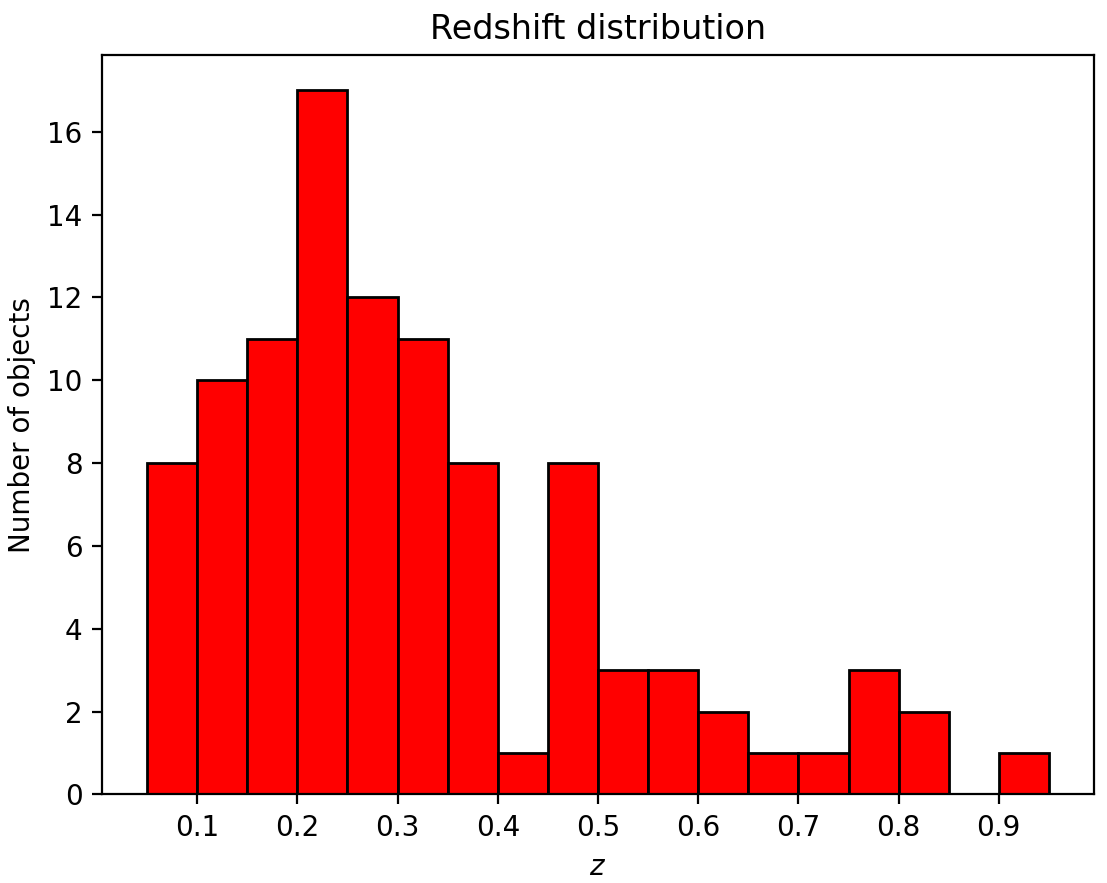} 
	\includegraphics[width=0.45\textwidth] {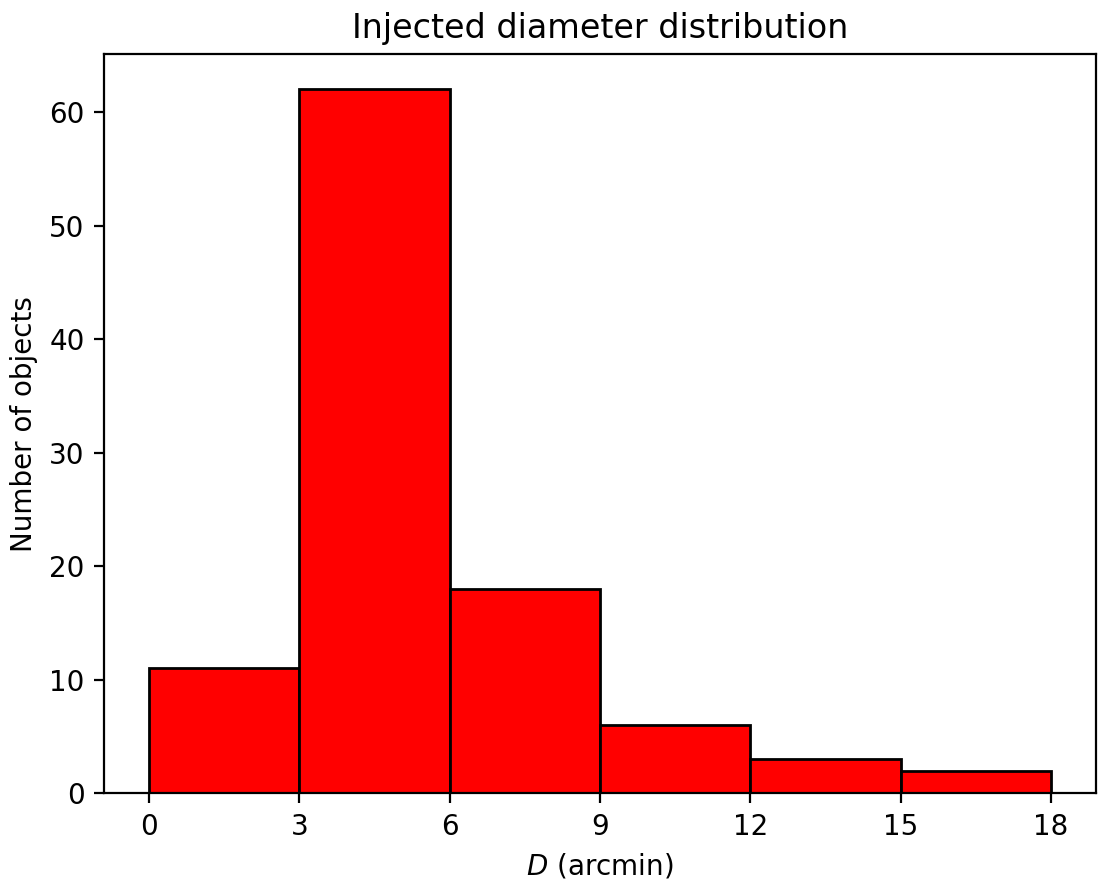}
	\caption{\textit{Top}: Distribution of redshifts of the 102 NDE galaxy clusters in our final sample. \textit{Bottom}: Distribution of the injected angular diameters under the assumption of $r_{\rm e,inj}=200$ kpc.}
	\label{sample_plots}%
\end{figure} 

From our initial sample of 140 NDE \textit{Planck} clusters, we first excluded 26 objects whose redshift is unknown, thus not allowing us to make assumptions on the angular size and limits to the radio power of a possible halo. For 11 additional galaxy clusters, upper limits would not be reliable due to the presence of contaminating AGN with extended emission and/or calibration artifacts in their central regions. Finally, we did not consider the lowest-redshift NDE cluster (PSZ2 G136.64-25.03, $z=0.016$) because a possible radio halo of $\sim$1 Mpc would have an angular size of $\sim 1^{\rm o}$ at the cluster redshift; sources of such large angular sizes require more specific calibration procedures than those adopted in LoTSS, making use of all the baselines instead of restricting to those $>80\lambda$ \citep[see e.g. the case of the Coma cluster in][]{bonafede21}. Therefore, we will focus on a sample of 102 NDE galaxy clusters. As shown in the upper panel of Fig. \ref{sample_plots}, these clusters lie in the redshift range $[0.062-0.9]$, where the mean and median are $\bar{z}=0.318$ and $\tilde{z}=0.267$, respectively. We refer to Table 1 in \cite{botteon22LoTSS} for additional information ($M_{500}$, $R_{\rm 500}$, image quality, X-ray data availability) on the full NDE cluster sample.

\subsection{Upper limit calculation}
\label{sect:Upper limit calculation}

\begin{figure*}
	\centering
	\includegraphics[width=0.31\textwidth] {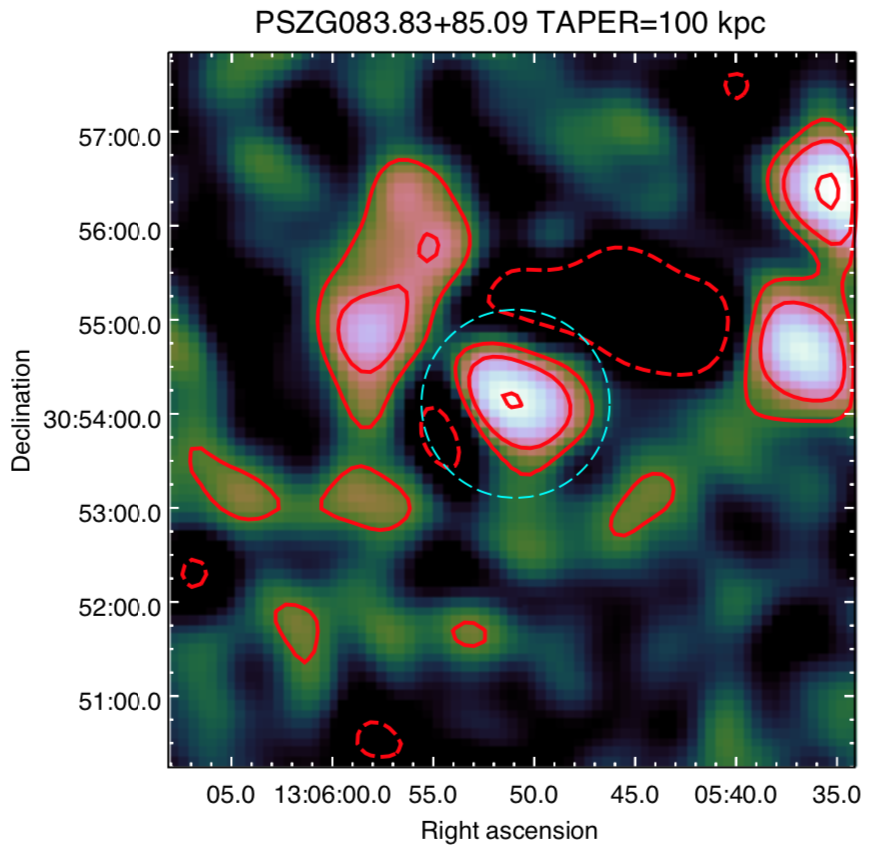} 
	\includegraphics[width=0.33\textwidth] {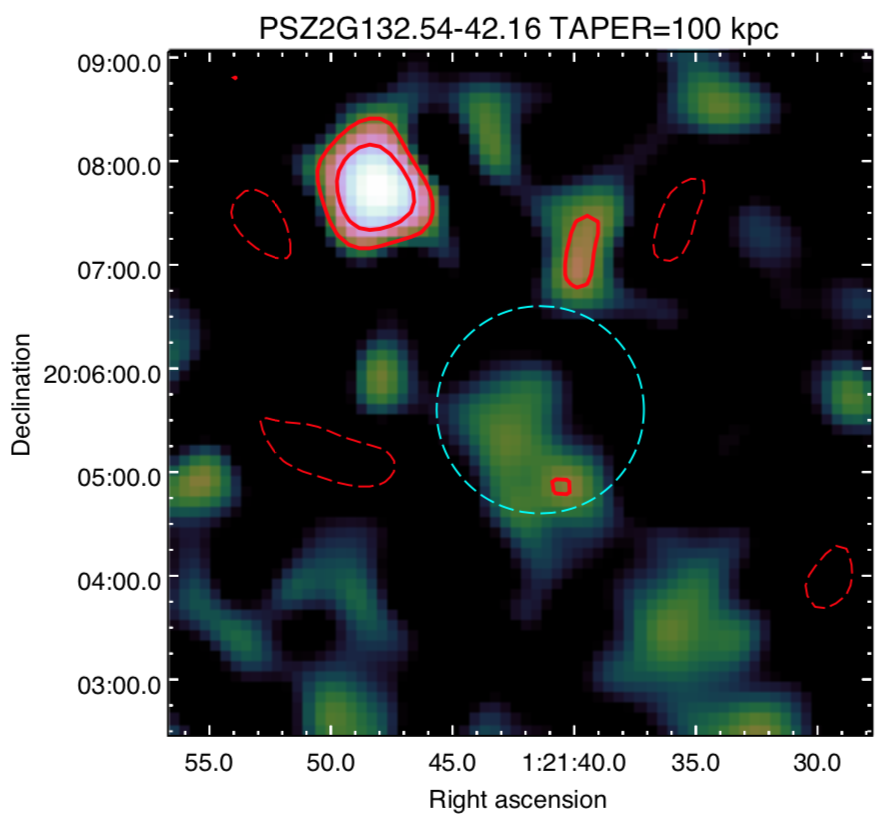}
	\includegraphics[width=0.34\textwidth] {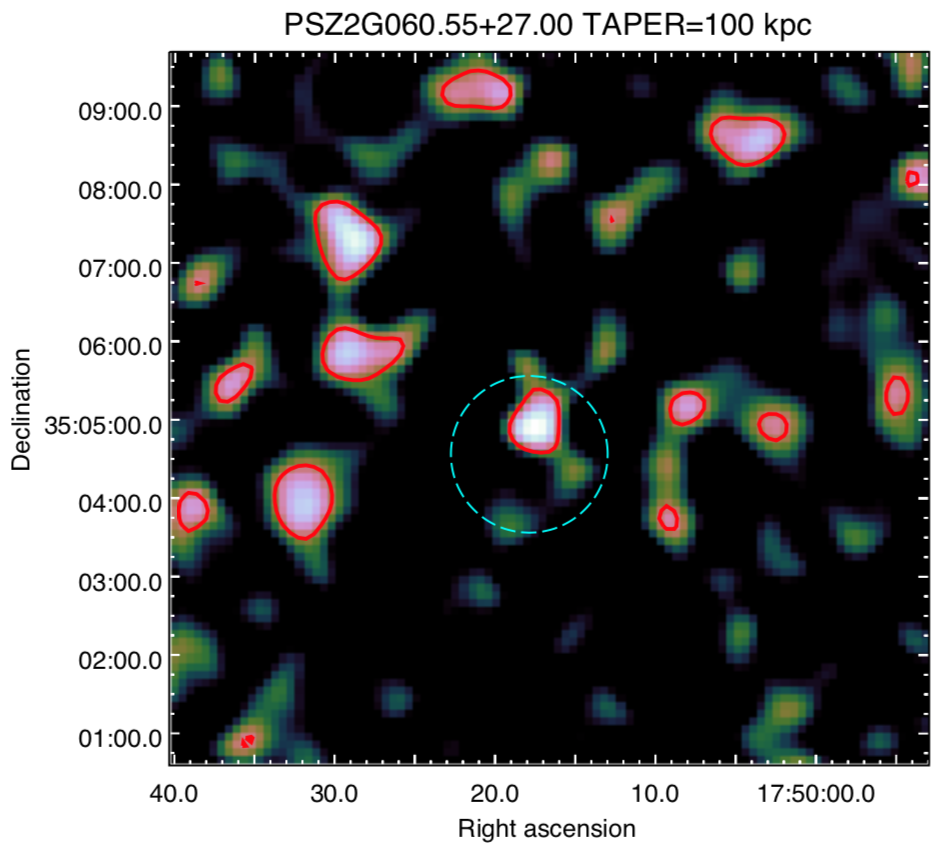}
	\caption{Representative examples of NDE clusters showing artifacts introduced by the subtraction of discrete sources near the cluster centre. The radio contours are drawn at $\pm 2\sigma$ and spaced by factors of 2. In all the panels, the dashed cyan circle indicates the cluster centre and has a fixed diameter of $2'$. \textit{Left}: NDE cluster excluded from our analysis due to severe (positive and/or negative) artifacts. \textit{Middle}: NDE cluster with ${\rm SQ=1}$ due to the absence of artifacts. \textit{Right}: NDE cluster with ${\rm SQ=2}$ due to the presence of moderate residuals. }
	\label{SQexample}%
\end{figure*} 

\begin{figure*}
	\centering
	\includegraphics[width=0.33\textwidth] {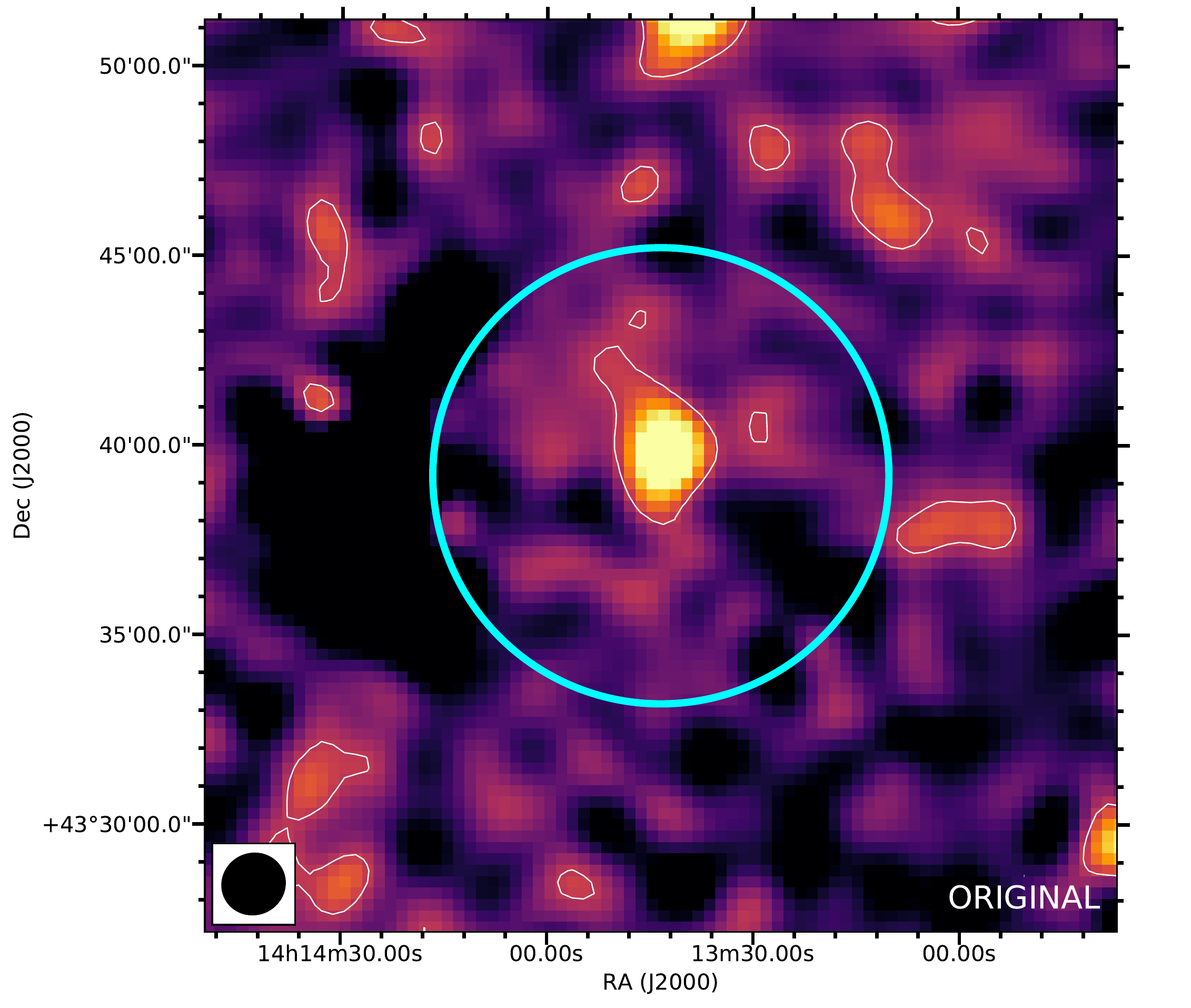} 
	\includegraphics[width=0.33\textwidth] {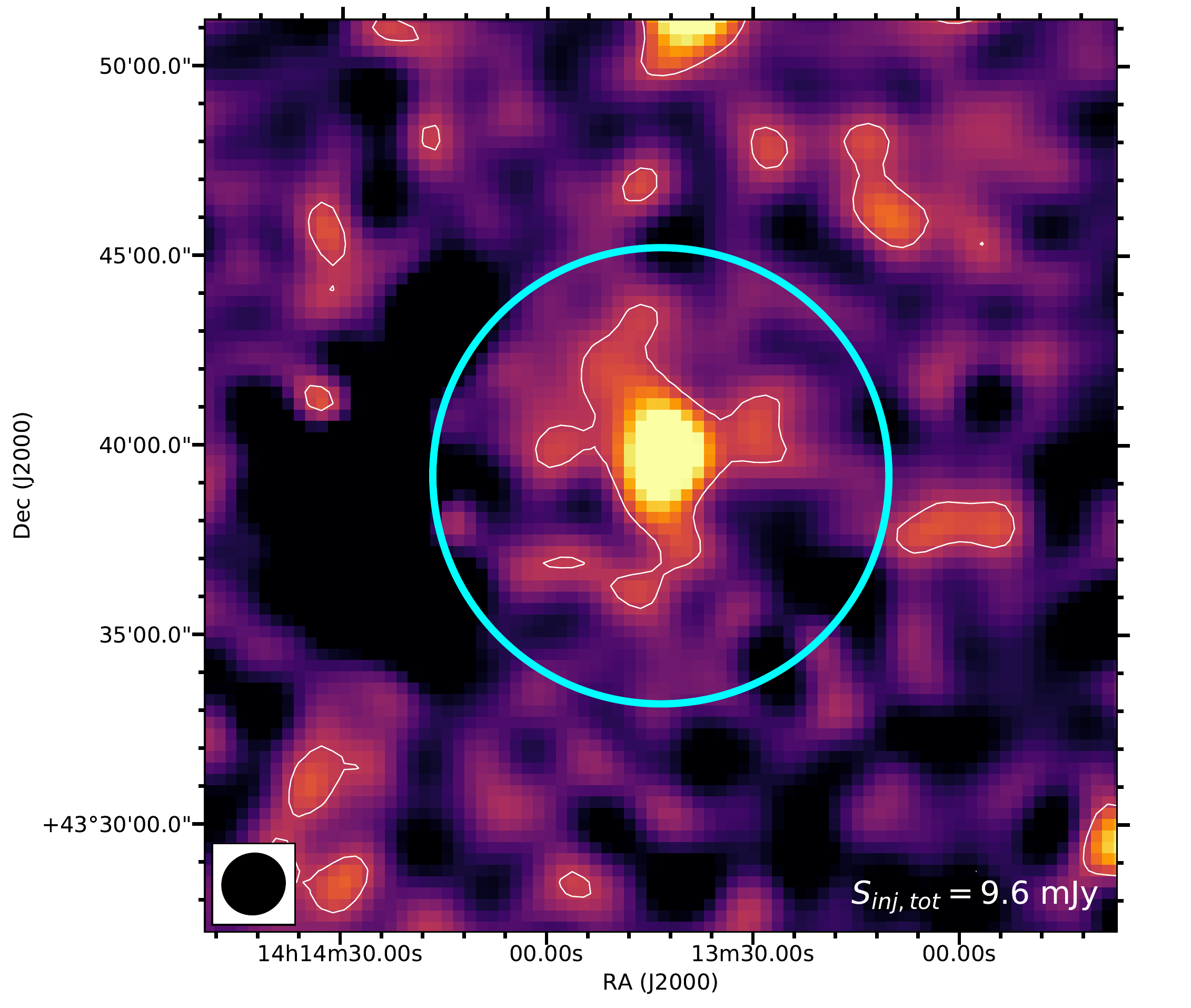}
	\includegraphics[width=0.33\textwidth] {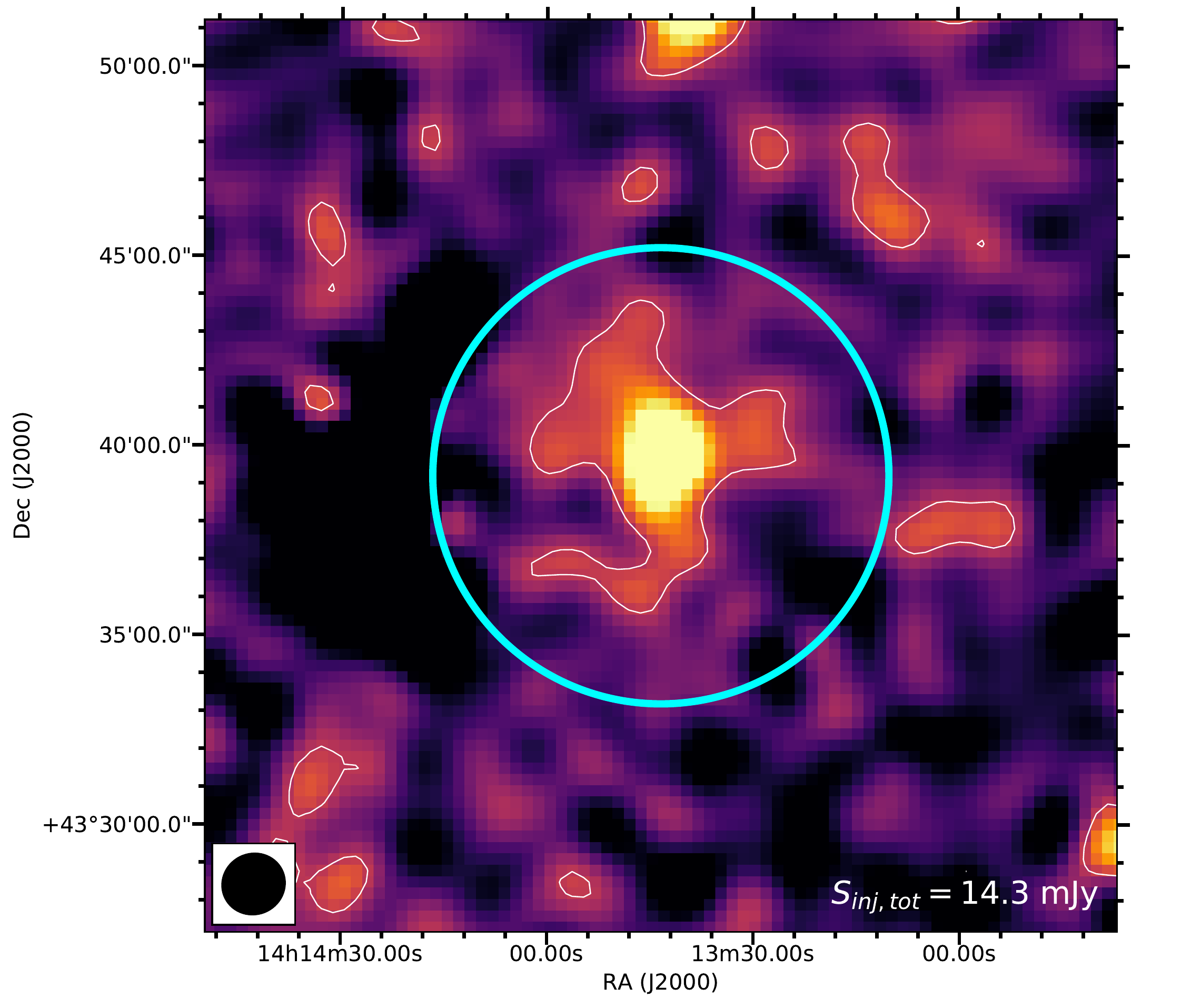}
	\caption{Examples of injections for a ${\rm SQ=2}$ cluster. Contours are drawn at $2\sigma$ of the pre-injection (left) image, and the cyan circle has a radius of $3r_{\rm e,inj}$. Two cycles of injections are performed with $S_{\rm UL}$ derived from Eq. (\ref{eq:fitUL}) and $1.5\times S_{\rm UL}$. The upper limit is obtained with the second cycle, which leaves extended excess $\sim 2$ times brighter with respect to the pre-injection image.}
	\label{SQ2limitexample}%
\end{figure*} 

\begin{figure}
	\centering
	\includegraphics[width=0.45\textwidth] {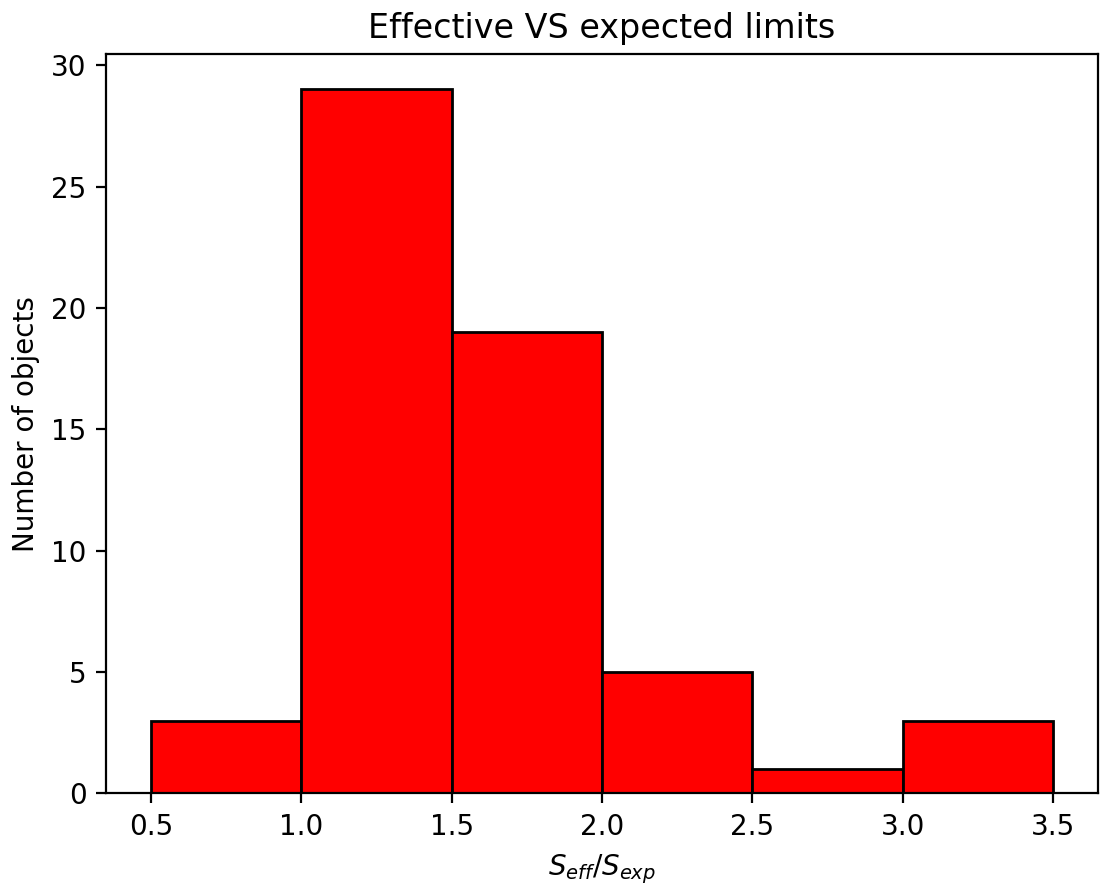} 
	\caption{Distribution of the ratio between the effective upper limit and the expected upper limit inferred from Eq. (\ref{eq:fitUL}) for the ${\rm SQ}=2$ clusters.}
	\label{trueVSexp}%
\end{figure}

The injection technique requires input values of $I_{\rm 0}$ and $r_{\rm e}$. As mentioned, the central brightness is obtained as $I_{0}=S_{\rm inj,tot}/{2\pi r_{\rm e}^2}$. In past works, the \textit{e}-folding radius was derived from scaling relations with the host cluster mass, but these are still poorly constrained and should be used with caution, as discussed in \cite{bonafede17}. To avoid the use of scaling relations, we consider mass-independent \textit{e}-folding radii. 

As discussed in Sections \ref{sect: The role of the uv-coverage} and \ref{sect:Upper limits dependencies}, our upper limits depend on the extent (in terms of number of beams) of the mock halo and the noise of the image, while the role of the \textit{uv}-coverage is negligible for the extension of halos in our LoTSS sample. Therefore, in principle, by means of Eq. (\ref{eq:fitUL}), we are able to immediately calculate the upper limits for the 102 NDE galaxy clusters of our sample as $S_{\rm UL}=10^q \sigma N_{\rm beam}^m$ (with $m=0.789$ and $q=-0.091$) without performing any additional injection. To this aim, values of $\sigma$ and $N_{\rm beam}$ are required, which can be obtained from images of the original datasets and with a choice of $r_{\rm e}$. Bearing in mind that the mean and median \textit{e}-folding radius of the PSZ2 radio halos in LoTSS-DR2 are $\bar{r}_{\rm e}=194$ kpc (with a standard deviation of 94 kpc) and $\tilde{r}_{\rm e}=186$ kpc, respectively, we assumed a nominal $r_{\rm e}=200$ kpc for all the NDE objects to derive the corresponding angular diameter ($D=6r_{\rm e}$). With this choice, the sizes of our mock halos are in the range [$2.6'$-$16.7'$]. The distribution of the angular diameters is shown in the lower panel of Fig. \ref{sample_plots}, where the mean and median are $\bar{D}=5.7'$ and $\tilde{D}=4.9'$, respectively; the $82\%$ of our clusters have $D<7.5'$, whereas only 2 clusters have $D>14'$. We then produced images by tapering the baselines to different convenient low resolutions to find the deepest $S_{\rm UL}$ from Eq. (\ref{eq:fitUL}) among the various combinations of $\sigma$ and $N_{\rm beam}$, under the condition of avoiding $N_{\rm beam}\ll 30$, where the slope of our correlation significantly flattens.

Having said that, since radio halos are centrally-located sources, physically-meaningful upper limits cannot ignore the local environment close to the cluster centre, which is usually dense of contaminating discrete sources. Therefore, at this stage it is also necessary to take into account the quality of the subtraction of the discrete sources close to the cluster centre, before blindly adopting Eq. (\ref{eq:fitUL}). The subtraction process is typically not perfect for many reasons; it assumes that the model of discrete sources obtained from the long baselines adequately describes their emission for the short baselines as well, but extended sources are primarily sampled by the short spacings, and calibration may not be homogeneous for all baselines, thus providing unequal levels of subtraction. These effects may introduce subtraction artifacts in the form of positive residual blobs and negative holes, which are enhanced at low resolution and contaminate the faint diffuse emission; in these cases, upper limits are not driven by the instrumental capabilities (i.e. the reached rms noise), but by the level of imaging artifacts, thus making the $S_{\rm UL}$ provided by Eq. (\ref{eq:fitUL}) not fully trustworthy. 

For these reasons, we inspected our non-subtracted and source-subtracted images at various resolutions, and excluded from our analysis 27 out of 102 objects which are affected by severe subtraction artifacts that prevent us to provide meaningful upper limits\footnote{Re-observations and/or more refined calibration and subtraction processes would be necessary to derive solid upper limits for these targets.} (see an example in Fig. \ref{SQexample}). We then assigned a subtraction-quality (SQ) parameter to each remaining NDE cluster, based on the presence and impact of subtraction artifacts close to the cluster centre\footnote{SQ is a qualitative and subjective parameter, but is rather reproducible by following the examples in Fig. \ref{SQ2limitexample}, and allows to easily find the best strategy to obtain upper limits.}; we assigned ${\rm SQ}=1$ if subtraction artifacts are absent or negligible (a sub-sample of these targets was used to derive Eq. (\ref{eq:fitUL})), and ${\rm SQ}=2$ if subtraction artifacts are not negligible (see examples in Fig. \ref{SQexample}) . We directly derived upper limits through Eq. (\ref{eq:fitUL}) only if ${\rm SQ}=1$ (15 out of 102).
If ${\rm SQ}=2$ (60 out of 102), the presence of artifacts is not dominant and more reliable limits can be determined through the injection process in the source-subtracted data (i.e. through the `Sub. \& Inj.' scheme). In these cases, we performed a first cycle of injection by using the flux density provided by Eq. (\ref{eq:fitUL}) with tapers corresponding to $N_{\rm beam}\sim30$, and then increased/decreased $S_{\rm inj,tot}$ in few additional cycles to further constrain the limit level. Pre-injection and post-injection images were then inspected and, guided by the $2\sigma$ contour levels, we considered as effective limits those cases where the mock emission leaves extended excess $\sim 2$ times brighter than the previous injection cycle (see an example in Fig. \ref{SQ2limitexample}). 

Based on the scatter around the fit of Eq. (\ref{eq:fitUL}) in the region $N_{\rm beam}\sim 30-50$, and the grid of values adopted to vary $S_{\rm inj,tot}$ in each cycle of injections, we claim conservative uncertainties of $\sim 10-15\%$ on the upper limits derived with the presented methods. As discussed in Sect. \ref{sect: The role of the uv-coverage}, the surface brightness is overestimated if emission is not fully deconvolved during imaging. This cleaning bias is expected to be higher for injections at the level of the upper limit and large angular size. Nevertheless, even in the case we were systematically biased by this cleaning effect, uncleaned excess would appear brighter, thus making the upper limits to be more conservative.

We now aim to evaluate the efficacy of our strategies based on Eq. (\ref{eq:fitUL}). For each of the 60 NDE clusters with ${\rm SQ}=2$, we obtained an estimate of the expected upper limit from Eq. (\ref{eq:fitUL}) by assuming $N_{\rm beam}=30$ and the noise corresponding to the adopted taper. In Fig. \ref{trueVSexp} we report the distribution of the ratio between the effective and expected upper limits; the mean and median ratios are 1.59 and 1.44, respectively. Most of the targets (51 out of 60) have ratios $\leq2$, with maximum ratios of $\sim 3.3$. The 9 clusters with ratios $>2$ have slightly higher levels of artifacts than those typical of ${\rm SQ=2}$, therefore they require higher injected flux densities to constrain the upper limit. This confirms that Eq. (\ref{eq:fitUL}) remains a good starting point even in the presence of moderate artifacts, thus allowing to perform fewer additional cycles of injections in order to constrain the final level of the limit. Moreover, even though the required time to obtain upper limits naturally depends on the available computing resources (see details in Sect. \ref{sect:injection algorithm}), we estimate that our strategies allowed us to remarkably reduce the computing time by factors $\sim$3-4\footnote{A single run of injection plus imaging takes $\sim2.0+0.5$ h of computing time for a single LoTSS pointing. With our methods, ${\rm SQ=1}$ objects only require imaging to obtain $\sigma$ and $\theta$, whereas 2-3 cycles of the full process are necessary for ${\rm SQ=2}$ clusters. With methods in the literature, at least 6 cycles of injections plus imaging are instead necessary.} with respect to standards methods in the literature.

\subsection{Upper limits and radio halos}

\begin{figure}
	\centering
	\includegraphics[width=0.45\textwidth] {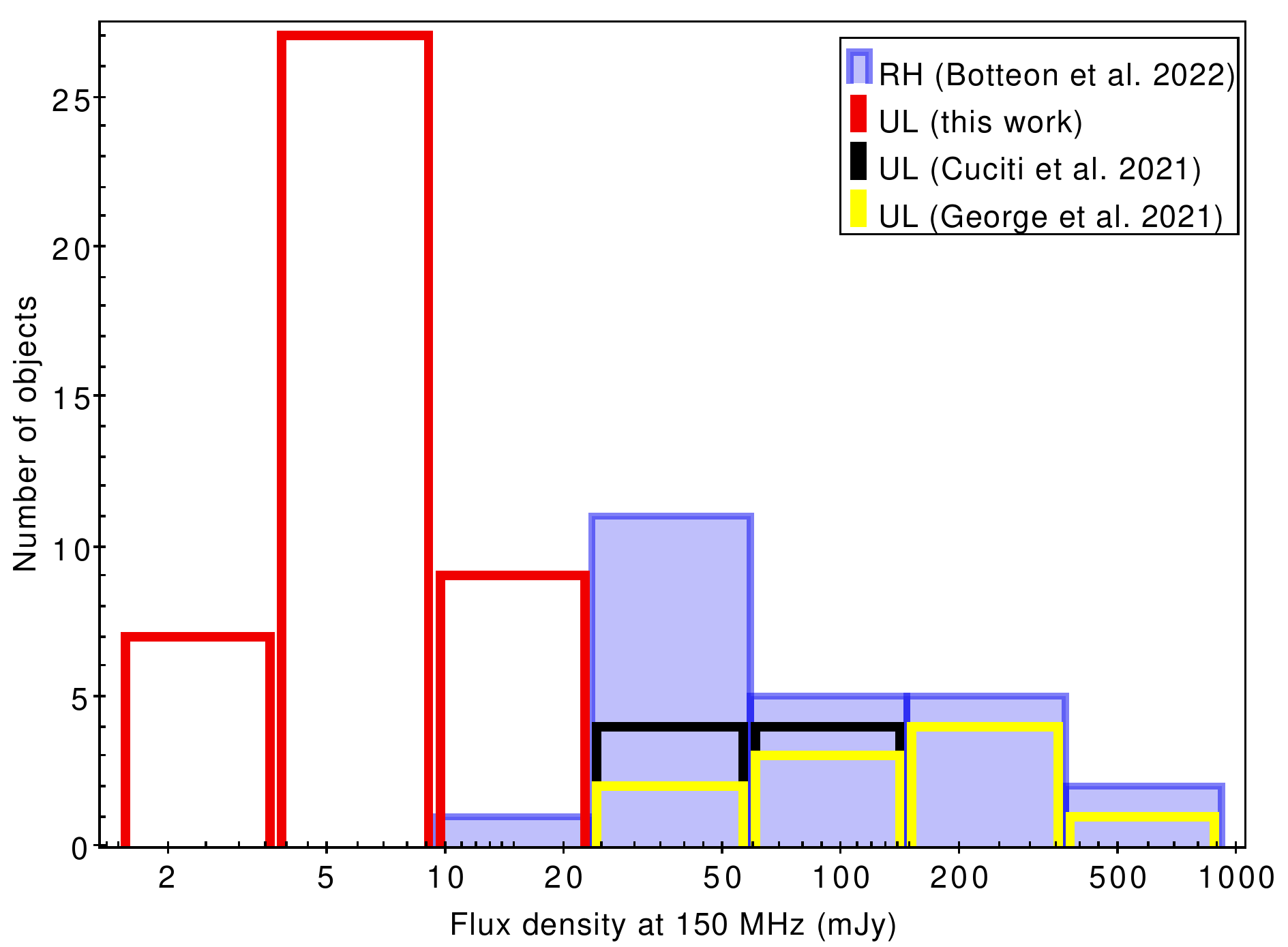} 
	\includegraphics[width=0.45\textwidth] {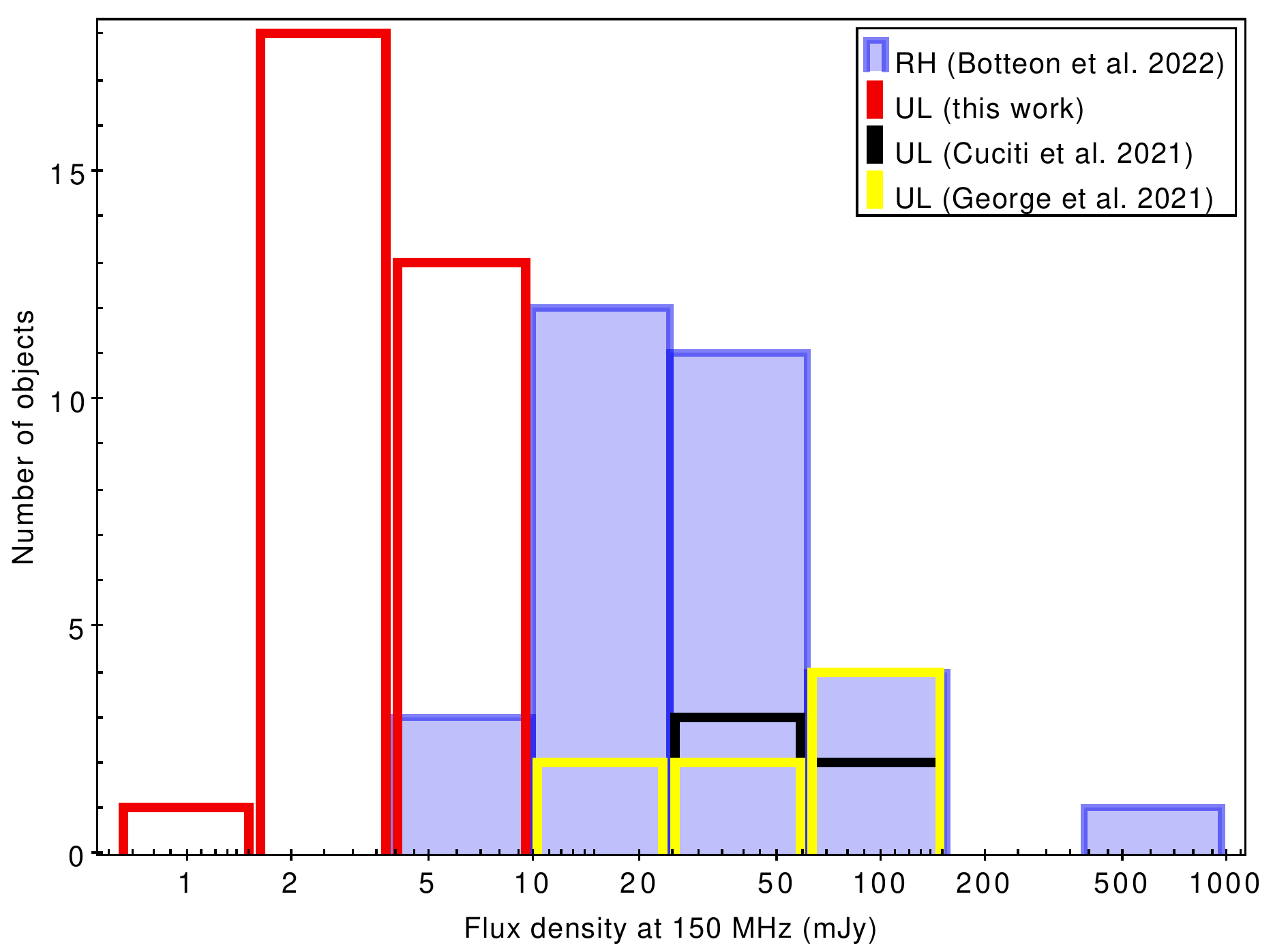} 
	\caption{Distributions of the flux density of confirmed and candidate radio halos (in blue) with $100 \leq r_{\rm e}\leq 400$ kpc from \cite{botteon22LoTSS}, our upper limits (in red), and upper limits from \cite{cuciti21a} (in black) and \cite{george21b} (in yellow). Upper limits from the literature are re-scaled to 150 MHz by assuming $\alpha=1.3$. The samples are split in redshift bins as $z<0.3$ (top panel) and $z>0.3$ (lower panel).}
	\label{limitiVSaloni}%
\end{figure} 

Uncertainties in the beam model of LOFAR HBA can introduce offsets in the flux density scale when amplitude solutions are transferred from the primary calibrator to the target \citep[e.g.][]{hardcastle16}. Therefore, all images need to be multiplied by a factor $f_{\rm LoTSS}$ (of order of unity), which is derived for each pointing after the data calibration \citep[see][]{botteon22LoTSS,shimwell22LoTSS} to align the flux density scale of LoTSS with that of \cite{roger73}. As our injections are performed in the \textit{uv}-data, we have to take into account the flux density scale correction by multiplying $S_{\rm UL}$ at 144 MHz by $f_{\rm LoTSS}$. 

Our upper limits will be exploited in statistical analysis in other papers of the series \citep[][Cuciti et al., in prep.]{zhang22,cassano23}, where the radio power of the detected halos is reported at 150 MHz, within radii of $3r_{\rm e}$ \citep[as done in][]{botteon22LoTSS}. To compare detections and upper limits, we therefore scaled $S_{\rm UL}$ from 144 to 150 MHz ($S_{\rm UL,150}=S_{\rm UL,144}(150/144)^{-\alpha}$, where $\alpha=1.3$), calculated the corresponding radio power, and finally considered the $80\%$ of this value which is expected within $3r_{\rm e,inj}$. In summary, the \textit{k}-corrected radio powers of the upper limits are obtained as:
\begin{equation}
P_{\rm UL,150}=0.8 \times f_{\rm LoTSS} \times 4 \pi D_{\rm L}^{2}S_{\rm UL,150}(1+z)^{{\alpha-1}} \; ,
\label{radiopower}
\end{equation}
where $D_{\rm L}$ is the luminosity distance at the cluster redshift. 

The main host properties (\textit{Planck} coordinates, redshift, mass), injection parameters (centre of injection, central brightness, \textit{e}-folding radius), flux density, and radio power at 150 MHz for the final sample of 75 NDE clusters are reported in Table \ref{tab: limits} in Appendix \ref{appendix:Upper limits}.

A comparison of the distributions of flux densities for our upper limits and the 56 (confirmed and candidate) radio halos detected in LoTSS-DR2 having $100 \leq r_{\rm e}\leq 400$ kpc \citep{botteon22LoTSS} is shown in Fig. \ref{limitiVSaloni}, where we also reported upper limits from the literature \citep{cuciti21a,george21b} that we re-scaled to 150 MHz. For a clearer inspection, we split the samples in two redshift bins at $z<0.3$ (upper panel) and $z>0.3$ (lower panel). Our limits range in 1.5-17 mJy, whereas radio halos have flux densities in the range $\sim$3-700 mJy. Except for few objects, the distributions of our limits are well separated from the radio halos in the lower redshift bin, whereas more overlapping is found in the higher redshift bin. Our limits are deeper than those from the literature obtained with other interferometers, thanks to the higher sensitivity of LOFAR (see also Sect. \ref{sect: Comparison}). The deepest limits at 150 MHz are 36.5 mJy (at $z=0.320$) and 11.3 mJy (at $z=0.396$) in samples from \cite{cuciti21b} and \cite{george21b}, respectively, which are factors 24.3 and 7.5 higher than our deepest limit of 1.5 mJy (at $z=0.830$), and factors 2.8 and 6.5 higher than our limits for objects at similar redshift ($z=0.397$ and $z=0.318$).

\section{Comparison with uGMRT and JVLA capabilities}

\label{sect: Comparison}
\begin{figure*}
	\centering
	\includegraphics[width=1.0\textwidth] {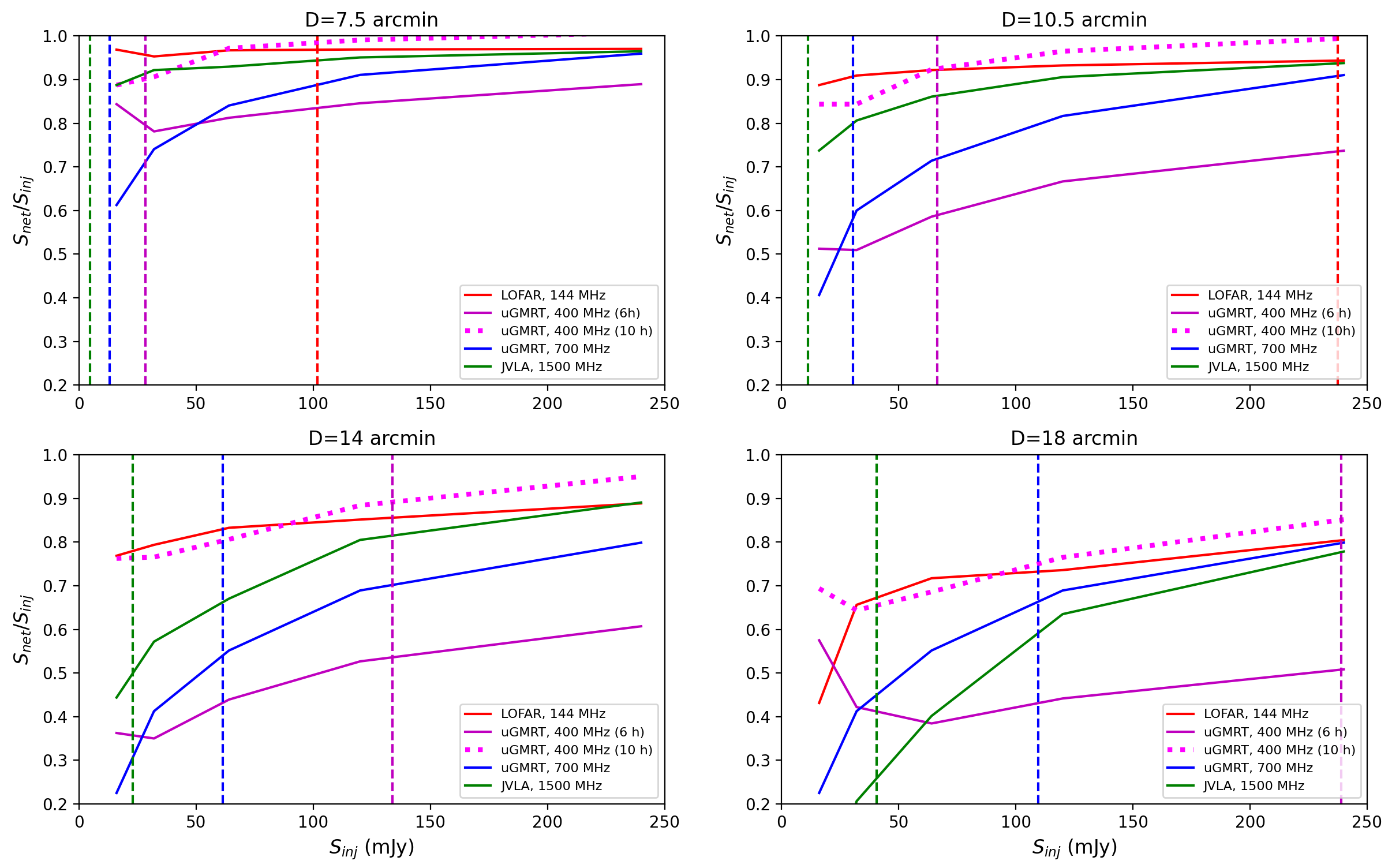}
	\caption{Recovered net flux density as a function of the injected flux density for various angular diameters and instruments. Red, magenta, blue, and green curves are injections in LOFAR (HBA at 144 MHz), uGMRT (band-3 at 400 MHz, 6 and 10 h on-source are shown with solid and dotted lines, respectively), uGMRT (band-4 at 700 MHz), and JVLA (DnC+BnC array, L-band at 1.5 GHz) datasets, respectively. The expected flux density integrated up to $3r_{\rm e}$ of a representative radio halo with $M_{\rm 500}=5\times10^{14} \; M_{\odot}$ is indicated by a dashed vertical line and can be exploited to estimate effective losses (see Table \ref{table:confronto} and discussion in Sect. \ref{sect: Comparison}).}
	\label{confrontoUV}%
\end{figure*}

 \begin{table*}
   \centering
   	\caption[]{Estimates of the percentage of recovered flux density for LOFAR (144 MHz), uGMRT (400 MHz), uGMRT (700 MHz), and JVLA (DnC+BnC-array, 1.5 GHz) based on our simulations. A representative radio halo with $M_{500}=5\times 10^{14} \; M_{\odot}$, $D=1.2$ Mpc, and $\alpha=1.3$ is considered as a general reference for flux densities at each frequency.}
   	\label{table:confronto}
   	\begin{tabular}{ccccccccc}
   	\hline
   	\noalign{\smallskip}
   	$D$ & $S_{\rm halo,144}$ & LOFAR & $S_{\rm halo,400}$ & ${\rm uGMRT^{\rm 400}}$ & $S_{\rm halo,700}$ & ${\rm uGMRT^{\rm 700}}$ & $S_{\rm halo,1500}$ & JVLA    \\
   	(arcmin) & (mJy) & ($\%$) &  (mJy) & ($\%$) & (mJy) & ($\%$) & (mJy) & ($\%$) \\
   	\noalign{\smallskip}
  	\hline
   	\noalign{\smallskip}

7.5 & 101.8 & 95 & 28.4 & 80-90 & 13.0 & 60 & 4.8 & 90 \\
10.5 & 237.5 & 95 & 66.4 & 60-90 & 30.4 & 60 & 11.3 & 70 \\       
14 & 479.1 & 90 & 133.8 & 55-90 & 61.4 & 55 & 22.8 & 50 \\
18 & 855.0 & 80 & 238.9 & 50-85 & 109.4 & 65 & 40.6 & 25 \\
   	\noalign{\smallskip}
   	\hline
   	\end{tabular}
   \end{table*}

\begin{figure}
	\centering
	\includegraphics[width=0.3\textwidth] {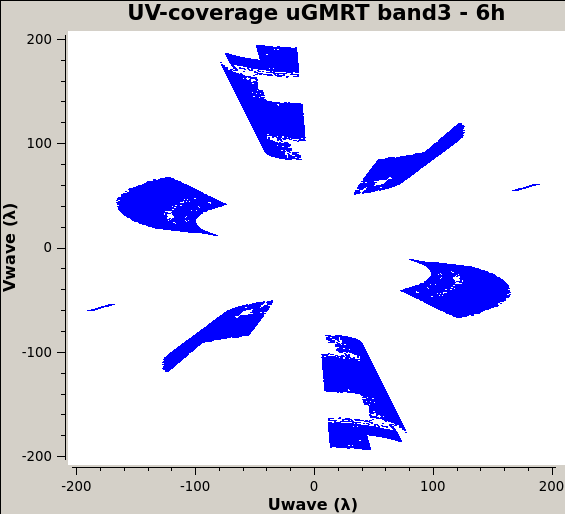}
	\includegraphics[width=0.3\textwidth] {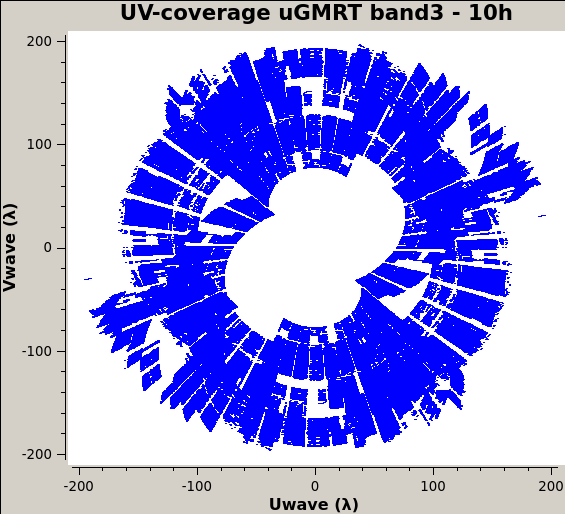}
	\includegraphics[width=0.3\textwidth] {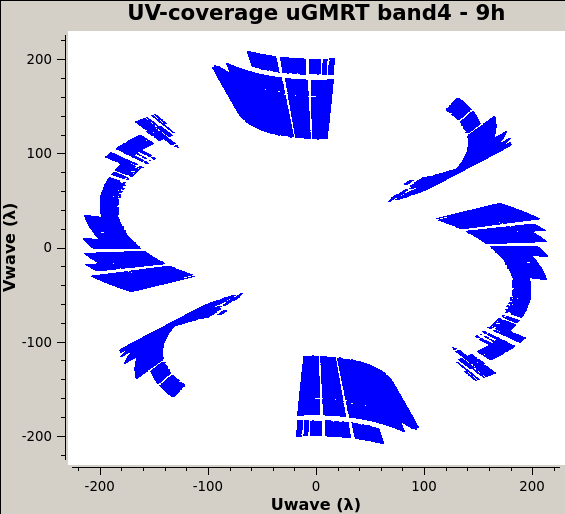}
	\caption{Inner ($\leq200\lambda$) \textit{uv}-coverage of the uGMRT observations used for injections: 6 hr in band-3 (${\rm Dec \sim -13^o}$, top panel), 10 hr in band-3 (${\rm Dec \sim +64^o}$, middle panel), 9 hr in band-4 (${\rm Dec \sim -29^o}$, lower panel). }
	\label{uvplot}%
\end{figure}  

To date, studies of radio halos have been mainly based on observations carried out with the upgraded Giant Metrewave Radio Telescope (uGMRT) and the Karl G. Jansky Very Large Array (JVLA). To quantitatively compare the capabilities of these instruments with LOFAR, we injected mock halos with $S_{\rm inj,tot}$ ranging from 20 to 300 mJy in LOFAR (8 hr observation), uGMRT (band-3 at 400 MHz, 6 and 10 hr observations), uGMRT (band-4 at 700 MHz, 9 hr observation), and JVLA (L-band in combined DnC and BnC array configurations at 1.5 GHz, 8 hr on-source in total) datasets. Even though the minimum baseline lengths of uGMRT and JVLA are similar to the inner \textit{uv}-cut considered for LOFAR at $80\lambda$, the density of the \textit{uv}-coverage at short spacings is notably different for each dataset.

As in Fig. \ref{fractionSloss}, we obtained the recovered flux density by considering the difference between $S_{\rm post}$ and $S_{\rm pre}$. Sources were not subtracted to avoid the introduction of heterogeneous artifacts of subtraction at the various frequencies. In Fig. \ref{confrontoUV} we show the fraction of the recovered flux density as a function of the injected flux density for the three considered instruments. Losses are negligible for all the three facilities up to $D=5'$ (and thus $D=3'$ and $D=5'$ are not shown), but different behaviours are seen for larger angular sizes. To roughly determine the effective losses, we consider a representative radio halo of $D=1.2 \; {\rm Mpc}$, host mass of $M_{\rm 500}=5\times 10^{14} \; M_{\odot}$, and spectral index $\alpha=1.3$; we infer its radio power at 150 MHz from the mass-power relation (with fitted slope and intercept from Cuciti et al., in prep.), and re-scale its flux density based on the required angular diameters and frequency; the corresponding flux density is indicated by a vertical dashed line in Fig. \ref{confrontoUV}. Our estimates on the effective recovered fractions are listed in Tab. \ref{table:confronto} (due to large differences, for uGMRT band-3 we both reported the minimum and maximum values, obtained with 6 and 10 hr observations, respectively).

Obtaining more accurate assessment of the performances for uGMRT and JVLA requires systematic injections in many datasets, as done for LOFAR, but this is beyond the scope of this work. However, important conclusions can be drawn through our simulations. Owing to the high number of short baselines of LOFAR, the 8-hr LoTSS pointings ensure similar densities of the \textit{uv}-coverage, independently on the specific observation. We found that uGMRT can provide very high performances in recovering extended emission, but the recovered trends are remarkably different for the uGMRT datasets that we considered. Indeed, Figs. \ref{confrontoUV}, \ref{uvplot} indicate that the density of uGMRT \textit{uv}-coverage is more dependent on the specific observation (e.g. total observing time, bandwidth, declination of the target, level of interference, flagging); in our tests, the lowest performances were obtained with the 6-hr band-3 dataset, which has not only the shortest duration, but also the highest number of flagged antennas, which largely contribute to compromise the recovery of the mock halos. Our simulations also suggest that losses for JVLA are high at $D\geq10.5'$. In summary, for uGMRT and JVLA data, the inspection of the \textit{uv}-coverage and classical derivation of upper limits through injection are recommended rather than the usage of scaling relations similar to ours in Eq. (\ref{eq:fitUL}).

\section{Summary and conclusions}

LoTSS-DR2 includes 140 non-detections of radio halos in \textit{Planck} clusters \citep{botteon22LoTSS}. We exploited these data to test the instrumental capabilities of LOFAR to recover diffuse extended emission and determine upper limits to the radio power of a possible halo. 

Through the injection of mock visibilities simulating radio halos into the observed datasets, we estimated the flux density losses due to insufficient short baselines. We find that they are negligible ($\lesssim 5$-$10\%$) for sources of sizes up to $D= 14'$, and reach fractions of $\sim 20\%$ at $D= 18'$. As common for LOFAR HBA data, our simulations were limited to baselines $>80\lambda$ as well, meaning that more flux density can be recovered by including shorter baselines. For the first time, we systematically carried out tests on a large sample of datasets of different quality, demonstrating that LOFAR is one of the facilities with the densest \textit{uv}-coverage in its inner part. It is thus able to recover large scale emission with lower flux density losses with respect to other instruments. Moreover, the low frequency range of LOFAR allows to explore host mass regimes that could be barely probed by facilities operating at higher frequencies.

We showed that non-detections of diffuse emission can be considered approximately independent on the \textit{uv}-coverage of LOFAR in LoTSS observations. We therefore explored the parameters which determine the flux density of the upper limits, and found a relation with the noise of the image and the number of beams within the injected mock halo (depending on the resolution of the image and angular size of the mock emission). Our relation can be safely adopted to infer the upper limit if the subtraction of the discrete sources close to the cluster centre does not leave contaminating artifacts, which are enhanced at low resolution. Otherwise, subtraction artifacts typically drive the level of the limit, thus making the injection procedure and visual inspection still necessary to provide more reliable limits.   

After excluding objects lacking redshift information, with extended radio galaxies, and with severe subtraction artifacts, we obtained upper limits for 75 \textit{Planck} clusters in LoTSS-DR2. Our limits will be exploited in forthcoming statistical analyses \citep[][Cuciti et al., in prep.]{zhang22,cassano23} and compared to the detected radio halos in LoTSS-DR2 to provide information on the populations, origin, and evolution of these sources.

\begin{acknowledgements}
We thank the referee for useful comments and suggestions that have improved the presentation of the paper. 
LB acknowledges Kamlesh Rajpurohit for providing us the JVLA data. GB and RC acknowledge support from INAF mainstream project `Galaxy Clusters Science with LOFAR', 1.05.01.86.05. ABotteon and RJvW acknowledge support from the VIDI research programme with project number 639.042.729, which is financed by the Netherlands Organisation for Scientific Research (NWO). VC acknowledges support from the Alexander von Humboldt Foundation. ABonafede and ABotteon acknowledge support from ERC Stg DRANOEL n. 714245 and MIUR FARE grant `SMS'. MB acknowledges funding by the Deutsche Forschungsgemeinschaft (DFG, German Research Foundation) under Germany's Excellence Strategy – EXC 2121 `Quantum Universe' – 390833306. DNH acknowledges support from the ERC through the grant ERC-Stg DRANOEL n. 714245. LOFAR \citep{vanhaarlem13} is the Low Frequency Array designed and constructed by ASTRON. It has observing, data processing, and data storage facilities in several countries, which are owned by various parties (each with their own funding sources), and that are collectively operated by the ILT foundation under a joint scientific policy. The ILT resources have benefited from the following recent major funding sources: CNRS-INSU, Observatoire de Paris and Universit\'e d’Orl\'eans, France; BMBF, MIWF- NRW, MPG, Germany; Science Foundation Ireland (SFI), Department of Business, Enterprise and Innovation (DBEI), Ireland; NWO, The Netherlands; The Science and Technology Facilities Council, UK; Ministry of Science and Higher Education, Poland; The Istituto Nazionale di Astrofisica (INAF), Italy. This research made use of the Dutch national e-infrastructure with support of the SURF Cooperative (e-infra 180169) and the LOFAR e-infra group. The J\"ulich LOFAR Long Term Archive and the German LOFAR network are both coordinated and operated by the J\"ulich Supercomputing Centre (JSC), and computing resources on the supercomputer JUWELS at JSC were provided by the Gauss Centre for Supercomputing e.V. (grant CHTB00) through the John von Neumann Institute for Computing (NIC). This research made use of the University of Hertfordshire high-performance computing facility and the LOFAR-UK computing facility located at the University of Hertfordshire and supported by STFC [ST/P000096/1], and of the Italian LOFAR IT computing infrastructure supported and operated by INAF, and by the Physics Department of Turin University (under an agreement with Consorzio Interuniversitario per la Fisica Spaziale) at the C3S Supercomputing Centre, Italy \citep{taffoni22}. This reaserch made use of APLpy, an open-source plotting package for Python \citep{robitaille&bressert12APLPY}, Astropy, a community-developed core Python package for Astronomy \citep{astropycollaboration13,astropycollaboration18}, Matplotlib \citep{hunter07MATPLOTLIB}, Numpy \citep{harris20NUMPY}, and TOPCAT \citep{taylor05TOPCAT}.

\end{acknowledgements}

\bibliographystyle{aa}
\bibliography{articolo}

\begin{appendix}

\section{Injections in PSZ2 G098.62+51.76}
\label{appendix:mock98}

\begin{figure*}
	\centering
	\includegraphics[width=0.35\textwidth]{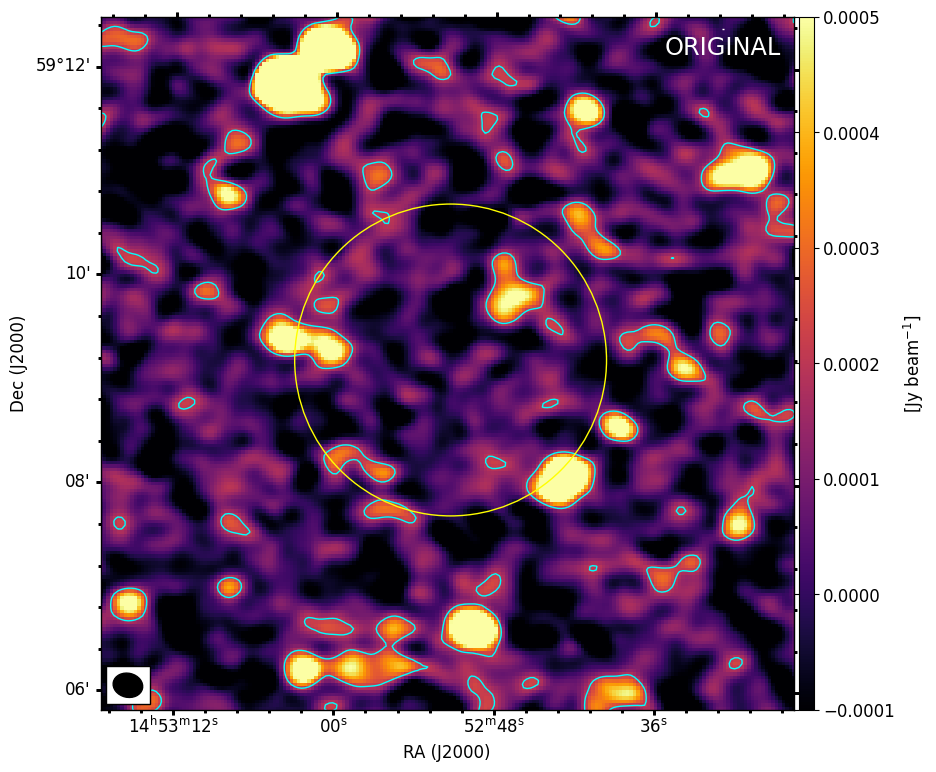} 
	\includegraphics[width=0.35\textwidth]{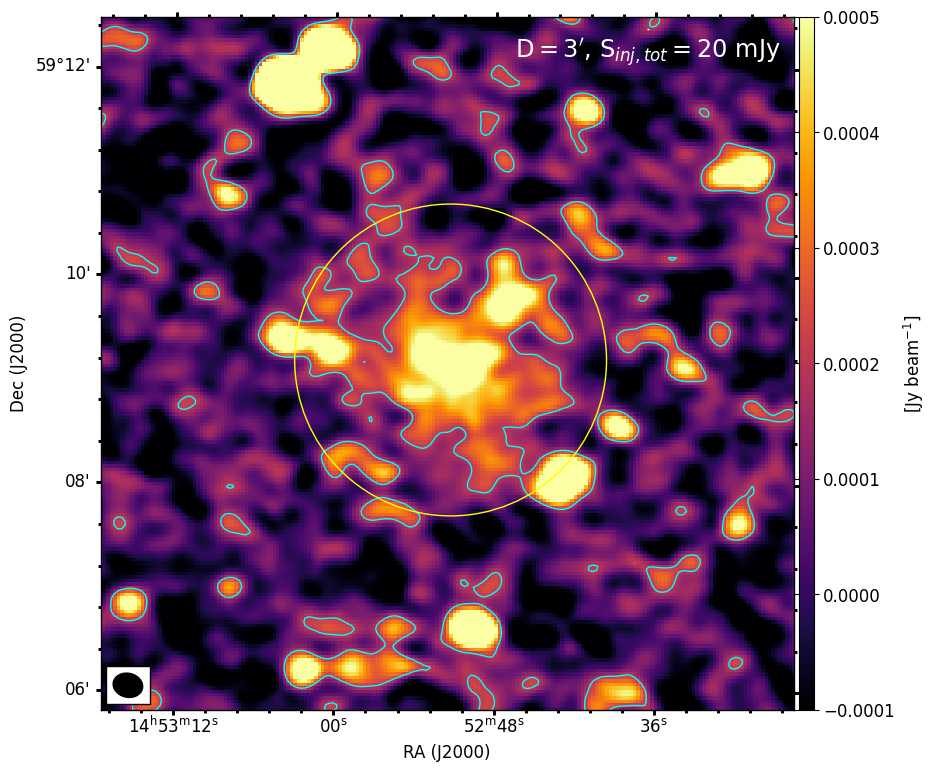}
	\includegraphics[width=0.35\textwidth] {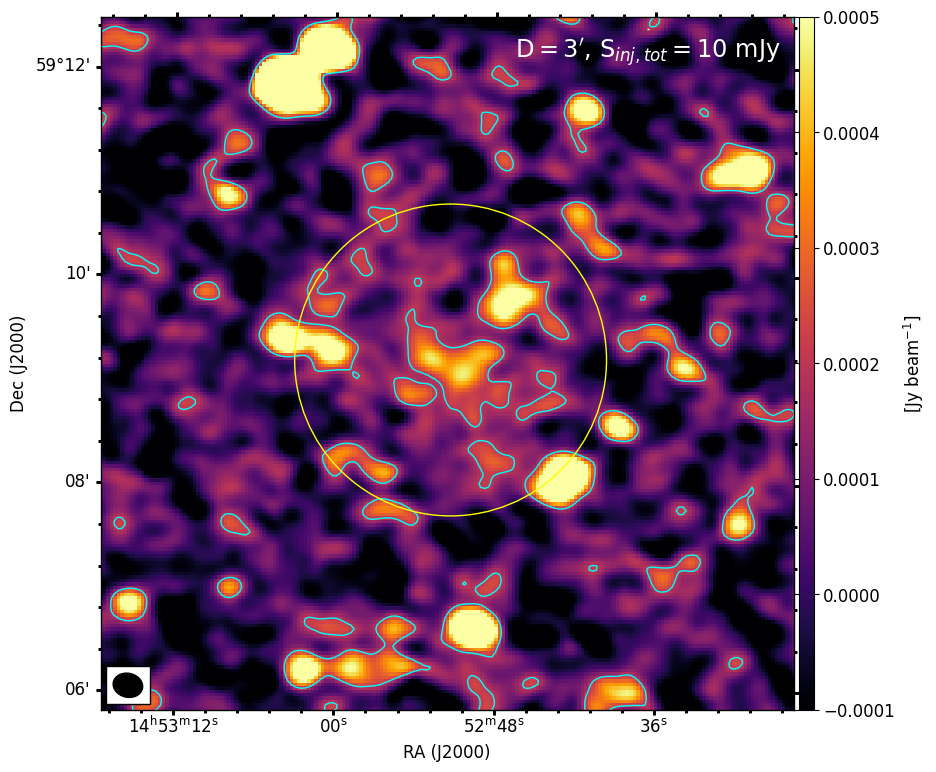}
	\includegraphics[width=0.35\textwidth] {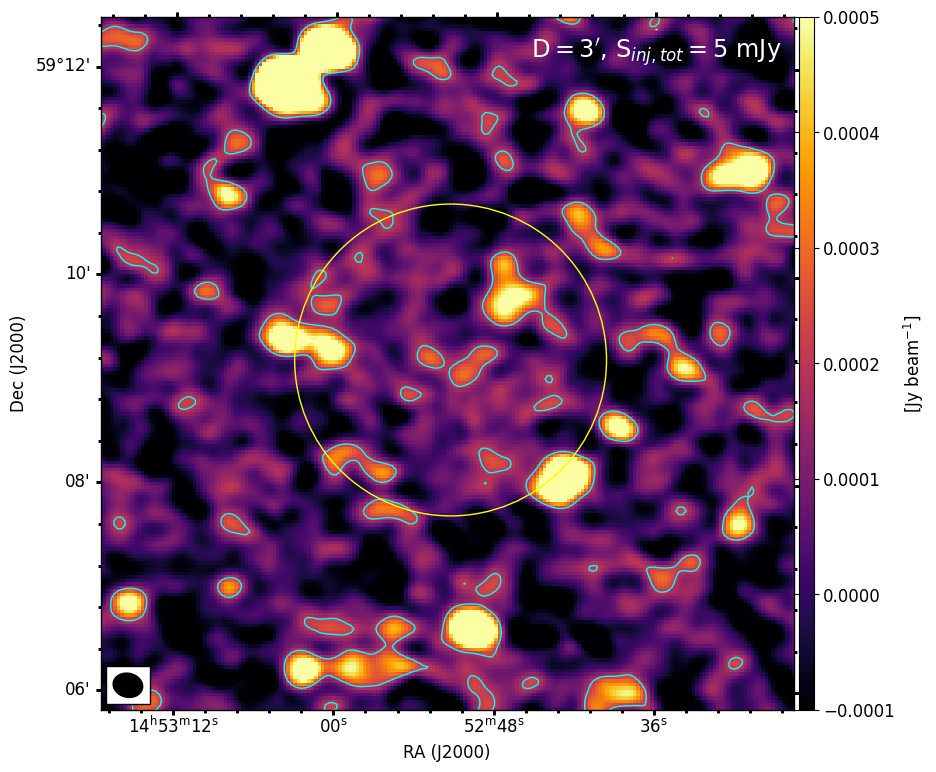}
	\caption{Injection of mock halos with $D=3'$ (corresponding to the diameter of the yellow circle). The $2\sigma$ contour level is reported, where $\sigma$ is the noise of the pre-injection map. The upper limit is obtained with $S_{\rm inj,tot}=5$ mJy.}
	\label{MOCK98_D3}%
\end{figure*}

\begin{figure*}
	\centering
	\includegraphics[width=0.35\textwidth] {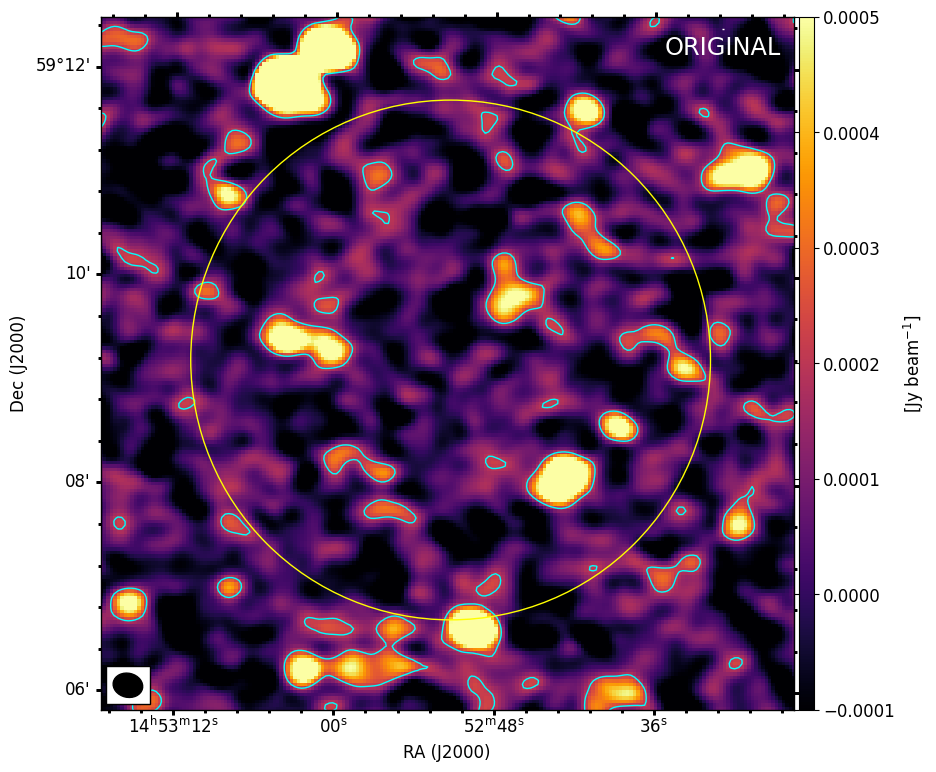} 
	\includegraphics[width=0.35\textwidth] {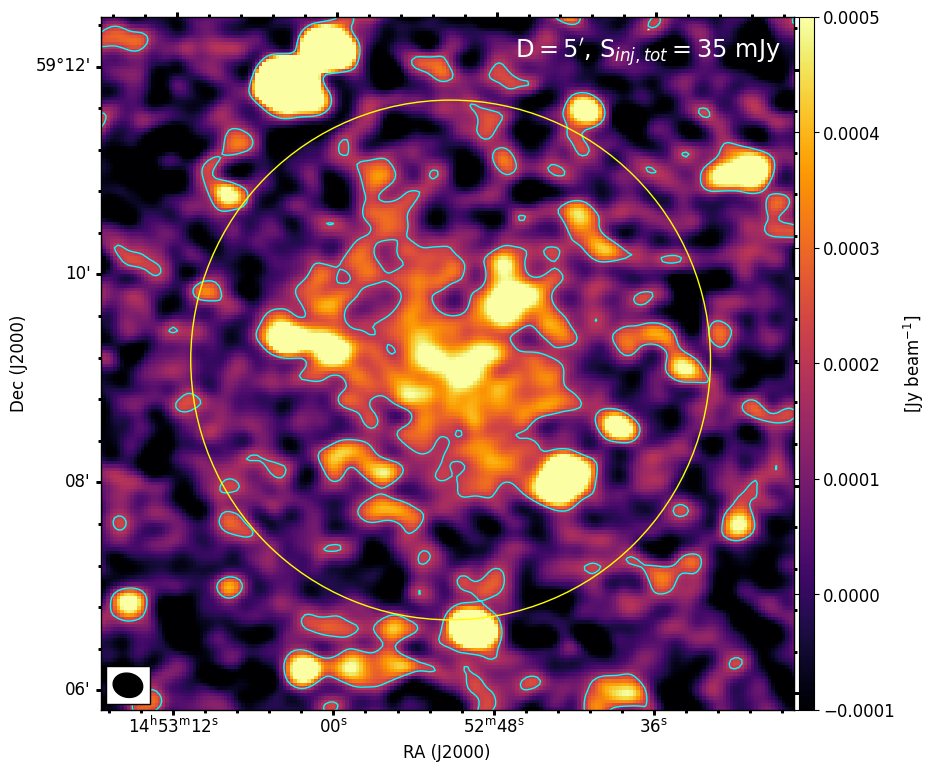}
	\includegraphics[width=0.35\textwidth] {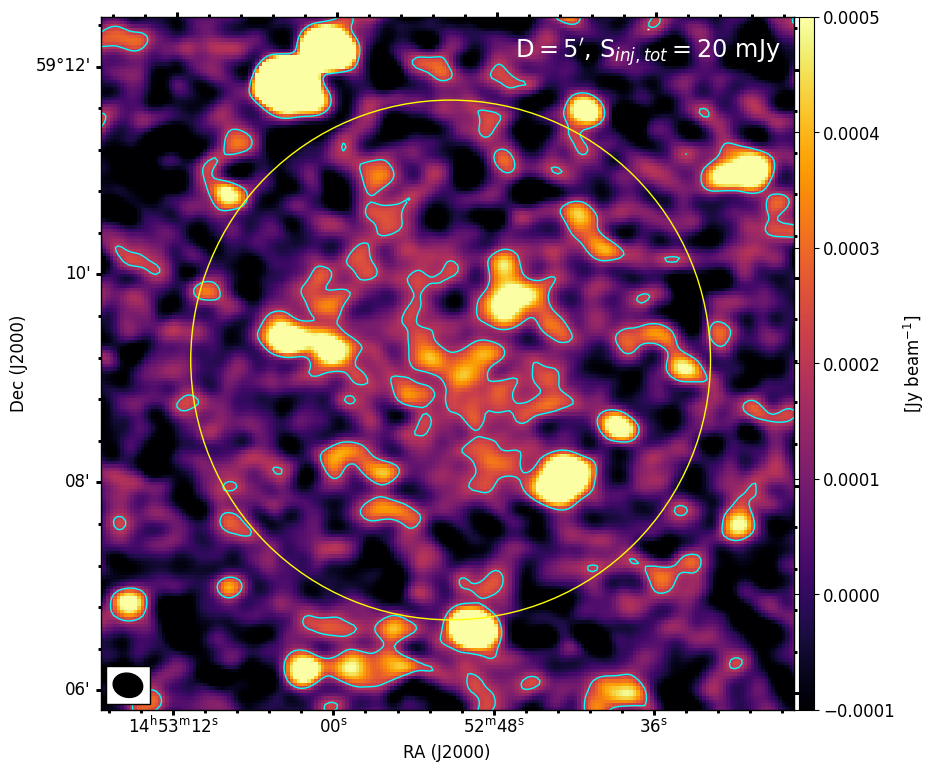}
	\includegraphics[width=0.35\textwidth] {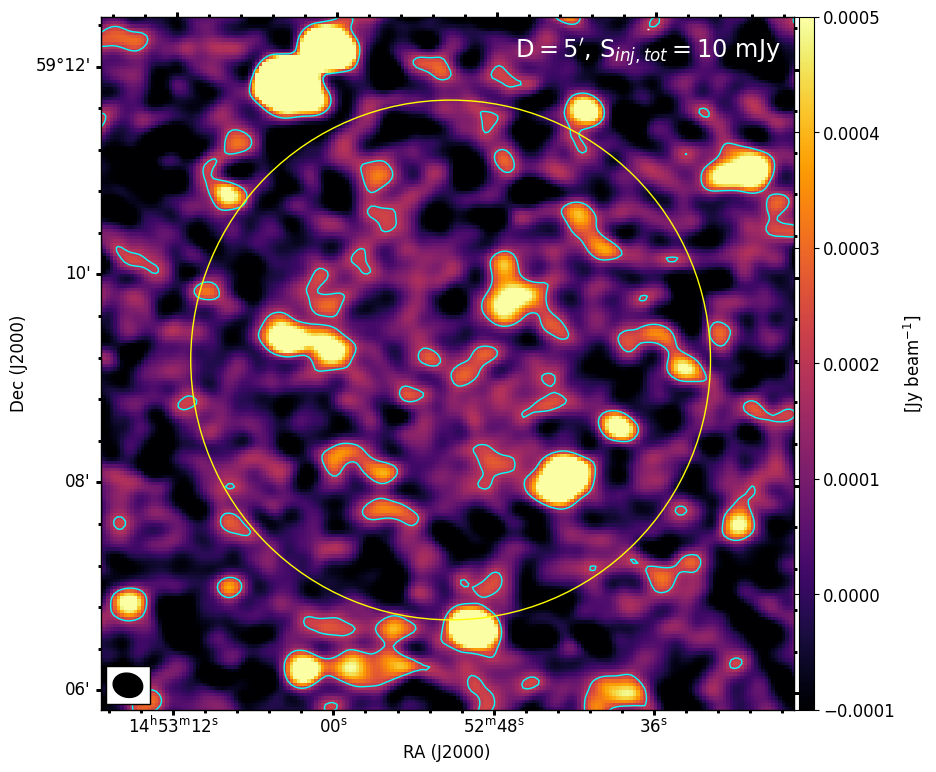}
	\caption{Injection of mock halos with $D=5'$ (corresponding to the diameter of the yellow circle). The $2\sigma$ contour level is reported, where $\sigma$ is the noise of the pre-injection map. The upper limit is obtained with $S_{\rm inj,tot}=10$ mJy.}
	\label{MOCK98_D5}%
\end{figure*}

\begin{figure*}
	\centering
	\includegraphics[width=0.35\textwidth] {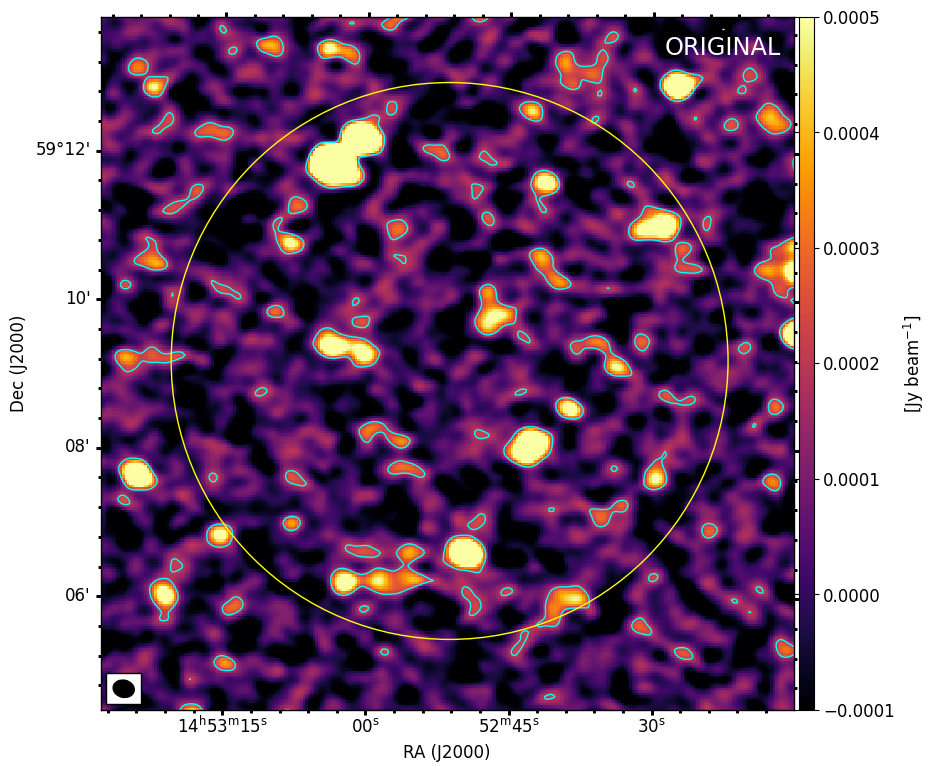} 
	\includegraphics[width=0.35\textwidth] {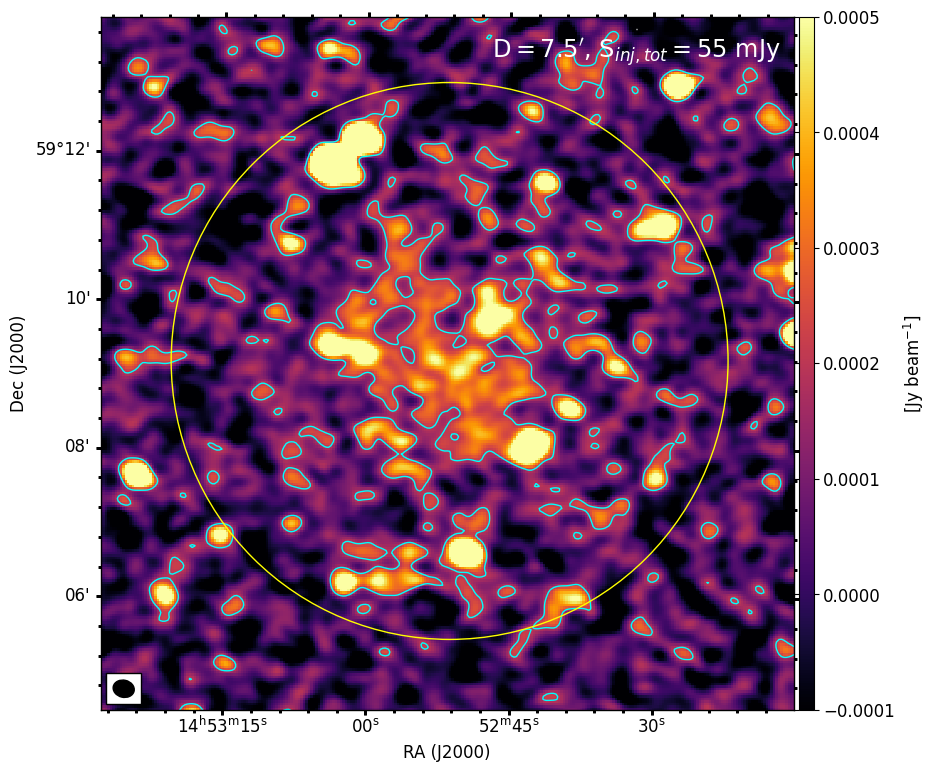}
	\includegraphics[width=0.35\textwidth] {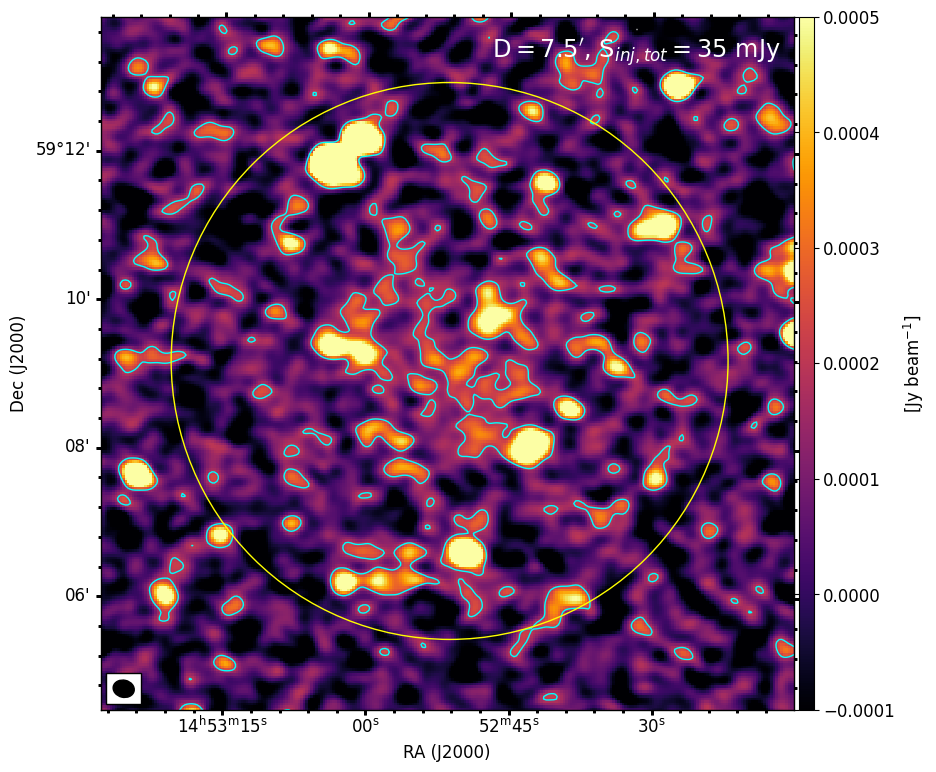}
	\includegraphics[width=0.35\textwidth] {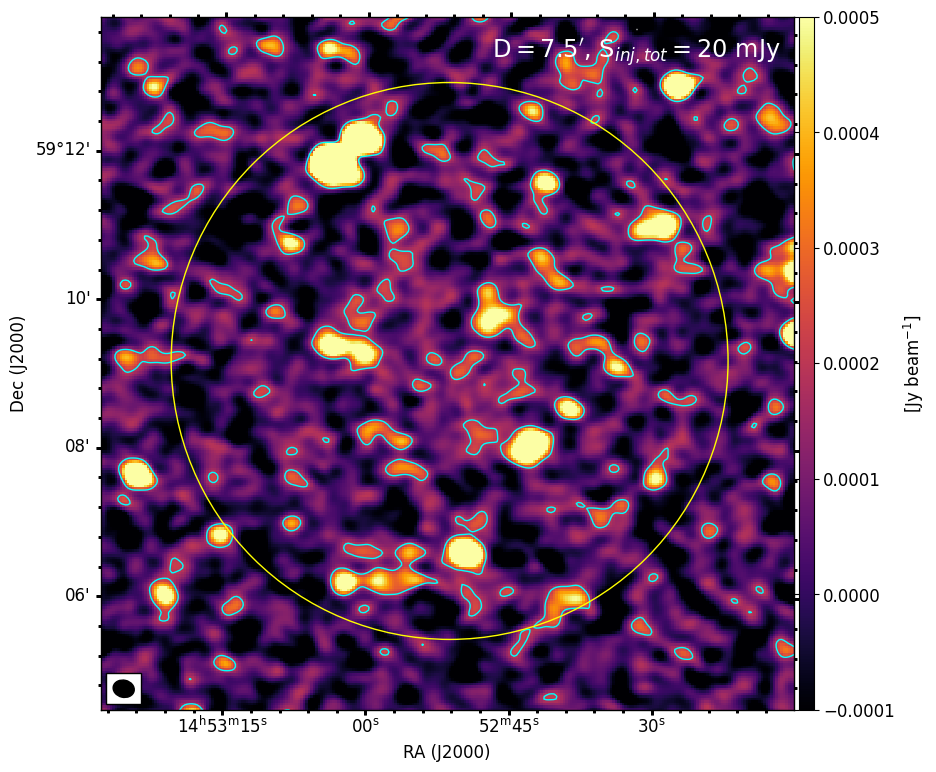}
	\caption{Injection of mock halos with $D=7.5'$ (corresponding to the diameter of the yellow circle). The $2\sigma$ contour level is reported, where $\sigma$ is the noise of the pre-injection map. The upper limit is obtained with $S_{\rm inj,tot}=20$ mJy.}
	\label{MOCK98_D7.5}%
\end{figure*}   

\begin{figure*}
	\centering
	\includegraphics[width=0.35\textwidth] {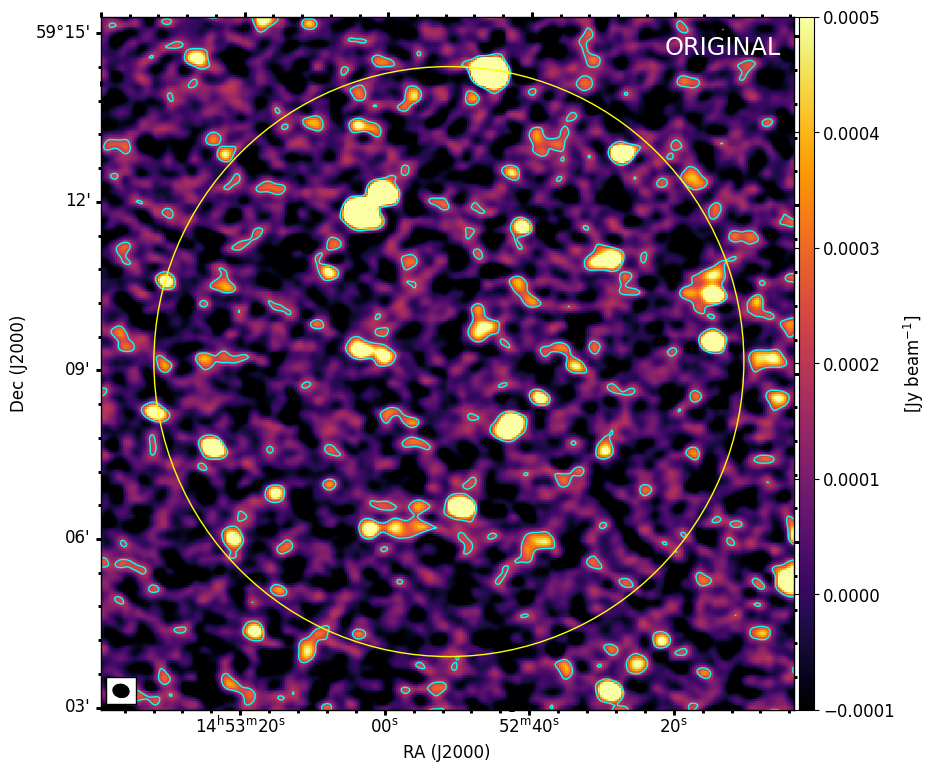} 
	\includegraphics[width=0.35\textwidth] {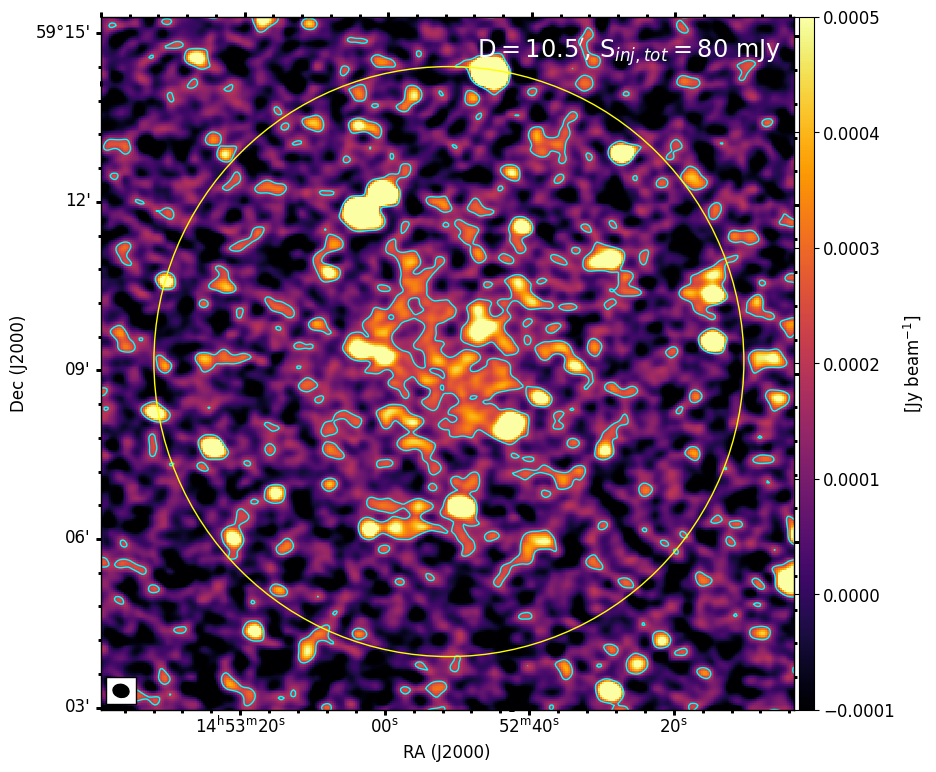}
	\includegraphics[width=0.35\textwidth] {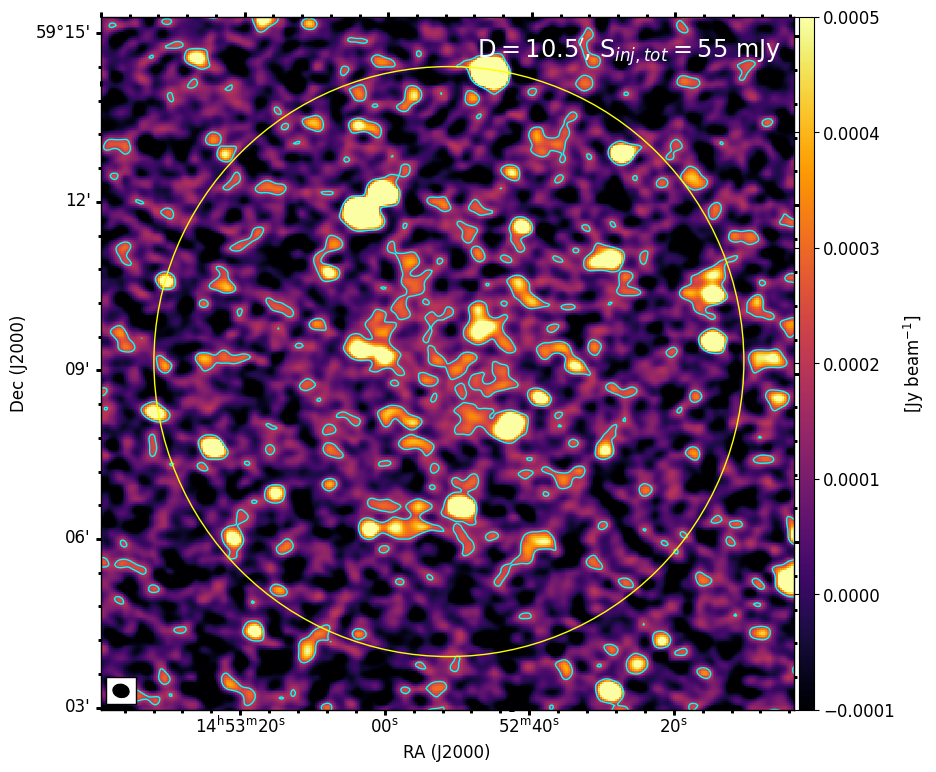}
	\includegraphics[width=0.35\textwidth] {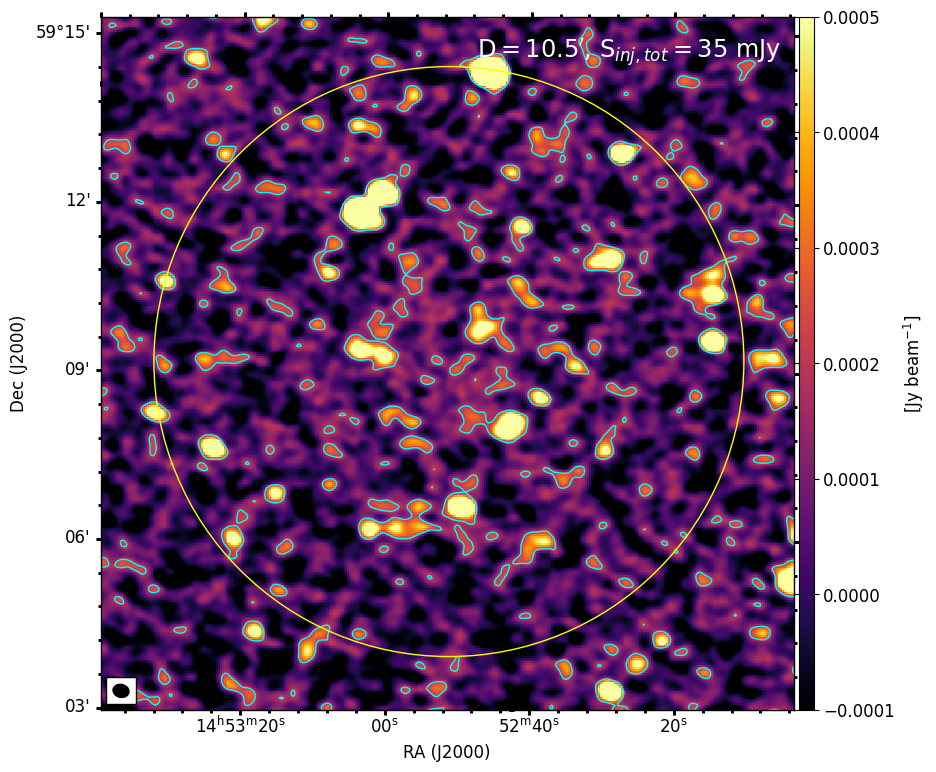}
	\caption{Injection of mock halos with $D=10.5'$ (corresponding to the diameter of the yellow circle). The $2\sigma$ contour level is reported, where $\sigma$ is the noise of the pre-injection map. The upper limit is obtained with $S_{\rm inj,tot}=35$ mJy.}
	\label{MOCK98_D10.5}%
\end{figure*}

\begin{figure*}
	\centering
	\includegraphics[width=0.35\textwidth] {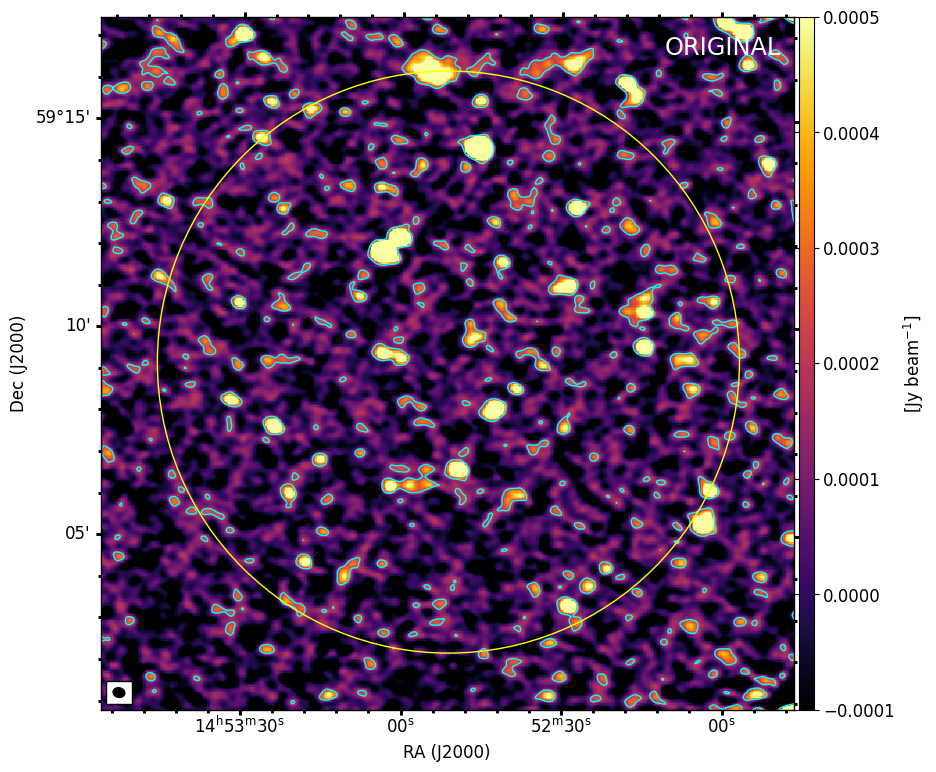} 
	\includegraphics[width=0.35\textwidth] {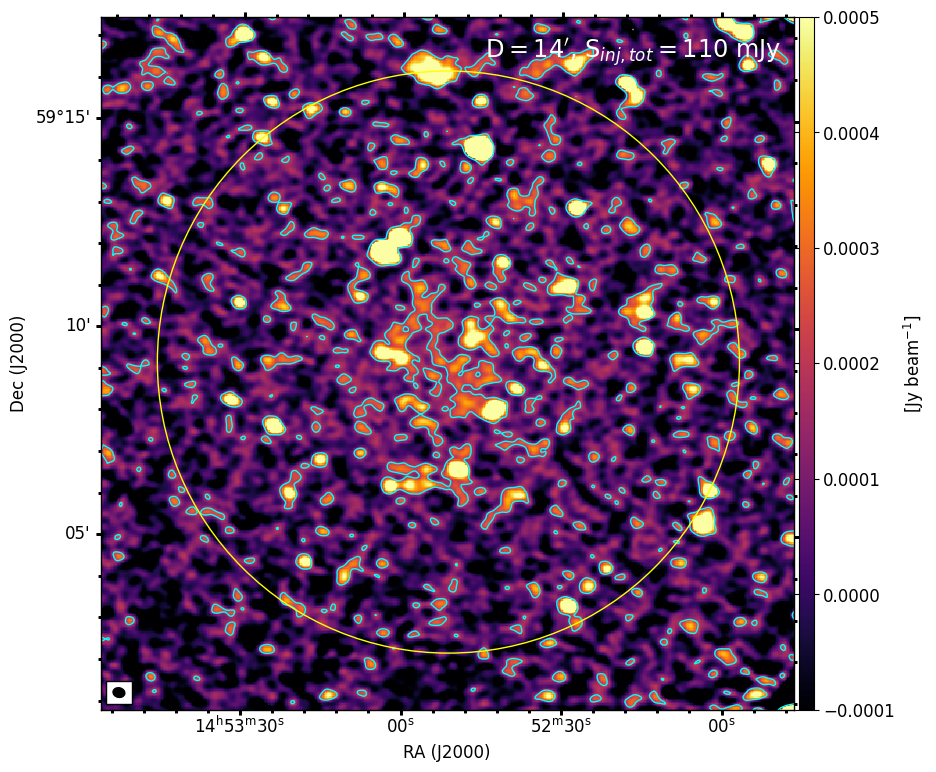}
	\includegraphics[width=0.35\textwidth] {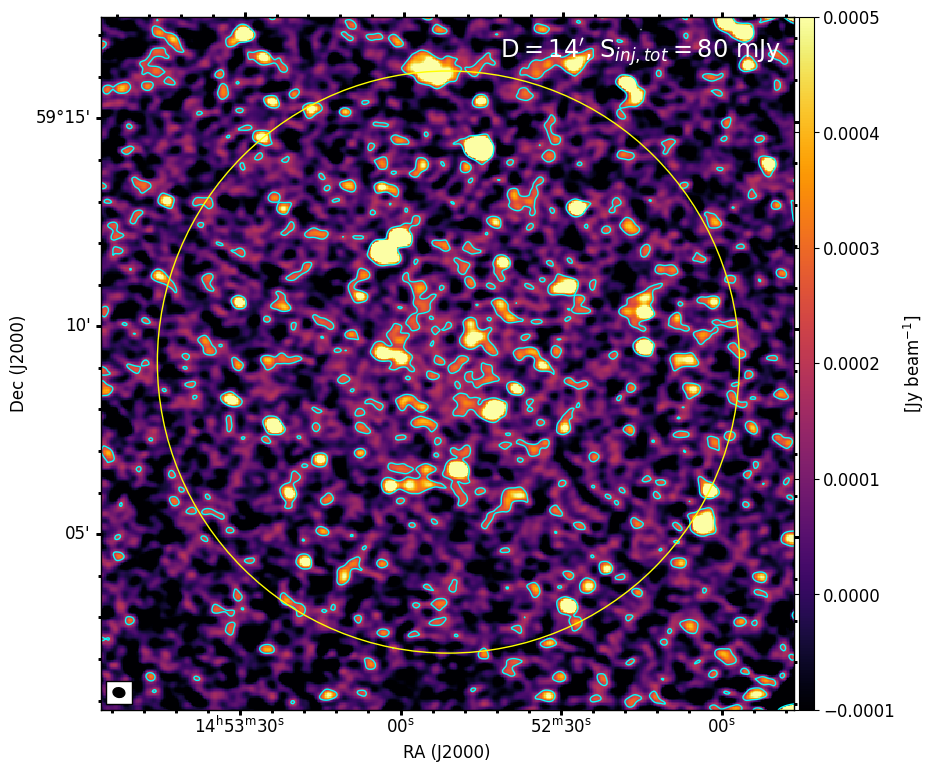}
	\includegraphics[width=0.35\textwidth] {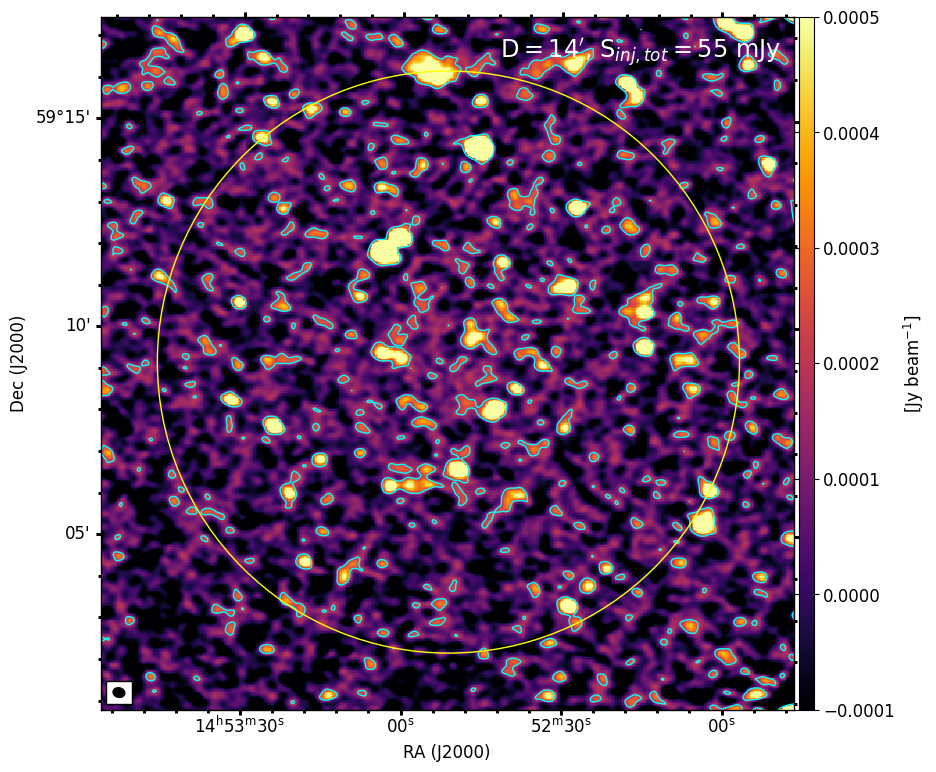}
	\caption{Injection of mock halos with $D=14'$ (corresponding to the diameter of the yellow circle). The $2\sigma$ contour level is reported, where $\sigma$ is the noise of the pre-injection map. The upper limit is obtained with $S_{\rm inj,tot}=55$ mJy.}
	\label{MOCK98_D14}%
\end{figure*}  

\begin{figure*}
	\centering
	\includegraphics[width=0.35\textwidth] {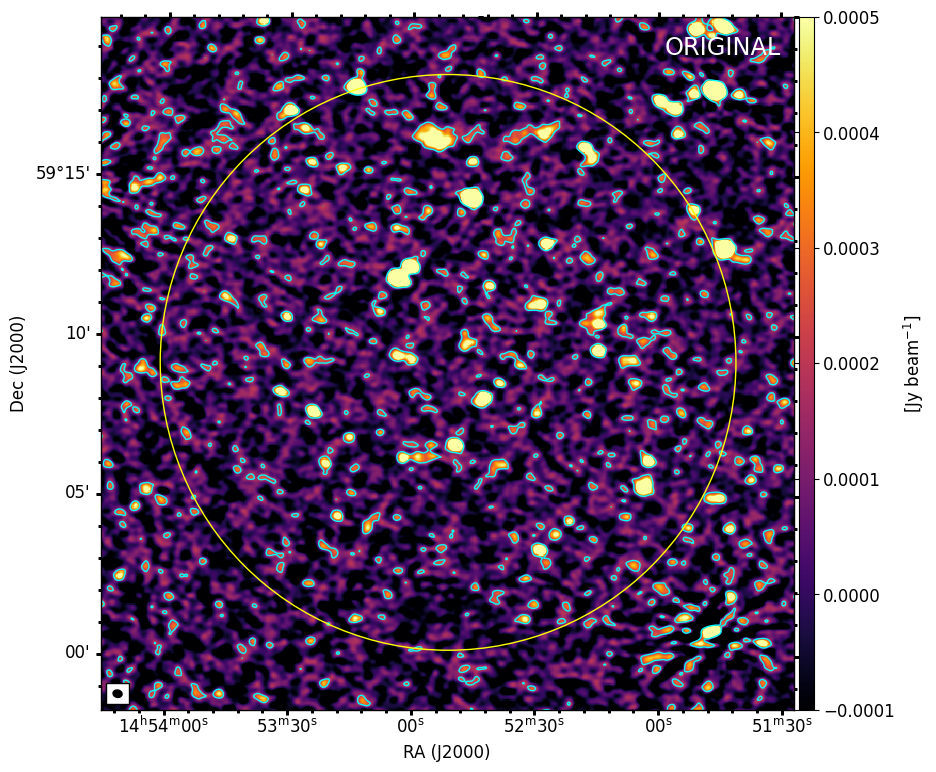} 
	\includegraphics[width=0.35\textwidth] {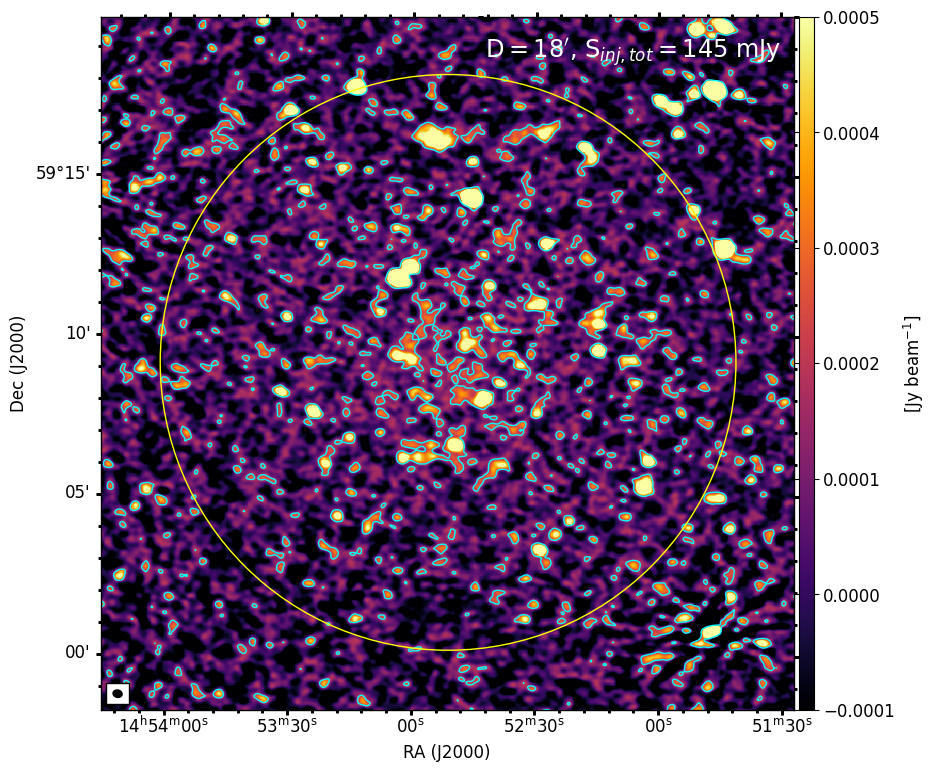}
	\includegraphics[width=0.35\textwidth] {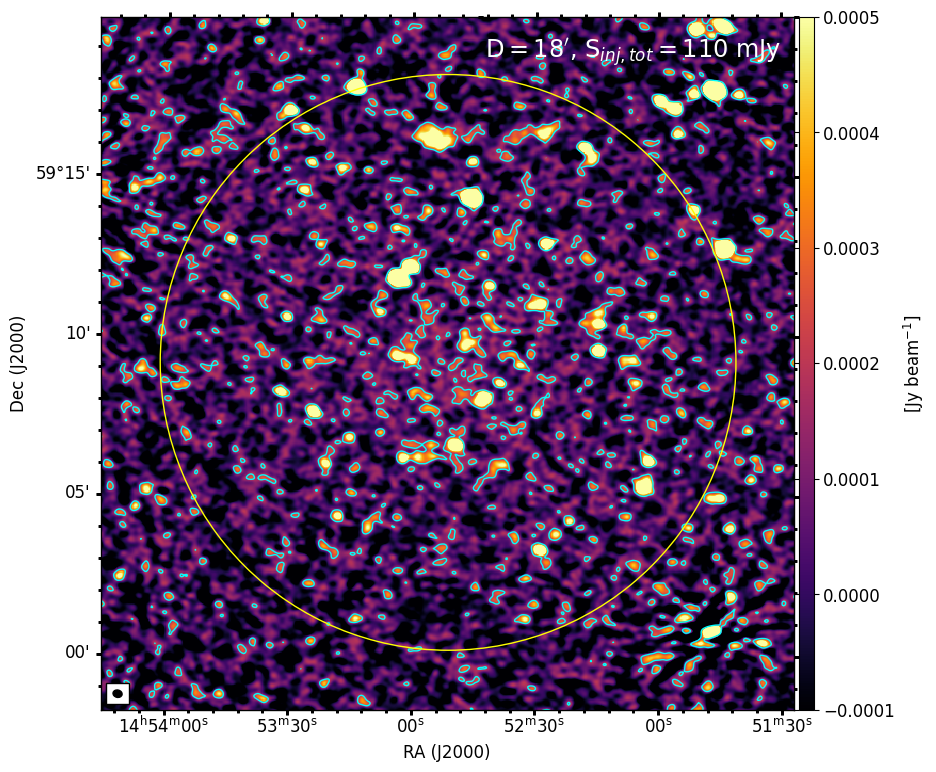}
	\includegraphics[width=0.35\textwidth] {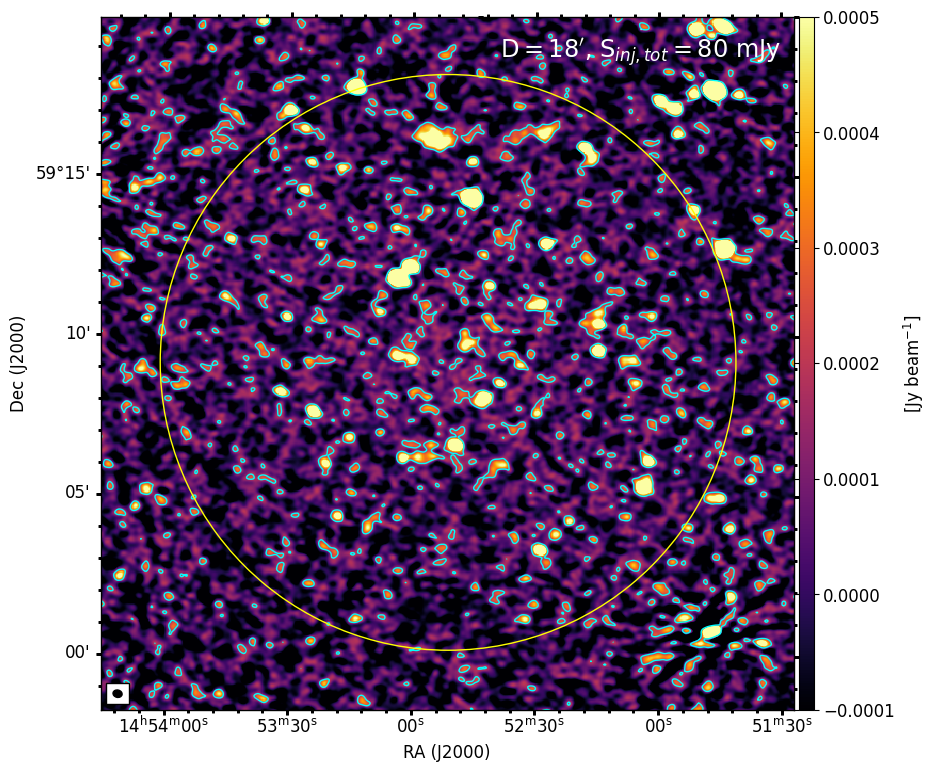}
	\caption{Injection of mock halos with $D=18'$ (corresponding to the diameter of the yellow circle). The $2\sigma$ contour level is reported, where $\sigma$ is the noise of the pre-injection map. The upper limit is obtained with $S_{\rm inj,tot}=80$ mJy.}
	\label{MOCK98_D18}%
\end{figure*}  

As a representative example of the various injections that we obtained in different NDE clusters, we report some of the images of PSZ2 G098.62+51.76, which were used to produce and discuss the plots shown in Sect. \ref{sect:LOFAR performances}

\section{Upper limits}
\label{appendix:Upper limits}
\onecolumn

In Table \ref{tab: limits} we summarise the main properties of the host cluster and the injected parameters used to obtain the upper limits.

\setlength{\LTcapwidth}{\textwidth}
\footnotesize
\fontsize{7.5}{7.5}\selectfont
\begin{longtable}{|c|c|c|c|c|c|c|c|c|c|c|c|c|}
\caption{Summary of the host cluster properties and upper limit parameters for the 75 considered targets. Columns from 1 to 5 report the \textit{Planck} pointing name, coordinates (RA, DEC), redshift ($z$), and mass ($M_{500}$) of the host. Columns 6 and 7 report the injection centre (${\rm RA_{\rm inj}, \; DEC_{\rm inj} }$). Column 8 reports the subtraction quality parameter (SQ) close to the injection centre. Columns 9 and 10 report the injected central brightness ($I_{\rm 0,inj}$) and \textit{e}-folding radius ($r_{\rm e,inj}$). Columns 11 and 12 report the level of the upper limit at 150 MHz in terms of flux density ($S_{\rm 150,UL}$) and radio power ($P_{\rm 150,UL}$)} \label{tab: limits} \\

\hline
  \multicolumn{1}{|c|}{Name} &
  \multicolumn{1}{c|}{RA} &
  \multicolumn{1}{c|}{DEC} &
  \multicolumn{1}{c|}{$z$} &
  \multicolumn{1}{c|}{$M_{500}$} &
  \multicolumn{1}{c|}{${\rm RA}_{\rm inj}$} &
  \multicolumn{1}{c|}{${\rm DEC}_{\rm inj}$} &
  \multicolumn{1}{c|}{${\rm SQ}$} &
  \multicolumn{1}{c|}{$I_{\rm 0, inj}$} &
  \multicolumn{1}{c|}{$r_{\rm e, inj}$} &
  \multicolumn{1}{c|}{$S_{\rm 150, UL}$} &
  \multicolumn{1}{c|}{$P_{\rm 150, UL}$} \\
    \multicolumn{1}{|c|}{} &
  \multicolumn{1}{c|}{(deg)} &
  \multicolumn{1}{c|}{(deg)} &
  \multicolumn{1}{c|}{} &
  \multicolumn{1}{c|}{($10^{14} \; M_{\odot}$)} &
  \multicolumn{1}{c|}{(deg)} &
  \multicolumn{1}{c|}{(deg)} &
  \multicolumn{1}{c|}{} &
  \multicolumn{1}{c|}{($ {\rm \mu Jy \; arcsec^{2}}$)} &
  \multicolumn{1}{c|}{(arcsec)} &
  \multicolumn{1}{c|}{(mJy)} &
  \multicolumn{1}{c|}{(${\rm 10^{23} \; W \; Hz^{-1}} $)} \\
\hline
   \endfirsthead
   \caption{\textit{Continued from previous page.}}\\
   \hline
  \multicolumn{1}{|c|}{Name} &
  \multicolumn{1}{c|}{RA} &
  \multicolumn{1}{c|}{DEC} &
  \multicolumn{1}{c|}{$z$} &
  \multicolumn{1}{c|}{$M_{500}$} &
  \multicolumn{1}{c|}{${\rm RA}_{\rm inj}$} &
  \multicolumn{1}{c|}{${\rm DEC}_{\rm inj}$} &
  \multicolumn{1}{c|}{${\rm SQ}$} &
  \multicolumn{1}{c|}{$I_{\rm 0, inj}$} &
  \multicolumn{1}{c|}{$r_{\rm e, inj}$} &
  \multicolumn{1}{c|}{$S_{\rm 150, UL}$} &
  \multicolumn{1}{c|}{$P_{\rm 150, UL}$} \\
    \multicolumn{1}{|c|}{} &
  \multicolumn{1}{c|}{(deg)} &
  \multicolumn{1}{c|}{(deg)} &
  \multicolumn{1}{c|}{} &
  \multicolumn{1}{c|}{($10^{14} \; M_{\odot}$)} &
  \multicolumn{1}{c|}{(deg)} &
  \multicolumn{1}{c|}{(deg)} &
  \multicolumn{1}{c|}{} &
  \multicolumn{1}{c|}{($ {\rm \mu Jy \; arcsec^{2}}$)} &
  \multicolumn{1}{c|}{(arcsec)} &
  \multicolumn{1}{c|}{(mJy)} &
  \multicolumn{1}{c|}{(${\rm 10^{23} \; W \; Hz^{-1}} $)} \\
\hline
   \endhead
   \hline
   \multicolumn{12}{r}{\textit{Continued on next page}}
   \endfoot
   \hline
   \endlastfoot

  PSZ2 G045.13+67.78 & 217.996 & 29.557 & 0.219 & 4.83 $\pm$ 0.45 & 217.996 & 29.557 & 2 & 0.323 & 56.5 & 5.0 & 7.5\\
  PSZ2 G048.75+53.18 & 234.967 & 30.696 & 0.098 & 2.53 $\pm$ 0.31 & 234.962 & 30.718 & 2 & 0.289 & 110.4 &  16.8 & 4.2\\
  PSZ2 G049.18+65.05 & 221.119 & 31.233 & 0.234 & 4.73 $\pm$ 0.49 & 221.133 & 31.227 & 2 & 0.688 & 53.73 & 9.5 & 16.5\\
  PSZ2 G050.46+67.54 & 218.168 & 31.588 & 0.131 & 2.92 $\pm$ 0.34 & 218.158 & 31.658 & 2 & 0.2 & 85.75 & 7.0 & 3.3\\
  PSZ2 G055.80+32.90 & 259.462 & 32.561 & 0.105 & 2.58 $\pm$ 0.31 & 259.481 & 32.578 & 2 & 0.154 & 103.87 & 7.9 & 2.3\\
  PSZ2 G056.14+28.06 & 265.075 & 31.603 & 0.426 & 5.53 $\pm$ 0.57 & 265.069 & 31.611 & 2 & 0.997 & 35.85 & 6.1 & 44.5\\
  PSZ2 G057.73+51.58 & 237.141 & 36.103 & 0.238 & 5.59 $\pm$ 0.51 & 237.144 & 36.096 & 2 & 0.452 & 53.05 & 6.1 & 11.0\\
  PSZ2 G057.78+52.32 & 236.215 & 36.122 & 0.065 & 2.38 $\pm$ 0.22 & 236.246 & 36.11 & 2 & 0.118 & 160.2 & 14.4 & 1.5\\
  PSZ2 G059.18+32.91 & 260.203 & 35.325 & 0.383 & 5.21 $\pm$ 0.56 & 260.195 & 35.324 & 2 & 0.485 & 38.23 & 3.4 & 19.0\\
  PSZ2 G059.29+44.49 & 245.99 & 36.973 & 0.343 & 5.76 $\pm$ 0.69 & 245.959 & 37.007 & 2 & 1.072 & 41.02 & 8.6 & 37.0\\
  PSZ2 G060.16+64.50 & 221.074 & 35.938 & 0.361 & 5.0 $\pm$ 0.63 & 221.065 & 35.957 & 2 & 0.411 & 39.69 & 3.1 & 15.0\\
  PSZ2 G060.55+27.00 & 267.574 & 35.076 & 0.171 & 3.48 $\pm$ 0.41 & 267.574 & 35.076 & 2 & 0.313 & 68.68 & 7.0 & 6.0\\
  PSZ2 G065.45+78.10 & 204.818 & 33.01 & 0.273 & 4.07 $\pm$ 0.53 & 204.766 & 32.966 & 1 & 0.347 & 47.98 & 3.8 & 9.5\\
  PSZ2 G065.79+41.80 & 249.717 & 41.599 & 0.336 & 5.22 $\pm$ 0.59 & 249.748 & 41.616 & 2 & 0.356 & 41.59 & 2.9 & 12.0\\
  PSZ2 G066.26+20.82 & 276.851 & 38.259 & 0.278 & 4.13 $\pm$ 0.5 & 276.846 & 38.238 & 2 & 1.614 & 47.37 & 17.3 & 45.0\\
  PSZ2 G066.68+68.44 & 215.432 & 37.282 & 0.163 & 3.79 $\pm$ 0.34 & 215.419 & 37.292 & 1 & 0.242 & 71.42 & 5.9 & 4.5\\
  PSZ2 G070.89+49.26 & 239.179 & 44.653 & 0.61 & 6.46 $\pm$ 0.69 & 239.212 & 44.621 & 2 & 0.89 & 29.69 & 3.7 & 67.0\\
  PSZ2 G071.63+29.78 & 266.826 & 45.19 & 0.157 & 4.13 $\pm$ 0.29 & 266.818 & 45.2 & 2 & 0.22 & 73.66 & 5.7 & 4.0\\
  PSZ2 G080.55-24.82 & 330.855 & 23.91 & 0.266 & 4.28 $\pm$ 0.53 & 330.844 & 23.901 & 2 & 0.486 & 48.89 & 5.5 & 13.0\\
  PSZ2 G083.14+66.57 & 213.445 & 43.652 & 0.089 & 2.07 $\pm$ 0.26 & 213.431 & 43.654 & 2 & 0.157 & 120.31 & 10.8 & 2.2\\
  PSZ2 G084.69+42.28 & 246.766 & 55.48 & 0.13 & 2.7 $\pm$ 0.26 & 246.746 & 55.474 & 2 & 0.164 & 86.31 & 5.8 & 2.7\\
  PSZ2 G086.43-24.95 & 335.558 & 27.143 & 0.231 & 3.81 $\pm$ 0.5 & 335.572 & 27.134 & 2 & 0.379 & 54.25 & 5.3 & 9.0\\
  PSZ2 G087.44-21.56 & 334.099 & 30.425 & 0.258 & 4.15 $\pm$ 0.51 & 334.099 & 30.425 & 2 & 0.461 & 49.98 & 5.5 & 12.0\\
  PSZ2 G091.27-38.62 & 347.335 & 17.895 & 0.105 & 3.14 $\pm$ 0.39 & 347.339 & 17.868 & 1 & 0.235 & 103.87 & 12.1 & 3.5\\
  PSZ2 G092.69+59.92 & 216.635 & 51.252 & 0.462 & 4.79 $\pm$ 0.6 & 216.634 & 51.266 & 2 & 0.533 & 34.21 & 3.0 & 26.5\\
  PSZ2 G093.04-32.38 & 345.428 & 24.04 & 0.512 & 6.34 $\pm$ 0.72 & 345.519 & 24.062 & 2 & 1.567 & 32.36 & 7.8 & 90.0\\
  PSZ2 G097.15+39.20 & 246.903 & 65.396 & 0.206 & 2.94 $\pm$ 0.32 & 246.944 & 65.421 & 2 & 0.212 & 59.23 & 3.5 & 4.6\\
  PSZ2 G098.38+77.22 & 199.606 & 38.585 & 0.78 & 6.62 $\pm$ 0.71 & 199.606 & 38.585 & 1 & 0.69 & 26.87 & 2.4 & 80.0\\
  PSZ2 G098.44+56.59 & 216.779 & 55.75 & 0.132 & 2.83 $\pm$ 0.27 & 216.844 & 55.749 & 2 & 0.175 & 85.2 & 6.1 & 2.9\\
  PSZ2 G098.62+51.76 & 222.827 & 59.331 & 0.298 & 3.35 $\pm$ 0.48 & 222.603 & 59.325 & 2 & 0.403 & 45.11 & 3.9 & 12.0\\
  PSZ2 G100.22+33.81 & 258.419 & 69.373 & 0.598 & 4.61 $\pm$ 0.47 & 258.414 & 69.358 & 2 & 1.029 & 29.96 & 4.4 & 75.0\\
  PSZ2 G101.52-29.98 & 351.596 & 29.326 & 0.227 & 4.88 $\pm$ 0.52 & 351.617 & 29.367 & 2 & 0.385 & 54.98 & 5.5 & 9.0\\
  PSZ2 G102.90-31.04 & 353.302 & 28.768 & 0.592 & 6.73 $\pm$ 0.72 & 353.302 & 28.768 & 1 & 0.446 & 30.1 & 1.9 & 32.0\\
  PSZ2 G105.76+54.73 & 212.59 & 59.68 & 0.316 & 4.41 $\pm$ 0.45 & 212.573 & 59.711 & 2 & 0.648 & 43.33 & 5.8 & 20.5\\
  PSZ2 G112.54+59.53 & 202.476 & 56.812 & 0.83 & 5.76 $\pm$ 0.66 & 202.476 & 56.812 & 1 & 0.456 & 26.31 & 1.5 & 60.0\\
  PSZ2 G114.83+57.25 & 201.446 & 59.33 & 0.17 & 3.27 $\pm$ 0.3 & 201.446 & 59.33 & 2 & 0.21 & 69.01 & 4.8 & 4.0\\
  PSZ2 G115.58-44.56 & 7.362 & 17.995 & 0.17 & 4.32 $\pm$ 0.4 & 7.362 & 17.995 & 1 & 0.21 & 69.01 & 4.8 & 4.0\\
  PSZ2 G120.08-44.41 & 10.715 & 18.407 & 0.267 & 4.65 $\pm$ 0.64 & 10.696 & 18.433 & 2 & 0.521 & 48.75 & 5.9 & 14.0\\
  PSZ2 G122.30+54.52 & 193.646 & 62.596 & 0.318 & 4.53 $\pm$ 0.45 & 193.689 & 62.566 & 2 & 0.628 & 43.15 & 5.6 & 20.0\\
  PSZ2 G123.66+67.25 & 192.422 & 49.872 & 0.284 & 4.38 $\pm$ 0.51 & 192.422 & 49.872 & 2 & 0.281 & 46.65 & 2.9 & 8.0\\
  PSZ2 G126.20-33.17 & 16.014 & 29.617 & 0.358 & 5.45 $\pm$ 0.74 & 16.062 & 29.564 & 2 & 1.243 & 39.9 & 9.4 & 45.0\\
  PSZ2 G126.72-21.03 & 17.603 & 41.692 & 0.22 & 4.11 $\pm$ 0.57 & 17.603 & 41.692 & 2 & 0.438 & 56.3 & 6.6 & 10.0\\
  PSZ2 G127.44-34.74 & 17.057 & 27.979 & 0.249 & 4.9 $\pm$ 0.71 & 17.057 & 27.979 & 2 & 0.733 & 51.3 & 9.2 & 18.5\\
  PSZ2 G127.50-30.52 & 17.516 & 32.183 & 0.353 & 5.29 $\pm$ 0.67 & 17.533 & 32.181 & 2 & 0.716 & 40.26 & 5.5 & 25.5\\
  PSZ2 G128.15-24.71 & 18.877 & 37.913 & 0.263 & 4.48 $\pm$ 0.6 & 18.877 & 37.914 & 1 & 0.302 & 49.29 & 3.5 & 8.0\\
  PSZ2 G130.25-26.50 & 20.952 & 35.903 & 0.216 & 4.49 $\pm$ 0.5 & 20.914 & 35.907 & 2 & 0.267 & 57.1 & 4.1 & 6.0\\
  PSZ2 G132.54-42.16 & 20.435 & 20.14 & 0.194 & 3.99 $\pm$ 0.51 & 20.422 & 20.094 & 1 & 0.288 & 62.08 & 5.3 & 6.0\\
  PSZ2 G133.92-42.73 & 21.401 & 19.405 & 0.636 & 7.24 $\pm$ 0.96 & 21.435 & 19.401 & 2 & 0.93 & 29.14 & 3.8 & 75.0\\
  PSZ2 G135.06+54.39 & 178.09 & 61.319 & 0.317 & 5.41 $\pm$ 0.42 & 178.058 & 61.334 & 1 & 0.268 & 43.24 & 2.4 & 8.5\\
  PSZ2 G136.31+54.67 & 176.96 & 60.766 & 0.477 & 5.98 $\pm$ 0.51 & 176.96 & 60.766 & 1 & 0.347 & 33.61 & 1.9 & 18.0\\
  PSZ2 G137.24+53.93 & 175.277 & 61.194 & 0.47 & 7.0 $\pm$ 0.48 & 175.277 & 61.194 & 1 & 0.373 & 33.89 & 2 & 19.0\\
  PSZ2 G137.74-27.08 & 28.784 & 33.944 & 0.087 & 2.83 $\pm$ 0.28 & 28.784 & 33.944 & 2 & 0.166 & 122.79 & 11.9 & 2.3\\
  PSZ2 G139.00+50.92 & 170.045 & 63.26 & 0.784 & 5.9 $\pm$ 0.7 & 170.065 & 63.242 & 2 & 1.282 & 26.82 & 4.4 & 150.0\\
  PSZ2 G139.72-17.13 & 34.981 & 42.832 & 0.155 & 3.61 $\pm$ 0.44 & 34.926 & 42.849 & 2 & 0.249 & 74.44 & 6.6 & 4.5\\
  PSZ2 G141.98+69.31 & 183.239 & 46.365 & 0.713 & 5.29 $\pm$ 0.68 & 183.169 & 46.356 & 2 & 0.519 & 27.79 & 1.9 & 51.0\\
  PSZ2 G144.33+62.85 & 177.306 & 51.609 & 0.132 & 2.66 $\pm$ 0.35 & 177.26 & 51.591 & 2 & 0.103 & 85.2 & 3.6 & 1.7\\
  PSZ2 G146.13+40.97 & 144.785 & 66.437 & 0.342 & 4.7 $\pm$ 0.65 & 144.792 & 66.417 & 2 & 0.233 & 41.1 & 1.9 & 8.0\\
  PSZ2 G146.82+40.97 & 144.185 & 65.975 & 0.259 & 4.49 $\pm$ 0.27 & 144.185 & 65.975 & 1 & 0.223 & 49.84 & 2.7 & 6.0\\
  PSZ2 G147.17+42.67 & 147.452 & 64.922 & 0.46 & 5.65 $\pm$ 0.72 & 147.504 & 64.925 & 2 & 0.364 & 34.3 & 2 & 18.0\\
  PSZ2 G150.24+48.72 & 155.851 & 59.808 & 0.205 & 3.56 $\pm$ 0.42 & 155.84 & 59.81 & 2 & 0.18 & 59.45 & 3.0 & 3.9\\
  PSZ2 G152.40+75.00 & 183.315 & 39.858 & 0.453 & 5.14 $\pm$ 0.7 & 183.315 & 39.858 & 1 & 0.372 & 34.59 & 2.1 & 18.0\\
  PSZ2 G152.47+42.11 & 142.465 & 61.658 & 0.9 & 6.58 $\pm$ 0.77 & 142.465 & 61.658 & 2 & 0.945 & 25.67 & 3.0 & 145.0\\
  PSZ2 G153.29+36.56 & 130.678 & 62.676 & 0.65 & 6.32 $\pm$ 1.27 & 130.678 & 62.676 & 2 & 0.574 & 28.86 & 2.3 & 48.0\\
  PSZ2 G153.57+36.26 & 129.942 & 62.515 & 0.132 & 3.37 $\pm$ 0.39 & 129.855 & 62.522 & 2 & 0.181 & 85.2 & 6.3 & 3.0\\
  PSZ2 G159.86+42.57 & 139.879 & 56.346 & 0.27 & 4.61 $\pm$ 0.57 & 139.917 & 56.366 & 2 & 0.442 & 48.36 & 4.9 & 12.0\\
  PSZ2 G163.87+48.54 & 148.201 & 51.888 & 0.214 & 3.62 $\pm$ 0.43 & 148.2 & 51.888 & 2 & 0.215 & 57.51 & 3.4 & 4.8\\
  PSZ2 G165.76+31.15 & 119.542 & 52.599 & 0.259 & 4.43 $\pm$ 0.66 & 119.496 & 52.638 & 2 & 0.44 & 49.84 & 5.2 & 11.5\\
  PSZ2 G165.95+41.01 & 135.723 & 52.202 & 0.062 & 1.79 $\pm$ 0.26 & 135.756 & 52.219 & 2 & 0.095 & 167.36 & 12.7 & 1.2\\
  PSZ2 G177.03+32.64 & 123.353 & 43.265 & 0.511 & 6.07 $\pm$ 0.76 & 123.336 & 43.226 & 2 & 0.873 & 32.39 & 4.4 & 50.0\\
  PSZ2 G178.00+42.32 & 136.684 & 43.136 & 0.237 & 3.97 $\pm$ 0.54 & 136.74 & 43.084 & 2 & 0.371 & 53.22 & 5.0 & 9.0\\
  PSZ2 G178.94+56.00 & 154.948 & 41.01 & 0.092 & 2.23 $\pm$ 0.3 & 154.954 & 40.976 & 2 & 0.219 & 116.79 & 14.2 & 3.1\\
  PSZ2 G184.24+43.69 & 138.602 & 38.602 & 0.397 & 5.41 $\pm$ 0.63 & 138.604 & 38.59 & 2 & 0.612 & 37.39 & 4.1 & 25.0\\
  PSZ2 G185.08+34.02 & 126.54 & 36.85 & 0.365 & 5.41 $\pm$ 0.66 & 126.492 & 36.874 & 2 & 0.865 & 39.41 & 6.4 & 32.0\\
  PSZ2 G189.23+20.55 & 112.196 & 29.673 & 0.398 & 5.46 $\pm$ 0.74 & 112.196 & 29.673 & 2 & 0.683 & 37.34 & 4.5 & 28.0\\
  PSZ2 G202.66+66.98 & 166.814 & 28.796 & 0.483 & 5.28 $\pm$ 0.7 & 166.814 & 28.796 & 1 & 0.549 & 33.38 & 2.9 & 29.0\\
\end{longtable}
 \tablefoot{We derived ${\rm RA_{\rm inj}, \; DEC_{\rm inj} }$ based on optical and X-ray images; when the cluster centre was ambiguous, we assumed the coordinates from \textit{Planck}. Flux densities and radio powers are integrated up to $3r_{\rm e,inj}$.}

\end{appendix}

\end{document}